\input amstex
\input amsppt.sty

\TagsOnRight \NoBlackBoxes

\define\A{\Bbb A}
\define\Y{\Bbb Y}
\define\Z{\Bbb Z}
\define\C{\Bbb C}
\define\R{\Bbb R}

\define\PP{\bold P}

\define\al{\alpha}
\define\be{\beta}
\define\ga{\gamma}
\define\Ga{\Gamma}
\define\de{\delta}
\define\La{\Lambda}
\define\la{\lambda}
\define\si{\sigma}
\define\th{\theta}
\define\epsi{\varepsilon}
\define\om{\omega}
\define\ze{\zeta}

\define\w{\tfrac{\xi}{\xi-1}}

\define\X{\frak X}
\define\M{\frak M}
\define\x{\frak X}
\define\y{\frak Y}

\define\ti{\tilde}
\define\wt{\widetilde}
\define\wh{\widehat}
\define\tht{\thetag}

\define\Prob{\operatorname{Prob}}
\define\supp{\operatorname{supp}}

\define\const{\operatorname{const}}

\define\Conf{\operatorname{Conf}}
\define\sgn{\operatorname{sgn}}

\define\up{\uparrow}
\define\down{\downarrow}

\define\unX{\underline X\,}
\define\unK{\underline K\,}
\define\hatK{\underline{\wh K}\,}

\define\g{{\text{gamma}}}
\define\m{{\text{Meixner}}}

\define\bxi{\xi(\cdot)}

\topmatter
\title Markov processes on partitions
\endtitle

\author
Alexei Borodin and Grigori Olshanski
\endauthor
\abstract We introduce and study a family of Markov processes on
partitions. The processes preserve the so-called z-measures on
partitions previously studied in connection with harmonic
analysis on the infinite symmetric group. We show that the
dynamical correlation functions of these processes have
determinantal structure and we explicitly compute their
correlation kernels. We also compute the scaling limits of the
kernels in two different regimes. The limit kernels describe the
asymptotic behavior of large rows and columns of the
corresponding random Young diagrams, and the behavior of the
Young diagrams near the diagonal.

Our results show that recently discovered analogy between random
partitions arising in representation theory and spectra of
random matrices extends to the associated time--dependent
models.
\endabstract

\endtopmatter
\document

\head Introduction
\endhead

In a series of papers (see \cite{BO1}, \cite{Ol2}, references
therein, and also \cite{BO5}) we have been studying a remarkable
family of probability distributions on partitions (equivalently,
Young diagrams) called {\it z-measures}. These objects have a
representation theoretic origin, they arise in harmonic analysis
on the infinite symmetric group, see \cite{KOV1}, \cite{KOV2}.
Surprisingly enough, the z-measures turned out to be related to
a number of probabilistic models of random matrix theory,
stochastic growth, random tilings, percolation theory, etc. In
this paper, we introduce and study a family of Markov processes
on partitions which preserve the z-measures. Our main result is
the computation of the dynamical correlation functions for these
Markov processes. We also compute the scaling limits of the
correlation functions corresponding to two different limit
regimes as the size of partitions tends to infinity. In the
first regime we look at the largest rows and columns of the
random Young diagram\footnote{This limit regime has a
representation theoretic meaning, see \cite{BO2}, \cite{Ol2}.}
while in the second one we focus on the boundary of the random
Young diagram near the diagonal.

Examples of dynamical models of random matrix type are well known.
The sources of dynamics may be very different: in the Gaussian
random matrix ensembles one allows the matrix elements to evolve
according to the stationary Ornstein--Uhlenbeck process (Dyson's
Brownian motion \cite{Dy}), in tiling models one reads the
two--dimensional picture section by section \cite{Jo2},
\cite{Jo5}, \cite{Jo6}, \cite{OkR}, in growth models the time
parameter is present from the very beginning \cite{PS}.

In our setting the construction of dynamics is different; it is
based on representation theory. We heavily rely on the fact that
the z-measures define characters of the infinite symmetric group
and thus possess a special {\it coherency property}. It reflects
the consistency of restrictions of a character of the infinite
symmetric group to various finite subgroups. The resulting
Markov processes are analogous to those arising in other models,
and in degenerations they even coincide with some of those, see
\cite{BO7}. It is rather surprising that the similarity of the
z-measures to measures of different origin extends to dynamics
associated with those models.

One of the elements of our construction is a special family of
birth and death processes associated with Meixner orthogonal
polynomials. Such birth and death processes, among many others,
were extensively studied by Karlin--McGregor \cite{KMG1},
\cite{KMG2}. Certain degenerations of our Markov processes admit
a natural description in the language of Karlin--McGregor, see
\S7.5 below.

Let us now describe our results in more detail.

Let $\Y$ denote the set of all Young diagrams. We consider a
family $M_{z,z',\xi}$ of probability measures on $\Y$ which
depend on two complex parameters $z$ and $z'$ and a real
parameter $\xi\in(0,1)$. The weight of a Young diagram $\la$
with respect to $M_{z,z',\xi}$ is given by
$$
M_{z,z',\xi}(\la)= (1-\xi)^{zz'}\,\xi^{|\la|}\,(z)_\la(z')_\la\,
\left(\frac{\dim\la}{|\la|!}\right)^2\,.
$$
Here
$$
(z)_\la=\prod_{(i,j)\in\la}(z+j-i)
$$
(product over the boxes of $\la$) is the generalized Pochhammer
symbol, and $\dim \la$ is the dimension of the irreducible
representation of the symmetric group of degree $|\la|$
associated to $\la$. In order for $M_{z,z',\xi}(\la)$ to be
nonnegative for all $\la\in\Y$, we need to impose certain
restrictions on $z$ and $z'$, for instance, $z'=\bar{z}$. All
possibilities for $(z,z')$ are given before Proposition 1.2
below.

Let $\Y_n$ denote the set of all Young diagrams with $n$ boxes.
Restricting $M_{z,z',\xi}$ to $\Y_n\subset \Y$ and renormalizing
it, we obtain a probability measure $M_{z,z'}^{(n)}$ on $\Y_n$,
which does not depend on $\xi$. The measure $M_{z,z',\xi}$ may
be viewed as a mixture of the finite level measures
$M_{z,z'}^{(n)}$.

The Markov processes that we are about to construct, are jump
processes with countable state space $\Y$ and continuous time
$t\in\R$. The jumps are of two types: one either adds a box to
the random Young diagram, or one removes a box from the diagram.

The event of adding or removing a box is governed by a birth and
death process $N_{c,\xi}(t):=|\la(t)|$ on $\Z_+$. This process
depends on $\xi$ and the product $c=zz'$, and its jump rates are
given by
$$
\align \Prob\{N_{c,\xi}(t+dt)=n+1&\mid N_{c,\xi}(t)=n\}
=\frac{\xi(c+n)dt}{1-\xi}\,,\\
\Prob\{N_{c,\xi}(t+dt)=n-1&\mid N_{c,\xi}(t)=n\}=\frac{n
\,dt}{1-\xi}\,.
\endalign
$$
This is special case of the birth and death processes considered
in \cite{KMG2}. Its invariant distribution, the so--called
negative binomial distribution, is the weight function for the
Meixner orthogonal polynomials.

Conditioned on the jump $n\to n+1$, the choice of the box
$(i,j)$ to be added to $\la$ is made according to the {\it
transition probabilities}
$$
p^\up(n,\la;n+1,\nu) = \frac{(z+j-i)(z'+j-i)
\dim\nu}{(zz'+n)(n+1)\dim\la}\,,
$$
and conditioned on the jump $n\to n-1$, the choice of the box
$(i,j)$ to be removed from $\la$ is made according to the {\it
cotransition probabilities}
$$
p^\down(n,\la;n-1,\mu)=\dfrac{\dim\mu}{\dim\la}\,.
$$

The transition and cotransition probabilities are naturally
associated with finite level measures $M_{z,z'}^{(n)}$. These
probabilities were introduced in \cite{VK} in the context of
general characters of the infinite symmetric group (see also
\cite{Ke2}).

The jump rates $\la\nearrow \nu$ and $\la\searrow \mu$ correctly
define a stationary Markov process $\La_{z,z',\xi}(t)$ on $\Y$.
The measure $M_{z,z',\xi}$ is the invariant measure for this
process. Moreover, $\La_{z,z',\xi}$ is reversible. In the
degenerate case of $z$ or $z'$ being an integer,
$\La_{z,z',\xi}$ can be interpreted in terms of finitely many
independent birth and death processes subject to a
nonintersection condition, see \S7.5 below.

One can also construct Markov chains which preserve the finite
level measures $M_{z,z'}^{(n)}$. The key idea is that finite
level measures are preserved by transition and cotransition
probabilities. Thus, adding a random box and removing a random
box afterwards leaves $M_{z,z'}^{(n)}$ invariant. Alternatively,
one can first remove a box and then add a box. These two
procedures yield two different Markov chains. They were
suggested by Kerov a long time ago (unpublished). The same idea
was independently exploited by Fulman \cite{Fu}. It should be
noted that our methods based on determinantal point processes
are not directly applicable to such Markov chains. The idea of
mixing all finite level measures together\footnote{which may be
viewed as a passage to the {\it grand canonical ensemble}, cf.
\cite{Ve}} is essential for us, it allows us to obtain explicit
formulas for dynamical correlation functions, as we explain
below.

It is well known that Young diagrams can be viewed as infinite
subsets (point configurations) in a one-dimensional lattice.
This parametrization of Young diagrams turns out to be very
useful.

Let $\Z'$ be the lattice of (proper) half--integers
$$
\Z'=\Z+\tfrac12=\{\dots,-\tfrac52,-\tfrac32,-\tfrac12,
\,\tfrac12,\,\tfrac32,\,\tfrac52,\dots\}.
$$
For any $\la\in\Y$ we set
$$
\unX(\la)=\{\la_i-i+\tfrac12\mid i=1,2,\dots\}\subset\Z'.
$$
For instance, for the empty diagram $\la=\varnothing$,
$\unX(\varnothing)=\{\dots,-\tfrac52,-\tfrac32,-\tfrac12\}$.
Using the correspondence $\la\mapsto \unX(\la)$ we interpret the
measure $M_{z,z',\xi}$ on $\Y$ as a probability measure on
$2^{\Z'}$. This makes it possible to speak about the {\it
dynamical correlation functions} of $\La_{z,z',\xi}$ which
uniquely determine the process. They are defined by
$$
\rho_n(t_1,x_1;t_2,x_2;\dots;t_n,x_n)=\Prob\left\{\unX(\la)\text{
at time $t_i$ contains } x_i \text{  for   } 1\le i\le
n\right\}.
$$
Here $n=1,2,\dots$, and the $n$th correlation function $\rho_n$
is a function of $n$ pairwise distinct arguments
$(t_1,x_1),\dots(t_n,x_n)\in \R\times\Z'$.

The notion of the dynamical correlation functions is a hybrid of
the finite-dimensional distributions of a stochastic process and
standard correlation functions of probability measures on point
configurations.

The reason why we are interested in dynamical correlation
functions is the same as in the ``static'' (fixed time) case: As
we take scaling limits of our processes, the notion of weight of
a point configuration ceases to make any sense because the space
of relevant point configurations becomes uncountable. On the
other hand, the scaling limits of the correlation functions do
exist, and they carry complete information about the asymptotic
behavior of our processes.

\proclaim{Theorem A (Part 1)} The dynamical correlation
functions of $\La_{z,z',\xi}$ have the determinantal form
$(n=1,2,\dots)$
$$
\rho_n(t_1,x_1;\dots;x_n,t_n)=
\det\bigl[\unK_{z,z',\xi}(t_i,x_i;t_j,x_j)\bigr]_{i,j=1}^n,
$$
where the correlation kernel $\unK_{z,z',\xi}(s,x;t,y)$ is a
function on $(\R\times\Z')^2$ which can be explicitly computed.

One way of writing the kernel is by a double contour integral
$$
\aligned &\unK_{z,z',\xi}(s,x;t,y)\\
&=\frac{e^{\frac
12(s-t)}\Ga(-z'-x+\frac12)\Ga(-z-y+\frac12)(-1)^{x+y+1}}
 {\bigl(\Ga(-z-x+\frac12)\Ga(-z'-x+\frac12)
 \Ga(-z-y+\frac12)\Ga(-z'-y+\frac12)\bigr)^{\frac12}}\\
 &\times\frac{1-\xi}{(2\pi
 i)^2}\oint\limits_{\{\om_1\}}\oint\limits_{\{\om_2\}}
\left(1-\sqrt{\xi}\om_1\right)^{-z'}\left(1-\sqrt{\xi}\,\om_1^{-1}\right)^{z}
\left(1-\sqrt{\xi}\om_2\right)^{-z}\left(1-\sqrt{\xi}\,\om_2^{-1}\right)^{z'}\\
&\qquad\qquad\qquad\qquad\times
\frac{\om_1^{-x-\frac12}\om_2^{-y-\frac12}\,d\om_1d\om_2}{e^{s-t}
\left(\om_1-\sqrt{\xi}\right)\left(\om_2-\sqrt{\xi}\right)-
\left(1-\sqrt{\xi}\om_1\right)\left(1-\sqrt{\xi}\om_2\right)} \,
\endaligned
$$
with the contours $\{\om_1\}$ and $\{\om_2\}$ of $\omega_1$ and
$\omega_2$ satisfying the following conditions:

$\bullet$ $\{\om_1\}$ and $\{\om_2\}$ go around 0 in positive
direction and pass between  $\sqrt{\xi}$ and $1/\sqrt{\xi}$;

$\bullet$ The contours are chosen so that the denominator in the
formula above does not vanish. There are two possibilities of
doing that; one of them is used for $s\ge t$, and the other one
is used for $s<t$, see Theorem 7.1 below for details.
\endproclaim

This integral representation is convenient for computing the
scaling limits of the correlation functions. However, it does
not reveal important structural features of the kernel. Let us
now present another way of writing the correlation kernel.

Consider a second order difference operator $D$ on $\Z'$,
depending on parameters $(z,z',\xi)$ and acting on functions
$f(\,\cdot\,)\in\ell^2(\Z')$  as follows
$$
\gather (Df)(x)=\sqrt{\xi(z+x+\tfrac12)(z'+x+\tfrac12)}\,f(x+1)\\
+\sqrt{\xi(z+x-\tfrac12)(z'+x-\tfrac12)}\,f(x-1)
-(x+\xi(z+z'+x))\,f(x).
\endgather
$$
This is a self-adjoint operator with discrete simple spectrum
$\operatorname{Sp} D=\{(1-\xi)\Z'\}$. Its eigenfunctions
$\psi_a$,
$$
D\psi_a=(1-\xi)a\cdot\psi_a,
$$
are explicitly written through the Gauss hypergeometric
function, see \tht{2.1} below. We normalize them by the
condition $\Vert \psi_a\Vert=1$.

\proclaim{Theorem A (Part 2)} The correlation kernel for the
dynamical correlation functions of the Markov process
$\La_{z,z',\xi}$ can also be written as
$$
\unK_{z,z',\xi}(s,x;t,y)=\pm\sum_{a=\frac 12,\frac 32,\frac
52,\dots}e^{- a|s-t|} \psi_{\pm a}(x)\,\psi_{\pm a}(y)
$$
with ``$+$'' taken for $s\ge t$ and ``$-$'' taken for $s<t$.
\endproclaim

The functions $\{\psi_a\}$ form an orthonormal basis in
$\ell^2(\Z')$. Thus, for $s=t$ the kernel $\unK_{z,z',\xi}$
defines a projection operator whose range is the span of the
eigenfunctions of $D$ corresponding to the positive part of the
spectrum of $D$. In this case (see Comments at the end of \S3)
the kernel can be written in a simpler, so-called {\it
integrable form\/}:
$$
\unK_{z,z',\xi}(x,y)=\frac{P(x)Q(y)-Q(x)P(y)}{x-y}\,,
$$
where $P$ and $Q$ are expressed through the Gauss hypergeometric
function.

The formula of Theorem A (Part 2) shows that our Markov process
is determined by the following data: a state space $\X$, a
Hilbert space $H$ of functions on $\X$, a self-adjoint operator
$D$ in $H$, and two complementary spectral projection operators
$P_\pm$ for $D$. In our case, $\X=\Z'$, $H=\ell^2(\Z')$, $D$ is
the difference operator given above, and $P_\pm$ are projections
on the positive and negative parts of the spectrum of $D$.

It seems that generating  Markov processes with determinantal
correlation functions by data $(\X,H,D,P_\pm)$ of this type is a
rather general phenomenon. Similar structures have appeared
earlier in the dynamics arising in polynuclear growth models
\cite{PS}, \cite{Jo3}, in tiling models \cite{Jo5}, \cite{Jo6},
and in random matrix theory \cite{NF}, \cite{Jo4}, \cite{TW}.
Following the terminology of those papers, we call the kernel of
Theorem A the {\it extended hypergeometric kernel}.

The reader might notice that in our Theorem A as well as in all
the papers cited above, the values of the extended (dynamical)
kernels are always given by somewhat different expressions
depending on the relative order of the time variables. This
dichotomy is unavoidable because of a certain discontinuity of
the dynamical correlation functions. For example, we must have
$$
s\approx t \; \Rightarrow \; \rho_2(s,x;t,y)\approx \cases
\rho_2(s,x;s,y), & x\ne y,\\
\rho_1(s,x), & x=y. \endcases
$$
If we assume the determinantal structure of the dynamical
correlation functions with a kernel $K(s,x;t,y)$ then
$$
\rho_1(s,x)=K(s,x;s,x),\qquad \rho_2(s,x;t,y)=\left|\matrix
K(s,x;s,x)&K(s,x;t,y)\\K(t,y;s,x)&K(t,y;t,y)\endmatrix\right|.
$$
If we further assume that the kernel is continuous in $s,t$
subject to the condition $s\ge t$,\footnote{We could have used
$s\le t$ equally well, this is a question of convention. For
instance, transposition of the kernel does not affect the
correlation functions, and this operation turns $s\ge t$ into
$s\le t$.} then the above relations imply
$$
K(s,x;s,y)=\lim_{\epsilon\to+0}
K(s,x;s-\epsilon,y)=\lim_{\epsilon\to+0}
K(s,x;s+\epsilon,y)+\delta_{xy}.
$$
The validity of the last relation for the extended
hypergeometric kernel can be immediately observed from Part 2 of
Theorem A using the fact that $\{\psi_a\}$ form an orthonormal
basis.

Let us now describe our results on scaling limits of the
dynamical correlation functions. In our previous works we
considered three asymptotic regimes for random Young diagrams
without dynamics: one for largest rows and columns, one for rows
and columns of intermediate growth, and one for the behavior of
the boundary of the Young diagrams near the diagonal, see
\cite{BO5} and references therein. In all three limit regimes
the parameter $\xi$ tends to 1, which makes the expected number
of boxes in the random Young diagram go to infinity.

In this paper we concentrate on the first and the third limit
regime, but with the presence of dynamics. Let us start with the
behavior of large rows and columns.

In order to catch the largest rows and columns in the limit
$\xi\nearrow 1$, we need to scale them by $(1-\xi)$. This leads
to scaling of the state space $\Z'$ by the same factor. That is,
$\Z'$ is replaced by $(1-\xi)\Z'$ which in the limit turns into
$\R^*=\R\setminus \{0\}$.

The parametrization of Young diagrams by point configurations
$\unX(\la)$ is not suitable for this limit transition. Or,
rather to say, the positive part of $\unX(\la)$ indeed reflects
the behavior of largest rows, while the behavior of the largest
columns is captured by the complement of the negative part of
$\unX(\la)$ in $\{\dots,-\frac52,-\frac 32,-\frac 12\}$. Thus,
instead of encoding $\la$ by $\unX(\la)$ we use the map
$$
\la\mapsto X(\la)=\Bigl(\unX(\la)\cap \{\tfrac 12,\tfrac
32,\tfrac 52,\dots\}\Bigr)\cup\Bigl(\{\dots,-\tfrac52,-\tfrac
32,-\tfrac12\} \setminus \unX(\la)\Bigr).
$$
We refer the reader to \cite{BO2}, \cite{Ol2} for representation
theoretic interpretation of this map and for further details.

\proclaim{Theorem B} The scaling limits, as $\xi\to 1$, of the
dynamical correlation functions of $\La_{z,z',\xi}$
corresponding to the map $\la\mapsto X(\la)$, under the
rescaling of $\Z'$ by $(1-\xi)$, have determinantal form with
the correlation kernel $K^W_{z,z'}(s,u;t,v)$ on $(\R\times
\R^*)^2$. This kernel has four blocks according to the choices
of signs of $u$ and $v$.

 The block with $u,v>0$ has an integral representation
$$
\multline
 K^W_{z,z'}(s,u;t,v)=e^{\pi i(z+z')}(u/v)^{\frac
{z-z'}2}e^{\frac 12(s-t)}\\
\times\frac 1{(2\pi i)^2}\oint\limits_{+\infty}^{0-}
\oint\limits_{+\infty}^{0-}\zeta_1^{-z'}(1+\zeta_1)^z\zeta_2^{-z}(1+\zeta_2)^{z'}
\frac{e^{-u(\zeta_1+\frac12)-v(\zeta_2+\frac12)}\,d\zeta_1
d\zeta_2} {e^{s-t}(1+\zeta_1)(1+\zeta_2)-\zeta_1\zeta_2}
\endmultline
$$
with different choices of contours for $s\ge t$ and $s<t$, see
Theorem 9.4 below.

The same block has a series representation
$$
K^W_{z,z'}(s,u;t,v)=\pm\sum_{a=\frac 12,\frac 32,\frac
52,\dots}e^{-a|s-t|} w_{\pm a}(u)\,w_{\pm a}(v),
$$
where ``$+$'' is taken for $s\ge t$, ``$-$'' is taken for $s<t$,
and
$$
w_a(u)=\lim_{\xi\to 1}\,(1-\xi)^{\frac
12}\psi_a\bigl([(1-\xi)^{-1}u]\bigr)
$$
are eigenfunctions of a second order differential operator on
$\R_+$:
$$
uw_a''(u)+w_a'(u)+\left(-\tfrac
u4+\tfrac{z+z'}2-\tfrac{(z-z')^2}{4u}\right)w_a(u)=aw_a(u),
$$
which are explicitly written through the Whittaker functions,
see \tht{9.1} below.

Similar expressions are available for three other blocks of
$K^W_{z,z'}(s,u;t,v)$, see Theorems 9.2 and 9.4 below.
\endproclaim

We call $K^W_{z,z'}(s,u;t,v)$ the {\it extended Whittaker
kernel}.

In the ``static'' case $s=t$ the kernel admits a simpler
``integrable'' form, see \cite{BO1}, \cite{BO2}, \cite{B1},
\cite{Ol2}, and \tht{9.4} below.

Let us now proceed to the other limit regime which describes the
behavior of the Young diagrams near the diagonal. This just means
that we stay on the lattice $\Z'$. For this asymptotic regime it
does not really matter whether we use $\unX(\la)$ or $X(\la)$ to
encode the Young diagrams. We refer to \cite{BO5} for a detailed
discussion of this regime.

In the following statement we will use a more detailed notation
$w_a(u;z,z')$ for the functions $w_a(u)$ introduced above.

\proclaim{Theorem C} The limits, as $\xi\to 1$, of the dynamical
correlation functions of $\La_{z,z',\xi}$ corresponding to the
map $\la\mapsto \unX(\la)$, under the rescaling of time by
$(1-\xi)^{-1}$, have determinantal form with the correlation
kernel $\unK^{\g}_{z,z'}(\sigma,x;\tau,y)$ on $(\R\times
\Z')^2$.

For $\sigma\ge \tau$, the correlation kernel can be written in
two different ways: as a double contour integral
$$
\multline
\unK^{\g}_{z,z'}(\sigma,x;\tau,y)\\=
 \frac{\Ga(-z'-x+\frac12)\Ga(-z-y+\frac12)e^{-\pi i(z+z')}(-1)^{x+y+1}}
 {\bigl(\Ga(-z-x+\frac12)\Ga(-z'-x+\frac12)
 \Ga(-z-y+\frac12)\Ga(-z'-y+\frac12)\bigr)^{\frac12}}\\
\times\frac 1{(2\pi i)^2}\oint\limits_{+\infty}^{0-}
\oint\limits_{+\infty}^{0-}
\frac{\zeta_1^{z'+x-\frac12}(1+\zeta_1)^{-z-x-\frac12}
\zeta_2^{z+y-\frac12}(1+\zeta_2)^{-z'-y-\frac12}\,d\zeta_1
d\zeta_2} {1+(\sigma-\tau)+\zeta_1+\zeta_2}
\endmultline
$$
and as a single integral
$$
\unK^{\g}_{z,z'}(\sigma,x;\tau,y)=\int_0^{+\infty}
e^{-u(\sigma-\tau)}w_x(u;-z,-z') w_y(u;-z,-z')du.
$$
The values of the kernel for $\sigma< \tau$ are obtained from
the above formulas using the symmetry property
$$
\unK^{\g}_{z,z'}(\sigma,x;\tau,y)
=(-1)^{x+y+1}\unK^{\g}_{-z,-z'}(\tau,-x;\sigma,-y),\qquad
\sigma\ne \tau.
$$
\endproclaim

For $\sigma=\tau$ the kernel admits a simpler expression of
``integrable'' type
$$
\frac{P(x)Q(y)-Q(x)P(y)}{x-y}
$$
where $P$ and $Q$ are are expressed through gamma functions
only, see \cite{BO5} and \tht{10.3} below. That kernel was
called the gamma kernel, and for this reason we call
$K^{\g}_{z,z'}(\sigma,x;\tau,y)$ the {\it extended gamma
kernel}.

Note that the extended gamma kernel fits into the same abstract
scheme as the extended hypergeometric kernel: one takes
$\X=\Z'$, $H=\ell^2(\Z')$, $D$ is the special case of the
difference operator given above corresponding to the limit value
$\xi=1$. The spectrum of this operator fills the whole real
axis, the eigenfunctions are $x\mapsto w_x(u;-z,-z')$, and the
spectral projections $P_\pm$ again correspond to the positive
and negative parts of the spectrum.

The functions $\psi_a(x)=\psi_a(x;z,z',\xi)$ used in the
discussion of the extended hypergeometric kernel have the
following symmetry:
$$
\psi_a(x;z,z',\xi)=\psi_x(a;-z,-z',\xi).
$$
This means, in particular, that $\psi_a(x)$ satisfies second
order difference equations both in $a$ and $x$ (the {\it
bispectrality property\/}, see \cite{Gr}). The two limit
transitions considered above (Theorems B and C) correspond to
taking continuous limits in $x$ and $a$, respectively. This
explains why we end up with the same functions $w_a(u)$ in
Theorems B and C.

 Let us make
some remarks about our proof of Theorem A. As a matter of fact,
we prove the theorem in a greater generality. We introduce
certain {\it time inhomogeneous} Markov processes on partitions.
Their fixed time distributions are also the measures
$M_{z,z',\xi}$, but now $\xi=\xi(t)$ varies with time $t$. The
construction of these processes is similar to the stationary
ones except that the birth and death process on $\Z_+$ becomes
time inhomogeneous. In particular, we consider pure birth and
pure death processes for which the Young diagrams either always
gain new boxes or always lose their boxes. These ``pure''
processes are simpler, their transition probabilities can be
evaluated explicitly. They can also be viewed as building blocks
of general processes, more exactly, the transition matrix
$P(s,t)$ for a general process can be represented as a product
$P(s,t)=P^\down(s,u)P^\up(u,t)$ of transition matrices of
``pure'' processes for a suitable intermediate time moment
$u\in(s,t)$.

This product representation of the transition matrix plays an
important role in the proof of Theorem A. We first prove the
theorem for a degenerate case, when one of the parameters $z,z'$
is an integer, and the process is ``finite-dimensional'', that
is, it lives on the Young diagrams with bounded number of rows
or columns. Then the needed formulas are derived from a version
of Eynard-Mehta theorem on spectral correlations of coupled
random matrices \cite{EM}.\footnote{Other proofs of this theorem
can be found in \cite{NF}, \cite{Jo3}, \cite{TW}, \cite{BR}.}
The passage from the degenerate case to the general one is based
on analytic continuation in the parameters $z$ and $z'$. This
passage is not trivial since we need to extrapolate from the
integer points to a complex domain. The needed analytic
properties of the dynamical correlation functions are derived
from the product formula for the transition matrix $P(s,t)$
mentioned above. Let us also emphasize that in our approach, the
introduction of time inhomogeneous processes is necessary for
handling the stationary case.

There is one more subtle issue that we would like to mention
here. Generally speaking, even for birth and death processes,
jump rates do not determine the transition matrix uniquely, see
e.g., \cite{Fe2, ch. XVII, \S10}. Since we want to define our
processes by their jump rates, we need to ensure the uniqueness.
We were unable to find suitable results in the literature and,
therefore, we were forced to invent a special sufficiency
condition which was suitable for our purposes, see \S4.

Let us point out that there exists another way of obtaining the
dynamical correlation functions of Theorem A, based on the
formalism of infinite wedge Fock space. In \cite{Ok2} Okounkov
gave an elegant derivation of static ($s=t$) correlation functions
(initially computed in \cite{BO2}) using a representation of
$SL(2)$ by the so-called Kerov operators. We can extend Okounkov's
approach to derive the formula of Theorem A. This alternative path
bears some similarity to the formalism of {\it Schur processes\/}
of \cite{OkR}, \cite{Ok3}. However, the Schur processes seem to be
not applicable in our situation. Note also that despite the beauty
of Okounkov's idea, a rigorous realization of this approach would
have to overcome certain nontrivial technical difficulties.

One more important subject that we do not touch upon in this
paper, is a family of Markov processes on partitions related to
Plancherel measures. In the limit $z,z'\to\infty$, $\xi\to 0$,
$zz'\xi\to \theta>0$, the measures $M_{z,z',\xi}$ tend to the
so-called {\it poissonized Plancherel measure\/} on $\Y$ with
Poisson parameter $\theta$. This connection was used in \cite{BOO}
to study the asymptotics of the Plancherel measures. Using the
general scheme presented in this paper, one constructs Markov
processes on $\Y$ which preserve the poissonized Plancherel
measures. These processes may be viewed as degenerations of the
processes considered in this paper. They are equivalent to the
droplet model of polynuclear growth. Our results on this other
family of Markov processes and their scaling limits are presented
in \cite{BO7}. Let us note that the analog of Theorem A for those
processes can be obtained either by limit transition from Theorem
A or by using the Schur process of \cite{OkR}.

The present paper is organized as follows. In Section 1 we
introduce the z-measures, the associated transition and
cotransition probabilities, and other notions related to the
Young graph. In Section 2 we study the eigenfunctions $\psi_a$
of the second order difference operator $D$ on $\Z'$. In Section
3 we prove the static variant of Theorem A using the method of
analytic continuation and reduction to the degenerate case of
integral parameters. In Section 4 we introduce time homogeneous
and inhomogeneous Markov processes on $\Y$, prove their
existence and uniqueness, and compute the transition
probabilities for ``pure'' ascending and descending processes.
In Section 5 we evaluate the transition matrices for integral
values of parameters. In Section 6 we study the analytic nature
of the dependence of the dynamical correlation functions  on the
parameters. In Section 7 we prove Theorem A first in the
degenerate case using Eynard--Mehta theorem and Meixner
polynomials, and then in the general case using analytic
continuation. In Section 8 we derive the dynamical correlation
functions of $\La_{z,z',\xi}$ corresponding to the map
$\la\mapsto X(\la)$ (as opposed to the map $\la\mapsto
\unX(\la)$ used in Theorem A). In Section 9 we prove Theorem B,
and in Section 10 we prove Theorem C.

\subhead Acknowledgements
\endsubhead
This research was partially conducted during the period the
first author (A.~B.) served as a Clay Mathematics Institute
Research Fellow. He was also partially supported by the NSF
grant DMS-0402047. The second author (G.~O.) was supported by
the CRDF grant RM1-2543-MO-03.

\head 1. Z-measures
\endhead

As in Macdonald \cite{Ma} we identify partitions and Young
diagrams. By $\Y_n$ we denote the set of partitions of a natural
number $n$, or equivalently, the set of Young diagrams with $n$
boxes. By $\Y$ we denote the set of all Young diagrams, that is,
the disjoint union of the finite sets $\Y_n$, where
$n=0,1,2,\dots$ (by convention, $\Y_0$ consists of a single
element, the empty diagram $\varnothing$). Given $\la\in\Y$, let
$|\la|$ denote the number of boxes of $\la$ (so that
$\la\in\Y_{|\la|}$),  let $\ell(\la)$ be the number of nonzero
rows in $\la$ (the length of the partition), and let $\la'$
denote the transposed diagram.

For two Young diagrams $\la$ and $\mu$ we write $\mu\nearrow\la$
(equivalently, $\la\searrow \mu$) if $\mu\subset \la$ and
$|\mu|=|\la|-1$, or, in other words, $\mu$ is obtained from
$\la$ by removing one box.

The {\it Young graph\/} is the graph whose vertices are the
elements of $\Y$ and the edges join all pairs $(\mu,\la)$ such
that $\mu\nearrow\la$. The Young graph will also be denoted by
$\Y$. Clearly, $\mu\nearrow\la$ implies $\mu'\nearrow\la'$, so
that the transposition operation $\la\mapsto\la'$ induces an
involutive automorphism of the Young graph.

For any $\la\in\Y_n$,  standard Young tableaux of shape $\la$
can be viewed as paths
$$
\varnothing\nearrow\la^{(1)}\nearrow\dots\nearrow\lambda^{(n)}=\la
$$
in $\Y$. Let $\dim\la$ be the number of all such paths. A
convenient explicit formula for $\dim\la$ is
$$
\dim\la=\frac{n!}{\prod_{i=1}^N(\la_i+N-i)!}\,\prod_{1\le i<j\le
N}(\lambda_i-i-\la_j+j),\qquad \la\in\Y_n,
$$
where $N$ is an arbitrary integer $\ge \ell(\la)$ (the above
expression is stable in $N$).

For $\la\in\Y_n$, $\mu\in\Y_{n-1}$ set
$$
p^\down(n,\la;n-1,\mu)=\cases \dfrac{\dim\mu}{\dim\la},&\mu\nearrow\la,\\
0,&{\text{otherwise}}
\endcases \tag1.1
$$
and note that
$$
\sum_{\mu\in\Y_{n-1}}p^\down(n,\la;n-1,\mu)=1.
$$
The numbers $p^\down(n,\la;n-1,\mu)$ are called the {\it
cotransition probabilities\/} of the Young graph.

A family $\{M^{(n)}\}$ of probability measures $M^{(n)}$ on
$\Y_n$, $n=0,1,2,\dots$, is called a {\it coherent system on
$\Y$} if the measures are consistent with the cotransition
probabilities in the following sense:
$$
M^{(n-1)}(\mu)=\sum_{\la\in\Y_n}M^{(n)}(\la)p^\down(n,\la;n-1,\mu),\qquad
\mu\in\Y_{n-1},\quad n=1,2,\dots\,. \tag1.2
$$
This concept has an important representation theoretic meaning.
Namely, there is a 1--1 correspondence between coherent systems on
$\Y$ and normalized positive definite class functions on the
infinite symmetric group, see  \cite{VK}, \cite{Ke2}, \cite{Ol2}.

Note that the cotransition probabilities are invariant under the
involution $\la\mapsto\la'$ of the Young graph. Consequently,
the push--forward of a coherent system under this involution is
again a coherent system.

\example{Example 1.1} The {\it Plancherel measures\/} defined by
$$
M^{(n)}_{Plancherel}(\la)=\frac{(\dim\la)^2}{n!}
$$
form a coherent family of probability measures, see \cite{VK}.
\endexample

Let $\Cal T$ be the set of all infinite paths in $\Y$ of the
form
$$
\varnothing \nearrow\la^{(1)}\nearrow \la^{(2)}\nearrow
\dots\nearrow\lambda^{(n)}\nearrow\dots,\qquad
\lambda^{(n)}\in\Y_n.
$$
This is a compact topological space (a closed subset of the
product space $\prod_{n\ge 0}\Y_n$).

A probability measure $\Cal M$ on $\Cal T$ is called {\it
central\/} if for any $n=1,2,\dots$ and any $\la\in\Y_n$, the
mass of each cylinder set consisting of all paths with fixed
$\la^{(1)},\dots,\la^{(n)}=\la$ depends on $\la$ only (and does
not depend on $\la^{(1)},\dots,\la^{(n-1)}$).

Any coherent system $\{M^{(n)}\}$ generates a central measure
$\Cal M$ on $\Cal T$. By definition, the mass of the cylinder
set mentioned above equals $M^{(n)}(\la)/\dim\la$. The relation
\tht{1.2} ensures that $\Cal M$ is correctly defined. This
defines a one-to-one correspondence between coherent systems
$\{M^{(n)}\}$ on $\Y$ and central measure $\Cal M$ on $\Cal T$,
see  \cite{VK}, \cite{Ke2}, \cite{Ol2}.

For any central measure $\Cal M$,
$$
p^\down(n,\la;n-1,\mu)=\Prob\{\la^{(n-1)}=\mu\mid
\la^{(n)}=\la\},
$$
which is a justification of the term ``cotransition
probability''.

Assuming $M^{(|\la|)}(\la)>0$ for all $\la\in\Y$, set
$$
p^\up(n,\la;n+1,\nu)=\Prob\{\la^{(n+1)}=\nu\mid \la^{(n)}=\la\},
\qquad n=|\la|.
$$
In contrast to $p^\down(n,\la;n-1,\mu)$, these numbers depend on
$\Cal M$. We call them the {\it transition probabilities} of the
central measure $\Cal M$ (or of the corresponding coherent
system $\{M^{(n)}\}$). The transition probabilities define $\Cal
M$ and $\{M^{(n)}\}$ uniquely.

Note an important relation between the transition and
cotransition probabilities:
$$
M^{(n)}(\la)p^\up(n,\la;n+1,\nu)=p^\down(n+1,\nu;n,\la)M^{(n+1)}(\nu).
\tag1.3
$$
It implies, in particular, that
$$
p^\up(n,\la;n+1,\nu) =\cases
\dfrac{M^{(n+1)}(\nu)\dim\la}{M^{(n)}(\la)\dim\nu}\,, &
\la\nearrow\nu,\\ 0, &\text{otherwise}\,.
\endcases \tag1.4
$$

If $M^{(|\la|)}(\la)$ vanishes for some  $\la\in\Y$ then the
definition has to be slightly modified. Namely, let $\supp\Cal
M$ be the set of those $\la\in\Y$ for which
$M^{(|\la|)}(\la)>0$. Equivalently, $\la\in\supp\Cal M$ if the
set of paths passing through $\la$ has positive mass with
respect to $\Cal M$. Note that $\la\in\supp\Cal M$ implies
$\mu\in\supp\Cal M$ for all $\mu\nearrow\la$. The set $\supp\Cal
M$ spans a subgraph of $\Y$ (which may be called the {\it
support of $\Cal M$\/}), and the transition probabilities are
correctly defined on this subgraph by the same formula
\tht{1.3}. Again, the initial central measure $\Cal M$ is
uniquely determined by its support and the transition
probabilities.

Note two useful equations
$$
\align
M^{(n-1)}(\mu)&=\sum_{\la}M^{(n)}(\la)p^\down(n,\la;n-1,\mu), \tag1.5\\
M^{(n+1)}(\nu)&=\sum_{\la}M^{(n)}(\la)p^\up(n,\la;n+1,\nu).
\tag1.6
\endalign
$$

We shall need the {\it generalized Pochhammer symbol\/}
$(z)_\la$:
$$
(z)_\la=\prod_{i=1}^{\ell(\la)}(z-i+1)_{\la_i}\,, \qquad z\in\C,
\quad \la\in\Y,
$$
where
$$
(x)_k=x(x+1)\dots(x+k-1)=\frac{\Ga(x+k)}{\Ga(x)}
$$
is the conventional Pochhammer symbol. Note that
$$
(z)_\la=\prod_{(i,j)\in\la}(z+j-i)
$$
(product over the boxes of $\la$), which implies at once the
symmetry relation
$$
(z)_\la=(-1)^{|\la|}(-z)_{\la'}.
$$

For two complex parameters $z,z'$ set
$$
M^{(n)}_{z,z'}(\la)=\frac{(z)_\la(z')_\la}{(zz')_n}\,\frac{(\dim\la)^2}{n!}\,,
\qquad n=0,1,\dots, \quad \la\in\Y_n\,, \tag1.7
$$
where $\dim\la$ was defined in the beginning of the section. The
expression \tht{1.7} makes sense if $(zz')_n$ does not vanish,
i.e., if $zz'\notin\{0,-1,-2,\dots\}$.  Obviously, \tht{1.7} is
symmetric in $z,z'$.

Note that (see Example 1.1)
$$
\lim_{z,z'\to\infty}M^{(n)}_{z,z'}(\la)=M^{(n)}_{Plancherel}(\la).
\tag1.8
$$

Let us say that two nonzero complex numbers $z,z'$ form an {\it
admissible\/} pair of parameters if one of the following three
conditions holds:

$\bullet$ The numbers $z,z'$ are not real and are conjugate to
each other.

$\bullet$ Both $z,z'$ are real and are contained in the same
open interval of the form $(m,m+1)$, where $m\in\Z$.

$\bullet$ One of the numbers $z,z'$ (say, $z$) is a nonzero
integer while $z'$ has the same sign and, moreover,
$|z'|>|z|-1$.

\proclaim{Proposition 1.2} If $(z,z')$ is an admissible pair of
parameters then $\{M^{(n)}_{z,z'}\}$ is a coherent family of
probability measures.
\endproclaim

\demo{Proof} It is readily checked that if (and only if) one of
the conditions above holds then $(z)_\la(z')_\la\ge0$ for all
$\la$, see  \cite{BO5, Proposition 1.8}. Moreover, $(zz')_n>0$
for all $n$. Hence \tht{1.7} is nonnegative. The fact that each
$M^{(n)}_{z,z'}$ is a probability measure and the coherency
property can be proved in several ways. See, e.g., \cite{Ol1},
 \cite{BO3}. \qed
\enddemo

We call the measures $M^{(n)}_{z,z'}$ the {\it z--measures\/} on
the floors $\Y_n$ of the Young graph. Depending on which of the
three conditions of Proposition 1.1 holds we will speak about
the {\it principal, complementary\/} or {\it degenerate\/}
series of z--measures, respectively. By virtue of \tht{1.8}, the
z--measures may be viewed as a deformation of the Plancherel
measure (for any fixed $n$). The principal series of z--measures
first appeared in  \cite{KOV1}, see also \cite{KOV2}. For more
information about the z--measures and their generalizations, see
\cite{BO2}, \cite{BO3}, \cite{BO4}, \cite{BO5}, \cite{BO6},
\cite{Ke1}.

Note that the involution $\la\mapsto\la'$ of the Young graph
takes $M^{(n)}_{z,z'}$ to $M^{(n)}_{-z,-z'}$.

Let $\Cal M_{z,z'}$ be the central measure corresponding to the
coherent family $\{M^{(n)}_{z,z'}\}_{n=0,1,\dots}$. In the case
of the principal or complementary series the support of $\Cal
M_{z,z'}$ is the whole $\Y$. For the degenerate series it is a
proper subset of $\Y$: if $z=k=1,2,\dots$ and $z'>k-1$ then
$\supp\Cal M_{z,z'}$ consists of diagrams with at most $k$ rows,
and if $z=-k=-1,-2,\dots$ and $z'<-(k-1)$ then $\supp\Cal
M_{z,z'}$ consists of diagrams with at most $k$ columns.

The transition probabilities of the z--measures are given by
$$
p^\up_{z,z'}(n,\la;n+1,\nu) =\frac{(z+c(\nu/\la))(z'+c(\nu/\la))
\dim\nu}{(zz'+n)(n+1)\dim\la}\,, \qquad \la\nearrow\nu, \tag1.9
$$
where $c(\nu/\la)$ denotes the {\it content\/} of the box
$(i,j)=\nu/\la$, that is, $c=j-i$. Indeed, \tht{1.9} follows
immediately from \tht{1.4} and \tht{1.7}. Note that if $\la$ is
in $\supp\Cal M_{z,z'}$ while $\nu$ is not (which may happen for
the degenerate series) then \tht{1.9} vanishes due to vanishing
of one of the factors $z+c(\nu/\la)$, $z'+c(\nu/\la)$.

For the Plancherel measure, the transition probabilities are
$$
p^\up_{Plancherel}(n,\la;n+1,\nu)
=\frac{\dim\nu}{(n+1)\dim\la}\,, \qquad \la\nearrow\nu,
$$
see \cite{VK}.

Consider a special case of the negative binomial distribution on
$\Z_+$ depending on two parameters $a>0$ and $\xi\in(0,1)$:
$$
\pi_{a,\xi}(n)=(1-\xi)^a \frac{(a)_n\xi^n}{n!}\,, \qquad
n=0,1,2,\dots \tag1.10
$$
The next formula defines a probability measure on $\Y$ which is
the mixture of all z--measures $M^{(n)}_{z,z'}$ with given fixed
parameters $z,z'$ and varying $n$ by means of the distribution
\tht{1.10} on $n$'s, with parameters $a=zz'$ and $\xi$:
$$
M_{z,z',\xi}(\la)=M^{(|\la|)}_{z,z'}(\la)\,\pi_{zz',\xi}(|\la|)
=(1-\xi)^{zz'}\,\xi^{|\la|}\,(z)_\la(z')_\la\,
\left(\frac{\dim\la}{|\la|!}\right)^2\,. \tag1.11
$$
We call \tht{1.11} the {\it mixed\/} z--measure. An
interpretation of formula \tht{1.11} is given in \cite{BO5,
Definition 1.4}.

Likewise, consider a mixture  of the Plancherel measures,
depending on a parameter $\th>0$:
$$
M_{Plancherel, \th}(\la)=M^{(|\la|)}_{Plancherel}(\la)\,e^{-\th}
\frac{\th^{|\la|}}{|\la|!}
=e^{-\th}\,\th^{|\la|}\left(\frac{\dim\la}{|\la|!}\right)^2\,.
\tag1.12
$$
We call \tht{1.12} the {\it poissonized Plancherel measure\/}.
Note that it can be obtained as a limit case of the mixed
z--measures:
$$
\lim\Sb z,z'\to\infty\\ \xi\to0\\zz'\xi\to\th\endSb
M_{z,z',\xi}(\la)=M_{Plancherel, \th}(\la).
$$

The main objects of this paper are the z-measures and related
Markov processes. One can also develop a parallel theory
associated with the Plancherel measure. We do not pursue this
goal in the present paper. An interested reader can found the
statements of the main results related to the Plancherel measure
in our paper \cite{BO7}.

\head 2. A basis in the $\ell^2$ space on the lattice and the
Meixner polynomials \endhead

In this section we examine a nice orthonormal basis in the
$\ell^2$ space on the 1--dimensional lattice. The elements of
this basis are eigenfunctions of a second order difference
operator. They can be obtained from the classical Meixner
polynomials via analytic continuation with respect to
parameters.

Throughout the section we will assume (unless otherwise stated)
that $(z,z')$ is in the principal series or in the complementary
series but not in the degenerate series. In particular, $z,z'$
are not integers.

Consider the lattice of (proper) half--integers
$$
\Z'=\Z+\tfrac12=\{\dots,-\tfrac52,-\tfrac32,-\tfrac12,
\,\tfrac12,\,\tfrac32,\,\tfrac52,\dots\}.
$$
We introduce a family of functions on $\Z'$ depending on a
parameter $a\in\Z'$ and also on our parameters $z,z',\xi$:
$$
\gather
\psi_a(x;z,z',\xi)=\left(\frac{\Ga(x+z+\tfrac12)\Ga(x+z'+\tfrac12)}
{\Ga(z-a+\tfrac12)\Ga(z'-a+\tfrac12)}\right)^{\frac12}
\xi^{\frac12(x+a)}(1-\xi)^{\tfrac12(z+z')-a}\\
\times\frac{F(-z+a+\tfrac12,-z'+a+\tfrac12;x+a+1;\w)}{\Ga(x+a+1)}\,,
\tag2.1
\endgather
$$
where $F(A,B;C;w)$ is the Gauss hypergeometric function.

Let us explain why this expression makes sense. Since, by
convention, parameters $z,z'$ do not take integral values,
$\Ga(x+z+\tfrac12)$ and $\Ga(x+z'+\tfrac12)$ have no
singularities for $x\in\Z'$. Moreover, the admissibilty
assumptions on $(z,z')$ (see \S1) imply that
$$
\Ga(x+z+\tfrac12)\Ga(x+z'+\tfrac12)>0, \qquad
\Ga(z-a+\tfrac12)\Ga(z'-a+\tfrac12)>0,
$$
so that we can take the positive value of the square root in
\tht{2.1}. Next, since $\xi\in(0,1)$, we have $\xi/(\xi-1)<0$,
and as is well known, the function $w\to F(A,B;C,w)$ is well
defined on the negative semi--axis $w<0$. Finally, although
$F(A,B;C,w)$ is not defined at $C=0,-1,-2,\dots$, the  ratio
$F(A,B;C,w)/\Ga(C)$ is well defined for all $C\in\C$.

Note also that the functions $\psi_a(x;z,z',\xi)$ are
real--valued. Their origin will be explained below.

Further, we introduce a second order difference operator
$D(z,z',\xi)$ on the lattice $\Z'$, depending on parameters
$z,z',\xi$ and acting on functions $f(x)$ (where $x$ ranges over
$\Z'$) as follows
$$
\gather D(z,z',\xi)f(x)=\sqrt{\xi(z+x+\tfrac12)(z'+x+\tfrac12)}\,f(x+1)\\
+\sqrt{\xi(z+x-\tfrac12)(z'+x-\tfrac12)}\,f(x-1)
-(x+\xi(z+z'+x))\,f(x).
\endgather
$$
Note that $D(z,z',\xi)$ is a symmetric operator in
$\ell^2(\Z')$.

\proclaim{Proposition 2.1} The functions $\psi_a(x;z,z',\xi)$,
where $a$ ranges over $\Z'$, are eigenfunctions of the operator
$D(z,z',\xi)$,
$$
D(z,z',\xi)\psi_a(x;z,z',\xi)=a(1-\xi)\psi_a(x;z,z',\xi).
\tag2.2
$$
\endproclaim

\demo{Proof} This equation can be verified using the relation
$$
\gather
w(C-A)(C-B)F(A,B;C+1;w)-(1-w)C(C-1)F(A,B;C-1;w)\\
+C[C-1-(2C-A-B-1)w]F(A,B;C;w)=0
\endgather
$$
for the Gauss hypergeometric function, see, e.g., \cite{Er, 2.8
(45)}. \qed
\enddemo

The next lemma  provides us a convenient integral representation
for functions $\psi_a$.

\proclaim{Lemma 2.2} For any $A,B\in\C$, $M\in\Z$, and
$\xi\in(0,1)$ we have
$$
\gathered
\frac{F(A,B;M+1;\w)}{\Ga(M+1)}=\frac{\Ga(-A+1)\xi^{-M/2}(1-\xi)^B}
{\Ga(-A+M+1)}\\
\times\, \frac1{2\pi i}\int\limits_{\{\om\}}
(1-\sqrt\xi\om)^{A-1}\left(1-\frac{\sqrt\xi}{\om}\right)^{-B}
\om^{-M}\,\frac{d\om}{\om}\,.
\endgathered \tag2.3
$$
Here $\xi\in(0,1)$ and $\{\om\}$ is an arbitrary simple contour
which goes around the points 0 and $\sqrt\xi$ in the positive
direction leaving $1/\sqrt\xi$ outside.
\endproclaim

\demo{Comments} 1. The branch of the function
$(1-\sqrt\xi\om)^{A-1}$ is specified by the convention that the
argument of $1-\sqrt\xi\om$ equals 0 for real negative values of
$\om$, and the same convention is used for the function
$\left(1-\frac{\sqrt\xi}{\om}\right)^{-B}$.

2. Like the Euler integral formula, formula \tht{2.3} does not
make evident the symmetry $A\leftrightarrow B$.

3. The right--hand side of formula \tht{2.3} makes sense for
$A=1,2,\dots$, when $\Ga(-A+1)$ has a singularity. Then the
whole expression can be understood, e.g., as the limit value as
$A$ approaches one of the points 1,2,\dots .
\enddemo

\demo{Proof} Since both sides of \tht{2.3} are real--analytic
functions of $\xi$ we may assume that $\xi$ is small enough.
Then we may apply the binomial formula which gives
$$
\gather
\xi^{-M/2}(1-\sqrt\xi\om)^{A-1}\left(1-\frac{\sqrt\xi}{\om}\right)^{-B}
\om^{-M}\\
= \sum_{k=0}^\infty\sum_{l=0}^\infty \frac{(-A+1)_k
(B)_l}{k!\,l!}\,\xi^{(k+l-M)/2}\,\om^{k-l-M}\,.
\endgather
$$
After integration only the terms with $k=l+M$ survive. It
follows that the right--hand side of \tht{2.3} is equal to
$$
\frac{(1-\xi)^B}{\Ga(-A+M+1)}\,\sum_{l\ge\max(0,-M)}
\frac{\Ga(-A+M+1+l)\,(B)_l} {\Ga(l+M+1)l!}\, \xi^l.
$$
We may replace the inequality $l\ge\max(0,-M)$ simply by $l\ge0$
because for negative integral values of $M$ (when we have to
start summation from $l=-M$), the terms with $l=0,\dots,-M-1$
automatically vanish due to the factor $\Ga(l+M+1)$ in the
denominator. Consequently, our expression is equal to
$$
\frac{(1-\xi)^B
F(-A+1+M,B;M+1;\xi)}{\Ga(M+1)}=\frac{F(A,B;M+1;\w)}{\Ga(M+1)}\,,
$$
where we used \cite{Er, 2.9 (4)}. \qed
\enddemo

\proclaim{Proposition 2.3} We have the following integral
representations
$$
\multline \psi_{a}(x;z,z',\xi)\\
=\left(\frac{\Ga(x+z+\tfrac12)\Ga(x+z'+\frac12)}
{\Ga(z-a+\tfrac12)\Ga(z'-a+\frac12)} \right)^\frac
12\, \frac{\Ga(z'-a+\tfrac12)}{\Ga(z'+ x+\tfrac12)} \,(1-\xi)^{\frac{z'-z+1}2}\\
\times\frac1{2\pi
i}\,\oint\limits_{\{\om\}}\left(1-\sqrt{\xi}\om\right)^{-z'+a-\tfrac12}
\left(1-\frac{\sqrt{\xi}}{\om}\right)^{z-a-\tfrac12}\om^{-x-a}
\,\frac{d\om}{\om}
\endmultline\tag2.4
$$
and
$$
\multline \psi_a(x;z,z',\xi)\psi_a(y;z,z',\xi)=\varphi_{z,z'}(x,y)\\
\times\,\frac{1-\xi}{(2\pi i)^2}
\oint\limits_{\{\om_1\}}\oint\limits_{\{\om_2\}}
\left(1-\sqrt{\xi}\om_1\right)^{-z'+a-\tfrac12}
\left(1-\frac{\sqrt{\xi}}{\om_1}\right)^{z-a-\tfrac12}\\
\times\left(1-\sqrt{\xi}\om_2\right)^{-z+a-\tfrac12}
\left(1-\frac{\sqrt{\xi}}{\om_2}\right)^{z'-a-\tfrac12}
\om_1^{-x-a}\om_2^{-y-a}
\,\frac{d\om_1}{\om_1}\frac{d\om_2}{\om_2}
\endmultline \tag2.5
$$
where
$$
\varphi_{z,z'}(x,y)=\frac{\sqrt{\Ga(x+z+\tfrac12)\Ga(x+z'+\tfrac12)
\Ga(y+z+\tfrac12)\Ga(y+z'+\tfrac12)}}
{\Ga(x+z'+\tfrac12)\Ga(y+z+\tfrac12)} \tag2.6
$$
Here each contour is an arbitrary simple loop, oriented in
positive direction, surrounding the points 0 and $\sqrt\xi$, and
leaving $1/\sqrt\xi$ outside. We also use the convention about
the choice of argument as in Comment 1 to Lemma 2.2.
\endproclaim

\demo{Proof} Indeed, \tht{2.4} immediately follows from
\tht{2.1} and \tht{2.3}. To prove \tht{2.5} we multiply out the
integral representation \tht{2.4} for the first function and the
same representation for the second function, but with $z$ and
$z'$ interchanged. The transposition $z\leftrightarrow z'$ in
\tht{2.4} is justified by the fact the initial formula \tht{2.1}
is symmetric with respect to $z\leftrightarrow z'$. As a result
of this trick the gamma prefactors involving $a$ are completely
cancelled out, and we obtain \tht{2.5}\qed
\enddemo

\proclaim{Proposition 2.4} The functions
$\psi_a=\psi_a(x;z,z',\xi)$, where $a$ ranges over $\Z'$, form
an orthonormal basis in the Hilbert space $\ell^2(\Z')$.
\endproclaim

\demo{Proof} {}From \tht{2.4} it is not difficult to see that
the function $\psi_a(x;z,z',\xi)$ has exponential decay as
$x\to\pm\infty$. In particular, it is square integrable. Since
$\psi_a$ is an eigenfunction of a symmetric difference operator
whose coefficients have linear growth at $\pm\infty$, and since
to different indices $a$ correspond different eigenvalues, we
conclude that these functions are pairwise orthogonal in
$\ell^2(\Z')$.

Let us show that $\Vert\psi_a\Vert^2=1$. Write \tht{2.5}, where
we set $x=y$. Then the whole expression simplifies because
\tht{2.6} becomes equal to 1. Next, in the double contour
integral, we replace the variable $\om_2$ by its inverse. We
obtain
$$
\gathered (\psi_{a}(x;z,z',\xi))^2=\frac{1-\xi}{(2\pi i)^2}
\oint\oint\left(1-\sqrt{\xi}\om_1\right)^{-z'+a-\tfrac12}
\left(1-\sqrt{\xi}\,\om_1^{-1}\right)^{z-a-\tfrac12}\\
\times\left(1-\sqrt{\xi}{\om_2}^{-1}\right)^{-z+a-\tfrac12}
\left(1-\sqrt{\xi}\,\om_2\right)^{z'-a-\tfrac12}
\left(\frac{\om_1}{\om_2}\right)^{-x-a}
\,\frac{d\om_1}{\om_1}\frac{d\om_2}{\om_2}
\endgathered
$$
To evaluate the squared norm we have to sum this expression over
$x\in\Z'$. We split the sum into two parts according to the
splitting $\Z'=\Z'_-\cup\Z'_+$. We take as the contours
concentric circles such that $|\om_1|<|\om_2|$ in the sum over
$\Z'_-$, and $|\om_1|>|\om_2|$ in the sum over $\Z'_+$. This
gives us
$$
\sum\limits_{x\in\Z'}(\psi_{a}(x;z,z',\xi))^2=
\underset{|\om_1|<|\om_2|}\to{\oint\oint}\frac{F(\om_1,\om_2)}{\om_2-\om_1}\,
\frac{d\om_1}{\om_1}\frac{d\om_2}{\om_2} +
\underset{|\om_1|>|\om_2|}\to{\oint\oint}\frac{F(\om_1,\om_2)}{\om_1-\om_2}\,
\frac{d\om_1}{\om_1}\frac{d\om_2}{\om_2}
$$
with
$$
\gather F(\om_1,\om_2)=\frac{1-\xi}{(2\pi
i)^2}\,\left(1-\sqrt{\xi}\om_1\right)^{-z'+a-\frac12}
\left(1-\sqrt{\xi}\,\om_1^{-1}\right)^{z-a-\frac12}\\
\times\left(1-\sqrt{\xi}{\om_2}^{-1}\right)^{-z+a-\frac12}
\left(1-\sqrt{\xi}\,\om_2\right)^{z'-a-\frac12}\,
\om_1^{\frac12-a}\om_2^{\frac12+a}
\endgather
$$
Recall that both contours go in positive direction.

Let us transform the second double--contour integral: keeping
the second contour fixed we move the first contour inside the
second contour. Then we obtain a double--contour integral which
is cancelled out with the first double--contour integral, plus a
single--contour integral arising from the residue of the
function $\om_1\to(\om_1-\om_2)^{-1}$:
$$
\frac{1-\xi}{2\pi i}\oint
F(\om,\om)\frac{d\om}{\om^2}=\frac{1-\xi}{2\pi
i}\oint\frac{d\om}{(1-\sqrt\xi\om)(\om-\sqrt\xi)}=1
$$

Thus, we have shown that the functions $\psi_a$ form an
orthonormal family in $\ell^2(\Z')$, and it remains to prove
that this family is complete. For $x\in\Z'$, let $\de_x$ stand
for the delta function at $x$. Since the functions $\de_x$ form
an orthonormal basis, it suffices to check that
$$
\sum_{a\in\Z'}\left((\de_x,\psi_a)_{\ell^2(\Z')}\right)^2 =
\sum_{a\in\Z'}(\psi_a(x;z,z',\xi))^2=1, \qquad \forall x\in\Z'.
$$
But this follows from the previous claim and  the symmetry
$a\leftrightarrow x$ established in the next proposition. \qed
\enddemo

\proclaim{Proposition 2.5} The following symmetry relation holds
$$
\psi_a(x;z,z',\xi)=\psi_x(a;-z,-z',\xi).
$$
\endproclaim

\demo{Proof} Using the classical formula
$$
\Ga(A+\tfrac12)\Ga(A-\tfrac12)=\frac{\pi}{\cos(\pi A)}
$$
and the fact that both $x+\tfrac12$ and $a+\tfrac12$ are
integers we check that
$$
\frac{\Ga(z+x+\tfrac12)\Ga(z'+x+\tfrac12)}
{\Ga(z-a+\tfrac12)\Ga(z'-a+\tfrac12)}
=\frac{\Ga(-z+a+\tfrac12)\Ga(-z'+a+\tfrac12)}
{\Ga(-z-x+\tfrac12)\Ga(-z'-x+\tfrac12)}\,.
$$
Applying this to \tht{2.1} and using another classical formula,
$$
F(A,B;C;w)=(1-w)^{C-A-B}F(C-A,C-B;C;w),
$$
we get the required relation.

Another way is to make a change of the variable in integral
\tht{2.4}:
$$
\om\mapsto \om'=\frac{\om-\sqrt\xi}{\sqrt\xi\om-1}\,.
$$
This is an involutive transformation such that
$0\leftrightarrow\sqrt\xi$ and $\infty\leftrightarrow
1/\sqrt\xi$. As is readily verified, it leads to transformation
$(a,x,z,z')\to(x,a,-z,-z')$. \qed
\enddemo

\proclaim{Corollary 2.6} The functions
$\psi_a=\psi_a(x;z,z',\xi)$ satisfy the following three--term
relation
$$
\gather (1-\xi)x \psi_a
=\sqrt{\xi(z-a+\tfrac12)(z'-a+\tfrac12)}\,\psi_{a-1}\\
+\sqrt{\xi(z-a-\tfrac12)(z'-a-\tfrac12)}\,\psi_{a+1}
+(-a+\xi(z+z'-a))\,\psi_a.
\endgather
$$
\endproclaim

\demo{Proof} Under symmetry $x\leftrightarrow a$ (Proposition
2.4), this turns into the formula stated in Proposition 2.1. Of
course, a direct verification is also possible. \qed
\enddemo

The formulas of Proposition 2.1 and Corollary 2.6 show that the
functions $\psi_a(x;z,z',\xi)$ possess the {\it bispectrality\/}
property in the sense of \cite{Gr}.

\proclaim{Proposition 2.7} One more symmetry relation holds:
$$
\psi_a(x;z,z',\xi)=(-1)^{x+a}\psi_{-a}(-x;-z,-z',\xi), \qquad
x,a\in\Z'.
$$
\endproclaim

\demo{Proof} This follows from the relation
$$
\gather\frac{F(A,B;C;w)}{\Ga(C)} =w^{1-C}\,
\frac{\Ga(A-C+1)\Ga(B-C+1)}{\Ga(A)\Ga(B)}\\
\times\frac{F(A-C+1,B-C+1;2-C;w)}{\Ga(2-C)}\,, \qquad C\in\Z,
\endgather
$$
see \cite{Er, 2.8 (19)}. Another way is to make a change of the
variable, $\om\mapsto1/\om$, in integral \tht{2.4}. \qed
\enddemo

In the remaining part of the section we will explain how the
functions $\psi_a$ are related to the Meixner polynomials.

Let $\Z_+=\{0,1,2,\dots\}$. Elements of $\Z_+$ will be denoted
by symbols $\ti x, \ti y$  (we reserve the notation $x$ and $y$
for points of the lattice $\Z'$). Recall that the Meixner
polynomials are the orthogonal polynomials with respect to the
weight function
$$
W_{\al,\xi}(\ti x)= \frac{(\al+1)_{\ti x}\xi^{\ti x}}{\ti x!}
=\frac{\Ga(\al+1+\ti x)\xi^{\ti x}}{\Ga(\al+1)\ti x!}\,,  \qquad
\wt x\in\Z_+ \tag2.7
$$
on $\Z_+$, where $\al>-1$ and, as before, $\xi\in(0,1)$. Our
notation for these polynomials is $\M_m(\ti x;\al;\xi)$. We use
the same normalization of the polynomials as in the handbook
\cite{KS} (there is only a minor difference in notation: our
parameter $\al$ corresponds to parameter $\be=\al+1$ in
\cite{KS}, while our $\xi$ is precisely parameter $c$ in
\cite{KS}).

Set
$$
\wt\M_n(\ti x;\al,\xi)=(-1)^n\,\frac{\M_n(\ti
x;\al,\xi)}{\Vert\M_n(\,\cdot\,;\al,\xi)\Vert}\,\sqrt{W_{\al,\xi}(\ti
x)}, \qquad \ti x\in\Z_+\,,\tag2.8
$$
where
$$
\Vert\M_n(\,\cdot\,;\al,\xi)\Vert^2 =\sum_{\ti x=0}^\infty
\M_n^2(\ti x;\al,\xi)W_{\al,\xi}(\ti x).
$$
The factor $(-1)^n$ is introduced for convenience: it will
compensate the same factor in formula \tht{2.10} below.

\proclaim{Proposition 2.8} Drop the assumption that $(z,z')$ is
not in the degenerate series, and assume, just on the contrary,
that $z=N$ and $z'=N+\al$, where $N=1,2,\dots$ and $\al>-1$.
Then expression \tht{2.1} for the functions $\psi_a(x;z,z',\xi)$
still makes sense provided that
$$
\ti x:=x+N-\tfrac12\,, \qquad n:=N-a-\tfrac12 \tag2.9
$$
are in $\Z_+$, and in this notation we  have
$$
\psi_a(x;z,z',\xi)=\wt\M_n(\ti x;\al,\xi).
$$
\endproclaim

\demo{Proof} As is well known, the Meixner polynomials can be
expressed through the Gauss hypergeometric function in two
different ways:
$$
\M_n(\ti x;\al,\xi)=F(-n,-\ti x;\al+1;\tfrac{\xi-1}{\xi})
$$
and
$$
\gather \M_n(\ti x;\al,\xi) =\frac{\Ga(\ti
x+1)\Ga(-\al-n)}{\Ga(-\al)}\, \left(
\frac{1-\xi}{\xi}\right)^n\, \frac{F(-n,-\al-n;\ti x+1-n;\w)}{\Ga(\ti x+1-n)}\\
=\frac{(-1)^n\Ga(\ti x+1)\Ga(\al+1)}{\Ga(\al+n+1)}\, \left(
\frac{1-\xi}{\xi}\right)^n\, \frac{F(-n,-\al-n;\ti
x+1-n;\w)}{\Ga(\ti x+1-n)}, \tag2.10
\endgather
$$
see \cite{KS}. Although the first expression looks simpler, it
turns out that only the second expression is suitable for our
purposes. Note that
$$
\Vert \M_n(\,\cdot\,;\al,\xi)\Vert^{-2}
=\frac{\xi^n(1-\xi)^{\al+1}\Ga(\al+n+1)}{\Ga(\al+1)\Ga(n+1)}\,.
$$

{}From the last two formulas and the definition of $\wt\M_n$ we
obtain
$$
\gather \wt\M_n(\ti x;\al,\xi)=\sqrt{\frac{\Ga(\ti x+1)\Ga(\ti
x+\al+1)}{\Ga(n+1)\Ga(n+\al+1)}}\,\xi^{(\ti
x-n)/2}(1-\xi)^{(\al+1+2n)/2}\\
\times \frac{F(-n,-\al-n;\ti x+1-n;\w)}{\Ga(\ti x+1-n)}\,.
\endgather
$$
Comparing this with \tht{2.1} and taking into account \tht{2.9}
we get the required equality. \qed
\enddemo

Thus, our functions $\psi_a$ can be obtained from the Meixner
polynomials by the following procedure:

$\bullet$ We replace the initial polynomials $\M_n$ by the
functions $\wt\M_n$. This step is quite clear: as a result we
get functions which form an orthonormal basis in the $\ell^2$
space on $\Z_+$ with respect to the weight function 1.

$\bullet$ Next, we make a change of the argument. Namely, we
introduce an additional parameter $N=1,2,\dots$ and we set
$x=\ti x-N+\tfrac12$. Then we get orthogonal functions on the
subset
$$
\{-N+\tfrac12, -N+\tfrac32, -N+\tfrac52, \dots\}\subset\Z'
$$
which exhausts the whole $\Z'$ in the limit as $N$ goes to
infinity.

$\bullet$ Then we also need a change of the index. Namely,
instead of $n$ we have to take $a=N-n-\tfrac12$. We cannot give
a conceptual explanation of this transformation, it is dictated
by the formulas. Again,  the range of the possible values for
$a$ becomes larger together with $N$, and in the limit as
$N\to+\infty$ we get the whole lattice $\Z'$.

$\bullet$ Finally, we make a (formal) analytic continuation in
parameters $N$ and $\al$, using an appropriate analytic
expression for the Meixner polynomials. Note that the difference
equation of Proposition 2.1 and the three--term relation of
Corollary 2.6 precisely correspond to similar relations for the
Meixner polynomials.

We hope that this detailed explanation will help the reader to
perceive the analytic continuation arguments in Sections 3 and
7.

Of course, instead of the lattice $\Z'$ we could equally well
deal with the lattice $\Z$, and then numerous ``$\frac12$''
would disappear. However, dealing with the lattice $\Z'$ makes
main formulas more symmetric.

\head 3. The discrete hypergeometric kernel
\endhead

Let $\X$ be a countable set. By a {\it point configuration\/} in
$\X$ we mean any subset $X\subseteq\X$. Let $\Conf(\X)$ be the
set of all point configurations; this is a compact space. Assume
we are given a probability measure on $\Conf(\X)$ so that we can
speak about the {\it random\/} point configuration in $\X$. The
$n$th {\it correlation function\/} of our probability measure
(where $n=1,2,\dots$) is defined by
$$
\rho_n(x_1,\dots,x_n)=\Prob\{\text{the random configuration
contains $x_1,\dots,x_n$}\},
$$
where $x_1,\dots,x_n$ are pairwise distinct points in $\X$. The
collection of all correlation functions determines the initial
probability measure uniquely.

We say that our probability measure is {\it determinantal\/} if
there exists a function $K(x,y)$ on $\X\times\X$ such that
$$
\rho_n(x_1,\dots,x_n)=\det\left[K(x_i,x_j)\right]_{i,j=1}^n\,,
\qquad n=1,2,\dots \tag3.1
$$
It is worth noting that if such a function $K(x,y)$ exists, then
it is not unique. Indeed, any ``gauge transformation'' of the
form
$$
K(x,y)\to \frac{f(x)}{f(y)}\,K(x,y), \tag3.2
$$
where $f$ is a nonvanishing function on $\X$, does not affect
the determinants in the right--hand side of \tht{3.1}.

Any function $K(x,y)$ satisfying \tht{3.1} will be called a {\it
correlation kernel\/} of the initial determinantal measure. Two
kernels giving the same system of correlation functions will be
called {\it equivalent}.

As in \S2, we are dealing with the lattice $\Z'$ of (proper)
half--integers. We split it into two parts,
$\Z'=\Z'_-\cup\Z'_+$, where $\Z'_-$ consists of all negative
half--integers and $\Z'_+$ consists of all positive
half--integers. For an arbitrary $\la\in\Y$ we set
$$
\unX(\la)=\{\la_i-i+\tfrac12\mid i=1,2,\dots\}\subset\Z'.
$$
For instance, $\unX(\varnothing)=\Z'_-$. The correspondence
$\la\mapsto \unX(\la)$ is a bijection between the Young diagrams
$\la$ and those (infinite) subsets $X\subset\Z'$ for which the
symmetric difference $X\triangle\,\Z'_-$ is a finite set with
equally many points in $\Z'_+$ and $\Z'_-$. Note that
$$
\unX(\la')=-(\Z'\setminus \unX(\la)).
$$

Using the correspondence  $\la\mapsto \unX(\la)$ we can
interpret any probability measure $M$ on $\Y$  as a probability
measure on $\Conf(\Z')$. This makes it possible to speak about
the correlation functions of $M$. Our goal is to compute them
explicitly for the z--measures.

Now we can state the main results of the section.
\footnote{These results are essentially not new, see \cite{BO2},
\cite{BO5}, \cite{Ok2}, \cite{BOk, Example 3}, and the comments
at the end of the section. However, the method of proof is new.
The same method, with suitable modifications, is applied in \S7
for the computation of the dynamical correlation functions.}

\proclaim{Theorem 3.1} For any admissible pair of parameters
$(z,z')$, see \S1, the corresponding mixed z--measure
$M_{z,z',\xi}$ is a determinantal measure.
\endproclaim

\proclaim{Theorem 3.2} If $(z,z')$ is not in the degenerate
series (so that $z$ and $z'$ are not integers) then the
correlation kernel of $M_{z,z',\xi}$ can be written in the form
$$
\unK_{z,z',\xi}(x,y)=\sum_{a\in\Z'_+}\psi_a(x;z,z',\xi)\psi_a(y;z,z',\xi),
\qquad x,y\in\Z', \tag3.3
$$
where the functions $\psi_a$ are defined in \tht{2.1}.
\endproclaim

Note that the series in the right--hand side is absolutely
convergent. Indeed, since $\{\psi_a\}$ is an orthonormal basis
in $\ell^2(\Z')$ (Proposition 2.4), this follows from the fact
that the series can be written as
$$
\sum_{a\in\Z'_+}(\de_x,\psi_a)(\psi_a,\de_y),
$$
where $\de_x$ stands for the delta--function at point $x$ on the
lattice $\Z'$, and $(\,\cdot\,,\,\cdot\,)$ denotes the inner
product in $\ell^2(\Z')$.

Formula \tht{3.3} simply means that $\unK_{z,z',\xi}(x,y)$ is
the matrix of the orthogonal projection operator in
$\ell^2(\Z')$ whose range is the subspace spanned by the basis
vectors $\psi_a$ with index $a\in\Z'_+\subset\Z'$.

\proclaim{Theorem 3.3} The correlation kernel \tht{3.3} can also
be written in the form
$$
\unK_{z,z',\xi}(x,y)=\varphi_{z,z'}(x,y)\,\hatK_{z,z',\xi}(x,y)\tag3.4
$$
where, as in \tht{2.6},
$$
\varphi_{z,z'}(x,y)=\frac{\sqrt{\Ga(x+z+\tfrac12)\Ga(x+z'+\tfrac12)
\Ga(y+z+\tfrac12)\Ga(y+z'+\tfrac12)}}
{\Ga(x+z'+\tfrac12)\Ga(y+z+\tfrac12)} \tag3.5
$$
and
$$
\multline \hatK_{z,z',\xi}(x,y)\\=\frac{1-\xi}{(2\pi
i)^2}\oint\limits_{\{\om_1\}}\oint\limits_{\{\om_2\}}
\dfrac{(1-\sqrt\xi\om_1)^{-z'}
\left(1-\dfrac{\sqrt\xi}{\om_1}\right)^{z}
(1-\sqrt\xi\om_2)^{-z}
\left(1-\dfrac{\sqrt\xi}{\om_2}\right)^{z'}}{\om_1\om_2-1}\\
\times \,\om_1^{-x-\tfrac12}\om_2^{-y-\tfrac12}d\om_1\,d\om_2
\endmultline\tag3.6
$$
where $\{\om_1\}$ and $\{\om_2\}$ are arbitrary simple contours
satisfying the following three conditions{\/\rm:}

$\bullet$ both contours go around 0 in positive
direction{\/\rm;}

$\bullet$ the point $\xi^{1/2}$ is in the interior of each of
the contours while the point $\xi^{-1/2}$ lies outside
them{\rm;}

$\bullet$ the contour $\{\om_1^{-1}\}$ is contained in the
interior of the contour $\{\om_2\}$ {\rm(}equivalently,
$\{\om_2^{-1}\}$ is contained in the interior of
$\{\om_1\}${\rm)}.

The kernels $\unK_{z,z',\xi}(x,y)$ and $\hatK_{z,z,\xi}(x,y)$
are equivalent. Namely, they are related by a ``gauge
transformation'',
$$
\hatK_{z,z',\xi}(x,y)
=\frac{f_{z,z'}(x)}{f_{z,z'}(y)}\,\unK_{z,z',\xi}(x,y), \qquad
x,y\in\Z',
$$
where
$$
f_{z,z'}(x)=\frac{\Ga(x+z'+\tfrac12)}
{\sqrt{\Ga(x+z+\tfrac12)\Ga(x+z'+\tfrac12)}}\tag3.7
$$
The kernel $\hatK_{z,z,\xi}(x,y)$ can serve as a correlation
kernel for all admissible values of parameters $(z,z')$,
including the degenerate series.
\endproclaim

\demo{Proof of Theorems 3.1--3.3} We prove these three theorems
simultaneously. Let $\rho_n^{(z,z',\xi)}(x_1,\dots,x_n)$ denote
the $n$--point correlation function of $M_{z,z',\xi}$. The proof
splits into two parts.

In the first part, we compute $\rho_n^{(z,z',\xi)}$ for special
values of the parameters (the degenerate series):
$z=N=1,2,\dots$ and $z'=N+\al$, where $\al>-1$. Here we use the
fact that for such $(z,z')$, the mixed z--measure can be
interpreted as the so--called $N$--particle Meixner ensemble. We
show that formula
$$
\rho_n^{(z,z',\xi)}(x_1,\dots,x_n)
=\det[\unK_{z,z',\xi}(x_i,x_j)]_{i,j=1}^n
$$
is valid (in particular, the values of the kernel in the
right--hand size are well defined) when $z=N$, $z'=z+\al$,
provided that $N$ is so large that the numbers $x_i+N-\tfrac12$
are nonnegative. Then we check that in that formula, the kernel
$\unK_{z,z',\xi}$ can be replaced by the kernel
$\hatK_{z,z',\xi}$:
$$
\rho_n^{(z,z',\xi)}(x_1,\dots,x_n)
=\det[\hatK_{z,z',\xi}(x_i,x_j)]_{i,j=1}^n
$$

In the second part, we extend the latter formula to arbitrary
admissible $(z,z')$. To do this we show that both sides are
analytic functions in parameters $(z,z',\xi)$. Moreover, these
functions are of such a kind that they are uniquely defined by
their values at points $(z=N, z'=N+\al,\xi)$.

We proceed to the detailed proof.

Let, as in \S2, $N$ be a natural number and $\al>-1$. Consider
the Meixner weight function $W_{\al,\xi}$ on $\Z_+$, see
\tht{2.7}. The $N$--point {\it Meixner ensemble\/} is formed by
random $N$--point configurations $\wt X=(\ti x_1>\dots>\ti x_N)$
in $\X=\Z_+$, where
$$
\Prob(\wt X) =\const\cdot\prod_{1\le i<j\le N}(\ti x_i-\ti
x_j)^2\prod_{i=1}^n W_{\al,\xi}(\ti x_i).
$$
By the $N$th {\it Meixner measure\/} we mean the corresponding
probability measure on $\Conf(\Z_+)$.

\proclaim{Lemma 3.4} The $N$th Meixner measure is a
determinantal measure. As its correlation kernel on
$\Z_+\times\Z_+$ one can take the ``Meixner kernel''
$$
K^{\m}_{N,\al, \xi}(\ti x,\ti y)=\sum_{m=0}^{N-1} \wt\M_m(\ti
x;\al,\xi)\,\wt\M_m(\ti y;\al,\xi), \qquad \ti x, \ti y\in\Z_+,
$$
where the functions $\wt\M_m(\ti x;\al,\xi)$ are defined in
\tht{2.8}.
\endproclaim

\demo{Proof} This is a special case of a well--known general
claim about orthogonal polynomial ensembles, see, e.g.,
\cite{De2}. \qed
\enddemo

Let $\Y(N)\subset\Y$ denote the set of Young diagrams $\la$ with
$\ell(\la)\le N$. Recall that the mixed z--measure with
parameters $z=N$, $z'>N-1$ is concentrated on $\Y(N)$. We define
a bijection between Young diagrams $\la\in\Y(N)$ and $N$--point
configurations $\wt X\subset\Z_+$ as follows
$$
\la\mapsto \wt X(\la)=(\ti x_1,\dots,\ti x_N), \qquad \ti
x_i=\la_i-i+N, \quad i=1,\dots,N.
$$

\proclaim{Lemma 3.5 (\cite{BO2, Proposition 4.1})} The
correspondence $\la\mapsto \wt X(\la)$ takes the z-measure
$M_{z,z',\xi}$ with parameters $z=N$, $z'=N+\al$ to the $N$th
Meixner measure with parameters $\al,\xi$.
\endproclaim

\demo{Proof} Direct verification. \qed \enddemo

Recall that we identify $M_{z,z',\xi}$ with its push--forward
under the correspondence $\la\mapsto X(\la)$.

\proclaim{Corollary 3.6} Let $z=N=1,2,\dots$ and $z'=z+\al$ with
$\al>-1$. Assume that $x_1,\dots,x_n$ lie in the subset
$\Z_+-N+\tfrac12\subset\Z'$, so that the points $\ti
x_i:=x_i+N-\tfrac12$ are in $\Z_+$.

Then
$$
\rho_n^{(z,z',\xi)}(x_1,\dots,x_n)=\det\left[K^\m_{N,\al,\xi}
(\ti x_i,\; \ti x_j)\right]_{i,j=1}^n
$$
\endproclaim

\demo{Proof} Let $\la\in\Y(N)$. Comparing the definition of the
infinite configuration $\unX(\la)\subset\Z'$ with that of the
$N$--point configuration $\wt X(\la)$ we see that
$$
\wt X(\la)=(\unX(\la)+N-\tfrac12)\cap\Z_+.
$$
Then the claim follows from Lemmas 3.4 and 3.5. \qed
\enddemo

We take \tht{3.3} as the definition of the kernel
$\unK_{z,z',\xi}(x,y)$.

\proclaim{Lemma 3.7} Let $z=N=1,2,\dots$ and $z'=z+\al$ with
$\al>-1$. Assume that $x$ and $y$ lie in the subset
$\Z_+-N+\tfrac12\subset\Z'$, so that $\ti x:=x+N-\tfrac12$ and
$\ti y:=y+N-\tfrac12$ are in $\Z_+$.

Then expression \tht{3.3} for the kernel $\unK_{z,z',\xi}(x,y)$
is well defined and we have
$$
\unK_{z,z',\xi}(x,y)=K^\m_{N,\al,\xi}(\ti x,\ti y).
$$
\endproclaim

\demo{Proof} We have to prove that
$$
\sum_{a\in\Z'_+}\psi_a(x;z,z',\xi)\psi_a(y;z,z',\xi)
=\sum_{m=0}^{N-1} \wt\M_m(\ti x;\al,\xi)\,\wt\M_m(\ti
y;\al,\xi)\tag3.8
$$

We recall that the functions $\psi_a(x;z,z',\xi)$ were defined
under the assumption that both $z,z'$ are not integers. However,
as it can be seen from \tht{2.1}, each summand in the left--hand
side of \tht{3.8} makes sense under the hypotheses of the lemma.

Set
$$
a(m)=N-m-\tfrac12, \qquad m=0,1,\dots,N-1
$$
By Proposition 2.8,
$$
\psi_{a(m)}(x;z,z',\xi)=\wt\M_m(\ti x;\al,\xi), \qquad
\psi_{a(m)}(y;z,z',\xi)=\wt\M_m(\ti y;\al,\xi),
$$
which implies that
$$
\sum_{a=\frac12}^{N-\frac12}\psi_a(x;z,z',\xi)\psi_a(y;z,z',\xi)
=\sum_{m=0}^{N-1} \wt\M_m(\ti x;\al,\xi)\,\wt\M_m(\ti y;\al,\xi)
\tag3.9
$$

Finally, observe that
$$
\left.\frac1{\Ga(z-a+\tfrac12)}\,\right|_{a=N+\tfrac12,\,
N+\tfrac32,
\dots}=\left.\frac1{\Ga(N-a+\tfrac12)}\,\right|_{a=N+\tfrac12,\,
N+\tfrac32, \dots}=0
$$

We conclude that the infinite sum in the left--hand side of
\tht{3.8} actually coincides with the finite sum in \tht{3.9}.
\qed
\enddemo

Together with Corollary 3.6 this implies

\proclaim{Corollary 3.8} Let $z=N=1,2,\dots$ and $z'=z+\al$ with
$\al>-1$. Assume that $x_1,\dots,x_n$ lie in the subset
$\Z_+-N+\tfrac12\subset\Z'$, so that the points $\ti
x_i:=x_i+N-\tfrac12$ are in $\Z_+$.

Then
$$
\rho_n^{(z,z',\xi)}(x_1,\dots,x_n)=\det\left[\unK_{z,z',\xi} (
x_i,\; x_j)\right]_{i,j=1}^n
$$
\endproclaim

\proclaim{Lemma 3.9} Assume that

$\bullet$ either $(z,z')$ is not in the degenerate series  and
$x,y\in\Z'$ are arbitrary

$\bullet$ or $z=N=1,2,\dots$, $z'>N-1$, and both $x,y$ are in
$\Z_+-N+\tfrac12$.

Then the kernel $\hatK_{z,z',\xi}(x,y)$ of Theorem 3.3 is
related to the kernel $\unK_{z,z',\xi}(x,y)$ by equality
\tht{3.4}. Equivalently, the kernels are related by the ``gauge
transformation'' \tht{3.2},
$$
\wh K_{z,z',\xi}(x,y)=\frac{f_{z,z'}(x)}{f_{z,z'}(y)}\,
\unK_{z,z',\xi}(x, y), \tag3.10
$$
where $f_{z,z'}$ is defined in \tht{3.7}.
\endproclaim

\demo{Proof} Let us start with expression \tht{3.3} of the
kernel $\unK_{z,z',\xi}$ and let us replace each summand by its
integral representation \tht{2.5}. It is convenient to set
$a-\tfrac12=k$ so that as $a$ ranges over $\Z'_+$,  $k$ ranges
over $\{0,1,2,\dots\}$. Then we obtain
$$
\gather K_{z,z',\xi} (x,y)=\varphi_{z,z'}(x,y)\\
 \times \frac{1-\xi}{(2\pi
i)^2}\sum_{k=0}^\infty\int\limits_{\{\om_1\}}\int\limits_{\{\om_2\}}
(1-\sqrt\xi\om_1)^{-z'}
\left(1-\dfrac{\sqrt\xi}{\om_1}\right)^{z-1}
(1-\sqrt\xi\om_2)^{-z}
\left(1-\dfrac{\sqrt\xi}{\om_2}\right)^{z'-1}\\
\times \,\om_1^{-x-\tfrac12}\om_2^{-y-\tfrac12}\,
\left(\frac{(1-\sqrt\xi\om_1)(1-\sqrt\xi\om_2)}
{(\om_1-\sqrt\xi)(\om_2-\sqrt\xi)}\right)^k
\frac{d\om_1\,d\om_2}{\om_1\om_2}\,.
\endgather
$$
We can choose the contours $\{\om_1\}$ and $\{\om_2\}$ so that
they are contained in the domain $|\om|>1$. Since the
fractional--linear transformation
$$
\om\mapsto\frac{1-\sqrt\xi\om}{\om-\sqrt\xi}
$$
preserves the unit circle $|\om|=1$ and maps its exterior
$|\om|>1$  into its interior $|\om|<1$, we  have on the product
of the contours a bound of the form
$$
\left|\frac{(1-\sqrt\xi\om_1)(1-\sqrt\xi\om_2)}
{(\om_1-\sqrt\xi)(\om_2-\sqrt\xi)}\right|\le q<1.
$$
Therefore, we can interchange summation and integration and then
sum the arising geometric progression in the integrand:
$$
\sum_{k=0}^\infty
\left(\frac{(1-\sqrt\xi\om_1)(1-\sqrt\xi\om_2)}
{(\om_1-\sqrt\xi)(\om_2-\sqrt\xi)}\right)^k
=\frac{\left(1-\dfrac{\sqrt\xi}{\om_1}\right)
\left(1-\dfrac{\sqrt\xi}{\om_1}\right)\om_1\om_2}
{(1-\xi)(\om_1\om_2-1)}
$$
Then we obtain equality \tht{3.4} with integral \tht{3.6}, as
desired. Finally, we can relax the assumption on the contour: it
suffices to assume that $\{\om_1^{-1}\}$ is strictly contained
inside $\{\om_2\}$, as in the formulation of Theorem 3.3.

It remains to show that \tht{3.4} is equivalent to \tht{3.10}.
According to \tht{3.5} consider the expression
$$
\frac1{\varphi_{z,z'}(x,y)}=\frac{\Ga(x+z'+\tfrac12)\Ga(y+z+\tfrac12)}
{\sqrt{\Ga(x+z+\tfrac12)\Ga(x+z'+\tfrac12)
\Ga(y+z+\tfrac12)\Ga(y+z'+\tfrac12)}}
$$
Let us show that
$$
\frac1{\varphi_{z,z'}(x,y)}=\frac{f_{z,z'}(x)}{f_{z,z'}(y)}
$$
Indeed, $1/\varphi_{z,z'}$ has the form
$$
\frac{a(x)b(y)}{\sqrt{a(x)b(x)a(y)b(y)}}\,,
$$
and our hypotheses imply that $a(x)b(x)$ and $a(y)b(y)$ are real
and strictly positive. We also have
$$
f_{z,z'}(x)=\frac{a(x)}{\sqrt{a(x)b(x)}}.
$$
Therefore, we get
$$
\gather
\frac{f_{z,z'}(x)}{f_{z,z'}(y)}=\frac{a(x)\sqrt{a(y)b(y)}}{\sqrt{a(x)b(x)}\,a(y)}
=\frac{a(x)a(y)b(y)}{\sqrt{a(x)b(x)a(y)b(y)}\,a(y)}\\
=\frac{a(x)b(y)}{\sqrt{a(x)b(x)a(y)b(y)}}=\frac1{\varphi_{z,z'}(x,y)}
\endgather
$$
\qed
\enddemo

\proclaim{Corollary 3.10} Let  $z=N=1,2,\dots$ and $z'>N-1$.
Then
$$
\rho_n^{(z,z',\xi)}(x_1,\dots,x_n)=\det\left[\hatK_{z,z',\xi}
(x_i,\,x_j)\right]_{i,j=1}^n \tag3.11
$$
provided that all the points $x_1,\dots,x_n\in\Z'$ lie in the
subset $\Z_+-N+\tfrac12\subset\Z'$.
\endproclaim

\demo{Proof} Indeed, this follows from Lemma 3.9 and Corollary
3.8. \qed
\enddemo

This completes the first part of the proof. Now we proceed to
the second part.

\proclaim{Lemma 3.11} {\rm(i)} Fix an arbitrary set of Young
diagrams $\Cal D\subset\Y$. For any fixed admissible pair of
parameters $(z,z')$, the function
$$
\xi\mapsto\sum_{\la\in \Cal D}M_{z,z',\xi}(\la),
$$
which is initially defined on the interval $(0,1)$, can be
extended to a holomorphic function in the unit disk $|\xi|<1$.

{\rm(ii)} Consider the Taylor expansion of this function at
$\xi=0$,
$$
\sum_{\la\in \Cal D}M_{z,z',\xi}(\la)=\sum_{k=0}^\infty
G_{k,\Cal D}(z,z')\xi^k.
$$
Then the coefficients $G_{k,\Cal D}(z,z')$ are polynomial
functions in $z,z'$. That is, they are restrictions of
polynomial functions to the set of admissible values $(z,z')$.
\endproclaim

\demo{Proof} (i) Set $\Cal D_n=\Cal D\cap\Y_n$. By the
definition of $M_{z,z',\xi}$,
$$
\align \sum_{\la\in \Cal D}M_{z,z',\xi}(\la)
&=\sum_{n=0}^\infty\left(\,\sum_{\la\in
\Cal D_n}M^{(n)}_{z,z'}(\la)\right) \pi_{zz',\xi}(n)\\
&=(1-\xi)^{zz'}\sum_{n=0}^\infty\left(\,\sum_{\la\in \Cal
D_n}M^{(n)}_{z,z'}(\la)\right)\frac{(zz')_n\,\xi^n}{n!}.
\endalign
$$
Each interior sum is nonnegative and does not exceed 1. On the
other hand,
$$
\sum_{n=0}^\infty |\pi_{zz',\xi}(n)|
=|1-\xi|^{zz'}\sum_{n=0}\frac{(zz')_n\,|\xi|^n}{n!}<\infty,
\qquad \xi\in\C, \quad |\xi|<1.
$$
This proves the first claim.

(ii) By \tht{1.11},
$$
\sum_{\la\in \Cal D}M_{z,z',\xi}(\la)
=(1-\xi)^{zz'}\sum_{n=0}^\infty\sum_{\la\in \Cal D_n}
(z)_\la(z')_\la\,\xi^n\left(\frac{\dim\la}{n!}\right)^2.
$$
It follows that
$$
G_{k,\Cal D}(z,z')=\sum_{n=0}^k
\frac{(-zz')_{k-n}}{(k-n)!}\sum_{\la\in \Cal D_n}
(z)_\la(z')_\la\,\left(\frac{\dim\la}{n!}\right)^2.
$$
Since each $\Cal D_n$ is a finite set, this expression is a
polynomial in $z,z'$. \qed
\enddemo

Now we can complete the proof of the theorems. Fix $n$ and an
arbitrary $n$--point subset $X=\{x_1,\dots,x_n\}\subset\Z'$, and
regard $\rho_n^{(z,z',\xi)}(x_1,\dots,x_n)$ as a function of
parameters $z,z',\xi$. We want to show that equality \tht{3.11}
holds for any admissible $(z,z')$. Apply Lemma 3.11 to the set
$\Cal D$ of those diagrams $\la$ for which $\unX(\la)$ contains
$X$, and observe that
$$
\rho_n^{(z,z',\xi)}(x_1,\dots,x_n)=\sum_{\la\in \Cal
D}M_{z,z',\xi}(\la).
$$
It follows that $\rho_n^{(z,z',\xi)}(x_1,\dots,x_n)$ is a
real--analytic function of $\xi\in(0,1)$ which admits a
holomorphic extension to the open unit disk $|\xi|<1$. Moreover,
the Taylor coefficients of this function depend on $z,z'$
polynomially.

On the other hand, from the expression \tht{3.6} for the kernel
$\hatK_{z,z',\xi}(x,y)$ it follows that this kernel (and hence
the right--hand side of \tht{3.11}) has the same property, with
$\xi$ replaced by $\sqrt\xi$.

Thus, both sides of \tht{3.11} can be viewed as (restrictions
of) holomorphic functions in $\sqrt\xi$ with polynomial Taylor
coefficients. Since the set
$$
\{(z,z')\mid \text{$z$ is a large natural number $N$ and
$z'>N-1$}\}
$$
is a set of uniqueness for polynomials in two variables, we
conclude that equality \tht{3.11} is true for any admissible
$(z,z')$.

This proves Theorem 3.1 and Theorem 3.3. Now, Theorem 3.2
follows from Theorem 3.3 and Lemma 3.9. \qed
\enddemo

\demo{Comments} 1. The correlation functions of the z--measures
$M_{z,z',\xi}$ were first computed in \cite{BO2} in a different
form: in that paper we dealt with another embedding of
partitions into the set of lattice point configurations (in the
notation of \S8, we used the map $\la\mapsto X(\la)$, instead of
$\la\mapsto\unX(\la)$). The kernel $\unK_{z,z',\xi}(x,y)$
coincides with one of the ``blocks'' of the kernel considered in
\cite{BO2}. The relation between both kernels is discussed in
detail in \cite{BO5} (see also \S8 below). The proofs in
\cite{BO2} and \cite{BO5} are very different from the arguments
of the present section.

2. Two other derivations of the kernel $\unK_{z,z',\xi}(x,y)$ are
given in Okounkov's papers \cite{Ok2} and \cite{Ok1}. In both
these papers, the correlation functions are expressed through the
vacuum state expectations of certain operators in the infinite
wedge Fock space. A (substantial) difference between the methods
of \cite{Ok2} and \cite{Ok1} consists in the concrete choice of
operators. The general formalism of Schur measures presented in
\cite{Ok1} is complemented by explicit computations in \cite{BOk,
\S4}.

3. As shown in the papers listed above, the kernel
$\unK_{z,z',\xi}(x,y)$ can be written in the form
$$
\unK_{z,z',\xi}(x,y)=\frac{P(x)Q(y)-Q(x)P(y)}{x-y}\,, \tag3.12
$$
where $P$ and $Q$ are certain functions on $\Z'$ depending on
parameters $z,z',\xi$. Since $P$ and $Q$ are expressed through the
Gauss hypergeometric function, we called $\unK_{z,z',\xi}(x,y)$
the {\it discrete hypergeometric kernel\/}. In general, kernels
admitting such an expression are called {\it integrable
kernels\/}, in accordance with the terminology of \cite{IIKS},
\cite{De1}, \cite{B2}.

4. The integrable form \tht{3.12} can be readily derived from
\tht{3.3} using the three--term relation for functions
$\psi_a(x)=\psi_a(x;z,z'\xi)$ given in Corollary 2.6.
Specifically, we obtain
$$
\unK_{z,z',\xi}(x,y)=\frac{\sqrt{zz'\xi}}{1-\xi}\;
\frac{\psi_{-\frac12}(x)\psi_{\frac12}(y)
-\psi_{\frac12}(x)\psi_{-\frac12}(y)}{x-y}\,, \tag3.13
$$
This  derivation of \tht{3.13} from \tht{3.3} is quite similar
to the standard derivation of the Christoffel--Darboux formula
for an arbitrary system of orthogonal polynomials. Since, as
explained in \S2, the functions $\psi_a$ are closely related to
the Meixner polynomials, this analogy is not surprising.

5. Once we know that the functions $\psi_a$ form an orthonormal
basis (Proposition 2.4), the series expression \tht{3.3} for the
kernel $\unK_{z,z',\xi}(x,y)$ immediately implies that it is a
projection kernel. This fact was first proved in \cite{BO5, \S5}
in a different way.

6. The series representation \tht{3.3} is equivalent to formula
\tht{3.16} in \cite{Ok2}. A double contour integral representation
of various correlation kernels related to Schur measures appeared
earlier in \cite{BOk}.
\enddemo

\head 4. Construction of Markov processes
\endhead

The goal of this section is to explain the construction of the
continuous time Markov processes on partitions which will be
studied in the rest of the paper. Their fixed time distributions
are the z-measures considered in the previous sections.

It is fairly easy to give the jump rates for these processes.
However, it is not {\it a priori} clear why these rates define
the process uniquely. Since we were unable to find suitable
uniqueness theorems in the literature, we will actually prove
that the rates define the process uniquely and compute the
transition probabilities for an underlying birth-death process.

\subhead 4.1. Preliminaries on Markov processes
\endsubhead
Let us recall some basic facts about continuous time Markov
processes and introduce the notation.

The time parameter $t$ always ranges over an open interval
$(t_{min}, t_{max})$ where $t_{min}\in\R\cup \{-\infty\}$ and
$t_{max}\in\R\cup\{+\infty\}$. Let us denote the state space by
$\A$, it is assumed to be either finite or countable.

We also denote by $P(s,t)$, $s\le t$, the matrix of transition
probabilities of a Markov process. This is a matrix with rows
and columns marked by elements of $\A$, its elements will be
denoted by $P_{ab}(s,t)$, $a,b\in\A$. By definition,
$P_{ab}(s,t)$ is the probability that the process will be in the
state $b$ at the time moment $t$ conditioned that it is in the
state $a$ at time $s$. Thus, all matrix elements of $P(s,t)$ are
nonnegative, and its sum is equal to one along any row. Such
matrices are called {\it stochastic}. The transition matrices
$P(s,t)$ also satisfy the {\it Chapman-Kolmogorov equation}
$$
P(s,t)P(t,u)=P(s,u),\qquad s\le t\le u. \tag4.1
$$

We assume that there exist $\A\times \A$ matrices $Q(t)$ with
continuously depending on $t$ entries, such that
$$
P_{ab}(s,t)=\delta_{ab}+Q_{ab}(t)(t-s)+o(|t-s|), \qquad |t-s|\to
0.
$$
This relation implies that $Q_{ab}(t)\ge 0$ for $a\ne b$ and
$Q_{aa}(t)\le 0$. Further, we assume that
$$
\sum_{b\ne a}Q_{ab}(t)=-Q_{aa}(t),\quad\text{for any  } a\in\A.
\tag4.2
$$
This is the infinitesimal analog of the condition
$\sum_{b\in\A}P_{ab}(s,t)=1$.

It is well known that \tht{4.1} then implies that $P(s,t)$ then
satisfies {\it Kolmogorov's backward equation}
$$
-\frac\partial{\partial s}\, P(s,t)=Q(s) P(s,t),\qquad s\le t,
\tag4.3
$$
with the initial condition
$$
P(t,t)\equiv \operatorname{Id}. \tag4.4
$$
 Under certain additional
conditions, $P(s,t)$ will also satisfy Kolmogorov's forward
equation
$$
\frac\partial{\partial t}\, P(s,t)= P(s,t) Q(t). \tag4.5
$$

In our concrete situation we would like to {\it define} a Markov
process by specifying the {\it transition rates} $Q(t)$.
However, it may happen that this does not specify the process
uniquely (then the backward equation has many solutions
$P(s,t)$). Uniqueness always holds if  $\A$ is finite or, more
generally, if $\A$ is infinite but  the functions $|Q_{aa}(t)|$
are bounded on any closed time interval (see, e.g. \cite{Fe1}).
However, these conditions are not satisfied in our case. There
exist other, more involved uniqueness conditions for time
homogeneous (stationary) Markov processes. However, in our
approach, even if we restrict our attention to stationary
processes, we still need to handle some non stationary processes
as auxiliary objects. For these reasons we had to find some more
special uniqueness condition.

Let us write $Q(t)$ in the form $Q(t)=-R(t)+\wt Q(t)$, where
$-R(t)$ is the diagonal part of $Q(t)$ and $\wt Q(t)$ is the
off-diagonal part of $Q(t)$. In other words,
$$
R_{ab}(t)=-\delta_{ab}Q_{aa}(t),\qquad \wt Q_{ab}(t)=\cases
Q_{ab}(t),&a\ne b,\\0,&a=b.\endcases
$$

For $s\le t$ set
$$
F(s,t)=\exp\left(-\int_s^t R(\tau)d\tau\right),\qquad
G(s,t)=F(s,t)\wt Q(s,t).
$$
Define $P^{[n]}(s,t)$ recursively by
$$
P^{[0]}(s,t)=F(s,t),\qquad P^{[n]}(s,t)=\int_s^t
G(s,\tau)P^{[n-1]}(\tau,t)d\tau,\quad n\ge 1,
$$
and set
$$
\overline{P}(s,t)=\sum_{n=0}^\infty P^{[n]}(s,t),\qquad s\le t.
$$

\proclaim{Theorem 4.1 \cite{Fe1}} (i) The matrix
$\overline{P}(s,t)$ is substochastic {\rm (}i.e., its elements
are nonnegative and $\sum_b P_{ab}(s,t)\le 1${\rm).} Its
elements are absolutely continuous and almost everywhere
differentiable with respect to both $s$ and $t$, and it provides
a solution of Kolmogorov's backward and forward equations
\tht{4.3}, \tht{4.5} with the initial condition \tht{4.4}.

(ii) $\overline{P}(s,t)$ also satisfies the Chapman-Kolmogorov
equation \tht{4.1}.

(iii) $\overline{P}(s,t)$ is the minimal solution of \tht{4.3}
{\rm (}or \tht{4.5}{\rm )} in the sense that for any other
solution $P(s,t)$ of \tht{4.3} {\rm (}or \tht{4.5}{\rm )} with
the initial condition \tht{4.4} in the class of substochastic
matrices, one has $P_{ab}(s,t)\ge \overline{P}_{ab}(s,t)$ for
any $a,b\in\A$.
\endproclaim

\proclaim{Corollary 4.2} If the minimal solution
$\overline{P}(s,t)$ is stochastic {\rm (}the sums of matrix
elements along the rows are all equal to 1\,{\rm )} then it is
the unique solution of \tht{4.3} {\rm (}or \tht{4.5}{\rm )} with
the initial condition \tht{4.4} in the class of substochastic
matrices.
\endproclaim

Let us note that the construction of $\overline{P}(s,t)$ is very
natural: the summands $P^{[n]}_{ab}(s,t)$ are the probabilities to
go from $a$ to $b$ in $n$ jumps. The condition of
$\overline{P}(s,t)$ being stochastic exactly means that we cannot
make infinitely many jumps in a finite amount of time.

Our next goal is to provide a convenient sufficient condition for
$\overline{P}(s,t)$ to be stochastic.

Fix $s\in(t_{min},t_{max})$ and $a\in \A$. For any finite $X$,
$X\subset \A$, $a\in X$, we denote by $T_{s,\,a,\,X}$ the time
of the first exit from $X$ under the condition that the process
is in $a$ at time $s$. Formally, we can modify $\A$ and $Q(t)$
by contracting all the states $b\in\A\setminus X$ into one
absorbing state $\wt{b}$ with $ Q_{\wt{b},c}\equiv0$ for any $
c\in X\cup \{\wt{b}\}$. We obtain a process with a finite number
of states for which the solution $\wt P(s,t)$ of the backward
equation is unique. Then $T_{s,\,a,\,X}$ is a random variable
with values in $(s,+\infty]$ defined by
$$
\Prob \{T_{s,\,a,\,X}\le t\}= \wt P_{a\wt b}(s,t).
$$
\proclaim{Proposition 4.3}  Assume that for any $a\in\A$ and any
$s<t$, $\varepsilon>0$, there exists a finite set
$X(\varepsilon)\subset \A$ such that
$$
\Prob \{T_{s,\,a,\,X(\varepsilon)}\le t\}\le \varepsilon.
$$
Then the minimal solution $\overline{P}(s,t)$ provided by
Theorem 4.1 is stochastic.
\endproclaim

\demo{Proof} Consider the modified process on the finite state
space $X(\epsi)\cup\{\wt{b}\}$ described above. Since its
transition matrix $\wt P(s,t)$ is stochastic,
$$
\sum_{b\in X(\epsi)} \wt P_{ab}(s,t)=1-\wt P_{a\wt b}(s,t)\ge
1-\epsi.
$$
The construction of the minimal solution as the sum of
$P^{[n]}$'s, see above, immediately implies that $P_{ab}(s,t)\ge
\wt P_{ab}(s,t)$. Thus, $\sum_{b} P_{ab}(s,t)\ge 1-\epsi$ for any
$\epsi>0$. \qed
\enddemo

\subhead 4.2. An application to birth-death processes
\endsubhead A {\it birth-death process} is a continuous time
Markov process on $\A=\Z_+=\{0,1,2,\dots\}$ such that the rates
$Q_{mn}(t)$ vanish if $|n-m|>1$. In other words, the process can
make jumps only of size 1. Our assumption \tht{4.2} means that
$$
-Q_{nn}(t)=R_{nn}(t)=Q_{n,n+1}(t)+Q_{n,n-1}(t).
$$
\proclaim{Proposition 4.4} Assume that for any closed segment
$[t',t'']\subset (t_{min},t_{max})$ there exists a sequence
$\{\ga_n\}_{n=0}^\infty$ of positive real numbers such that
$R_{nn}(t)\le \ga_n$ for any $t\in [t',t'']$, and
$\sum_n\gamma_n^{-1}=\infty$. The the minimal solution
$\overline P(s,t)$ is stochastic.
\endproclaim

\demo{Proof} We will apply Proposition 4.3. Let us fix $a\in
\A=\Z_+$. As $X=X(\varepsilon)$ we will take a set of the form
$\{0,1,\dots,n-1\}$ for a suitable $n$. Then $T_{s,a,X}$ is the
moment of the first arrival at $n$ given that we start at $a$ at
the time moment $s$. To simplify the notation, set
$T_n=T_{s,a,\{0,1,\dots,n-1\}}$.

In order to estimate $T_n$ we will compare our inhomogeneous
birth-death process to the pure birth homogeneous process with
transition rates $\widehat Q_{n,n+1}=\ga_n$, $\widehat
Q_{n+1,n}=0$, for all $n\in\Z_+$. Let $\widehat T_n$ denote the
time of reaching $n$ given that we start at $a$ at time $s$. Note
that since this is a pure birth process, once the process leaves
$\{0,1,\dots,n-1\}$ it never comes back.

It is known that for the pure birth process with rates $\ga_n$
the minimal solution is stochastic if and only if
$\sum_n\ga_n^{-1}=\infty$, see \cite{Fe1}, \cite{Fe3, ch. XIV,
\S8}. By our hypothesis, $\sum_n\ga_n^{-1}=\infty$, and we may
denote by $\widehat P(s,t)$ the unique stochastic solution of
the backward and forward equations. Clearly,
$$
\Prob\{\widehat T_n\le t\}=\sum_{b=n}^\infty \widehat P_{ab}(s,t),
$$
which tends to zero as $n\to\infty$. Thus, the statement of this
proposition will follow from Proposition 4.3 if we show that
$\Prob\{T_n\le t\}\le \Prob\{\widehat T_n\le t\}$ for any $t>s$.

For any $n=a+1,a+2,\dots$, set
$$
F_n(t)=\Prob\{T_n\le t\},\qquad\widehat F_n(t)=\Prob\{\widehat
T_n\le t\}.
$$
We will prove that $F_n(t)\le \widehat F_n(t)$ for all $n>a$ using
induction on $n$.

Let us start with $n=a+1$. Clearly,
$$
\widehat F_{a+1}(t)=1-e^{-\ga_a (t-s)}
$$
because the time of the jump $a\to a+1$ is exponentially
distributed with parameter $\ga_a$. On the other hand, the
probability that our birth-death process will jump from $a$ to
either $a-1$ or $a+1$ before time $t$ equals
$$
1-e^{-\int_s^{t}(Q_{a,a+1}(t)+Q_{a,a-1}(t))dt}= 1-e^{-\int_s^{t}
R_{aa}(t) dt}.
$$
Thus, $F_{a+1}(t)\le 1-\exp({-\int_s^{t} R_{aa}(t)dt})$, and since
$R_{aa}(t)\le \ga_a$ by hypothesis, the estimate follows.

In order to prove the induction step, note that for $n\ge a+2$ we
have
$$
\gathered
 F_n(t)=\int_s^tdF_{n-1}(\tau)\Prob\{T_n-T_{n-1}\le
t-\tau\mid T_{n-1}=\tau\},\\
 \widehat F_n(t)=\int_s^td\widehat F_{n-1}(\tau)\Prob
 \{\widehat T_n-\widehat T_{n-1}\le
t-\tau\mid\widehat T_{n-1}=\tau\}.
\endgathered
$$
Arguing exactly as in the case $n=a+1$ above, we see that
$$
\gathered \Prob
 \{\widehat T_n-\widehat T_{n-1}\le
t-\tau\mid\widehat T_{n-1}=\tau\}=1-e^{-\gamma_{n-1}(t-\tau)},\\
\Prob\{T_n-T_{n-1}\le t-\tau\mid T_{n-1}=\tau\}\le
1-e^{-\gamma_{n-1}(t-\tau)}.
\endgathered
$$
Hence, it suffices to verify that
$$
\int_s^tdF_{n-1}(\tau)(1-e^{-\gamma_{n-1}(t-\tau)})\le
\int_s^td\widehat F_{n-1}(\tau)(1-e^{-\gamma_{n-1}(t-\tau)}).
$$
When we integrate by parts both sides of this inequality, we
notice that the non-integral terms vanish (because
$F_{n-1}(s)=\widehat F_{n-1}(s)=0$). Thus, we obtain the
equivalent inequality
$$
\gamma_{n-1}\int_s^t F_{n-1}(\tau)e^{-\gamma_{n-1}(t-\tau)}dt\le
\gamma_{n-1}\int_s^t \widehat
F_{n-1}(\tau)e^{-\gamma_{n-1}(t-\tau)}dt
$$
which immediately follows from $F_{n-1}(\tau)\le \widehat
F_{n-1}(\tau)$. \qed
\enddemo

\example{Remark 4.5} It is very plausible that Proposition 4.4
holds under the weaker assumption $Q_{n,n+1}(t)\le \ga_n$.
However, the proof of such a statement would require additional
considerations.
\endexample

\subhead 4.3. Birth-death process associated with Meixner
polynomials \endsubhead From now on we restrict our attention to
birth-death processes with
$$
Q_{n,n+1}=\al(t)(c+n),\qquad Q_{n,n-1}=\be(t)\,n, \tag4.6
$$
where $\al(t)\ge 0$, $\be(t)\ge 0$ are continuous functions on
$(t_{min},t_{max})$ and $c>0$ is a constant. Proposition 4.4
implies that for any process of this kind there exists a unique
stochastic solution $P(s,t)$ of Kolmogorov's backward and
forward equations. By breaking the time interval into finitely
many subintervals, we may as well assume that $\alpha(t)$ and
$\beta(t)$ are piecewise continuous functions with finitely many
points of discontinuity at which they have finite left and right
limits.

The {\it negative binomial distribution} $\pi_{c,\xi}$ on $\Z_+$
with parameters $c>0$ and $\xi\in(0,1)$ is defined by
$$
\pi_{c,\xi}(n)=(1-\xi)^c\,\frac{(c)_n}{n!}\,\xi^n,\qquad
n=0,1,2,\dots\,.
$$
It will be convenient to interpret $\pi_{c,\xi}$ as an infinite
row-vector.

\proclaim{Proposition 4.5} Let $\xi(t)$ be a continuous,
piecewise continuously differentiable function in $t$ with
values in $(0,1)$.  Assume that $\xi(t)$ solves the differential
equation
$$
\frac{\Dot\xi(t)}{\xi(t)(1-\xi(t))}=\frac{\alpha(t)}{\xi(t)}-\beta(t),
\qquad t\in(t_{min},t_{max}). \tag4.7
$$
Then the row vector $\pi_{c,\xi(t)}$ solves
$\pi_{c,\xi(s)}P(s,t)=\pi_{c,\xi(t)}$ for any $s\le t$.
\endproclaim

\demo{Proof} Let us differentiate $\pi_{c,\xi(s)}P(s,t)$ with
respect to $s$ and use Kolmogorov's backward equation.
Collecting the coefficients of $P_{xy}(s,t)$ in the $y$th
coordinate gives
$$
\multline \pi_{c,\xi(s)}(x)P_{xy}(s,t)\Biggl( -\frac
{c\,\Dot\xi}{1-\xi} +\frac {x\Dot\xi}\xi+(\alpha(c+x)+\beta
x)\\-\alpha(c+x-1)
\frac{\pi_{c,\xi(s)}(x-1)}{\pi_{c,\xi(s)}(x)}-\beta(x+1)
\frac{\pi_{c,\xi(s)}(x+1)}{\pi_{c,\xi(s)}(x)}
 \Biggr).
\endmultline
$$
Simplifications show that this expression is zero for all $m,n$
if \tht{4.7} holds. The initial condition
$\pi_{c,\xi(s)}P(s,t)\bigl|_{s=t}=\pi_{c,\xi(t)}$ is obviously
satisfied. \qed
\enddemo

Once we have a family of distributions $\pi_{c,\xi(t)}$
satisfying $ \pi_{c,\xi(s)}P(s,t)=\pi_{c,\xi(t)}$, we can define
a birth-death process by the matrix of transition probabilities
$P(s,t)$ (which is uniquely determined by the jump rates) and
one-dimensional distributions $\pi_{c,\xi(t)}$.

It is not {\it a priori} clear what is a convenient way to
parametrize these processes. In particular, multiplying both
$\al(t)$ and $\be(t)$ by the same function of $t$ leads only to
a reparametrization of time in our process. In order to
eliminate this freedom, we will always use one specific choice
of time in our processes which we call {\it interior\/} or {\it
canonical\/} time of the corresponding process. The convenience
of this choice will soon become clear.

The interior time is uniquely determined by the condition that
$\alpha(t)$ and $\beta(t)$ are expressed through $\xi(t)$ by
$$
\alpha(t)=\left(1+\frac{\Dot
\xi(t)}{2\xi(t)}\right)\frac{\xi(t)}{1-\xi(t)}\,,\qquad
\beta(t)=\left(1-\frac{\Dot \xi(t)}{2\xi(t)}\right)\frac
1{1-\xi(t)}\,. \tag4.8
$$
Evidently, these formulas imply \tht{4.7}. Moreover, for any
$(\alpha(t),\beta(t),\xi(t))$ satisfying \tht{4.7}, if
$\alpha(t)$ and $\beta(t)$ do not vanish simultaneously, we can
choose a new time variable $\tau(t)$ with
$$
\Dot\tau=\frac 12\left(\frac\alpha\xi+\beta\right)(1-\xi)
$$
so that $\bigl(\Dot\tau^{-1}\alpha(t(\tau)),\Dot\tau^{-
1}\beta(t(\tau)),\xi(t(\tau)\bigr)$ satisfy both \tht{4.7} and
\tht{4.8} as functions in $\tau$.

Thus, from now on we will parametrize our processes by
continuous, piecewise continuously differentiable functions
$\xi(t)$ taking values in $(0,1)$ such that
$|\Dot\xi(t)/\xi(t)|\le 2$ (this condition is necessary to
guarantee the nonnegativity of $\alpha$ and $\beta$). Such
curves $\xi(t)$ will be called {\it admissible}.
 Then the
corresponding birth--death process is determined by jump rates
given by \tht{4.6}, \tht{4.8} and one--dimensional distributions
$\pi_{c,\xi(t)}$. We will denote this process by $N_{c,\bxi}$.

In other words, if we set $A(t)=-\frac 12\ln \xi(t)$ then $A(t)$
has to satisfy three conditions: $A(t)\ge 0$, for all $t$; $|\Dot
A(t)|\le 1$ for all $t$; and $A(t)$ is continuous and piecewise
continuously differentiable.

In terms of $A(t)$ it is convenient to single out important
special cases: $A(t)\equiv\const$ corresponds to the homogeneous
birth--death process; $A(t)=t+\const$ corresponds to pure death
processes; and $A(t)=-t+\const$ corresponds to pure birth
processes.

Note that in case of a pure birth process $A(t)$ will hit zero in
finite time which means that in terms of the canonical time
parametrization, the process reaches infinity in a finite amount
of time.

The connection of the processes $N_{c,\bxi}$ with Meixner
polynomials discussed in the previous section is already obvious
from the fact that the distributions $\pi_{c,\xi}$ coincide, up
to a constant factor, with the weight functions $W_{c-1,\xi}$,
see \tht{2.7}. Our next goal is to express $P(s,t)$ in terms of
the Meixner polynomials. We will use the notation \tht{2.8}.

\proclaim{Proposition 4.6} The matrix $P(s,t)$ of transition
probabilities for the birth--death process $N_{c,\bxi}$  has the
form
$$
P_{xy}(s,t)=\left(\frac{\pi_{c,\xi(t)}(y)}{\pi_{c,\xi(s)}(x)}\right)^{\frac
12}\sum_{n=0}^\infty
e^{n(s-t)}\wt\M_n(x;c,\xi(s))\,\wt\M_n(y;c,\xi(t)), \tag4.9
$$
where $s\le t$ and $x,y\in\Z_+$.
\endproclaim

\demo{Comments} 1. In the stationary case $\xi(t)\equiv\const$
this formula was derived by Karlin and McGregor \cite{KMG2} as a
part of a much more general formalism, see also \cite{KMG1}.

2. The formula implies that $P(s,t)$ depends on the initial value
$\xi(s)$, final value $\xi(t)$ and the length $t-s$ of the time
interval. However, $P(s,t)$ does {\it not} depend on the behavior
of the curve $\xi(\cdot)$ inside this time interval, as one might
expect.

3. The simplicity of the factor $e^{n(s-t)}$ is a consequence of
our choice of the interior time of the process.

4. Since $\wt\M_0(x;c,\xi)=(\pi_{c,\xi}(x))^{\frac12}$, the
prefactor may be rewritten as
$$
\left(\frac{\pi_{c,\xi(t)}(y)}{\pi_{c,\xi(s)}(x)}\right)^{\frac
12}=\frac{\wt \M_0(y;c,\xi(t))}{\wt \M_0(x;c,\xi(s))}\,.
$$

5. The formula implies that the kernel on $\Z_+\times\Z_+$
$$
(x,y)\mapsto \sum_{n=0}^\infty q^n\,
\wt\M_n(x;c,\zeta)\,\wt\M_n(y;c,\eta) \tag4.10
$$
 takes nonnegative values for
$\zeta,\eta\in(0,1)$ and $ 0<q\le
\min\left\{\sqrt{\frac\zeta\eta},\sqrt{\frac\eta\zeta}\right\}$.
The bound on $q$ follows from the inequality $|\Dot \xi/\xi|\le
2$.
\enddemo

Our proof of Proposition 4.6 consists of few steps. Let us
denote the right--hand side of \tht{4.9} by $\widehat
P_{xy}(s,t)$.

First, we show that $\widehat P(s,t)$ satisfies Kolmogorov's
backward equation. Since we know that there exists only one
stochastic solution, it remains to prove that $\widehat P(s,t)$ is
a stochastic matrix.

The fact that the sum of the matrix elements along any row is
equal to 1 is obvious (only $n=0$ term gives a nonzero
contribution due to orthogonality of nonconstant Meixner
polynomials to constants). The fact that $\widehat P(s,t)$ is
always nonnegative is not so obvious. In order to prove that we
explicitly evaluate $\widehat P(s,t)$ in the cases of pure birth
and pure death processes, and then show that in the general case,
$\widehat P(s,t)$ is always a product of a ``pure death'' and a
``pure birth'' transition matrices.

\proclaim{Lemma 4.7} The following relations hold
$$
\align \sqrt{\xi (x+1)(x+c)}\,\wt\M_n(x+1;c,\xi)&+\sqrt{\xi
x(x+c-1)}\,\wt\M_n(x-1;c,\xi)\\
-(x(1+\xi)&+c\xi)\wt\M_n(x;c,\xi)=-n(1-\xi)\wt\M_n(x;c,\xi),
\tag4.11\\
2\xi(1-\xi)\frac{\partial}{\partial\xi}\,\wt\M_n(x;c,\xi)&+
\sqrt{\xi (x+1)(x+c)}\,\wt\M_n(x+1;c,\xi)\\&-\sqrt{\xi
x(x+c-1)}\,\wt\M_n(x-1;c,\xi)=0. \tag4.12
\endalign
$$
\endproclaim

\demo{Proof} Straightforward computation using
$$
\xi^2\frac{\partial}{\partial\xi}\,\M_n(x;c,\xi)=\frac{nx}{c}\,\M_{n-1}(x-
1,c+1,\xi)
$$
and \cite{KS,~1.9.5,~1.9.6,~1.9.8}.\qed
\enddemo

\demo{Proof of Proposition 4.6} First of all, we need to verify
that $\widehat P(s,t)$ (the right--hand side of \tht{4.9})
satisfies the backward equation. This is the equality
$$
-\frac{\partial}{\partial s} \widehat P_{xy}(s,t)=
Q_{xx}(s)\widehat P_{xy}(s,t)+Q_{x,x+1}(s)\widehat
P_{x+1,y}(s,t)+Q_{x,x-1}(s)\widehat P_{x-1,y}(s,t) \tag4.13
$$
with $Q_{xx}(s)=-Q_{x,x+1}(s)-Q_{x,x-1}(s)$ and
$$
Q_{x,x+1}(s)=(c+x)\left(1+\frac{\Dot
\xi(s)}{2\xi(s)}\right)\,\frac{\xi(s)}{1-\xi(s)},\quad
Q_{x,x-1}(s)=x\left(1+\frac{\Dot
\xi(s)}{2\xi(s)}\right)\,\frac{1}{1-\xi(s)}.
$$

The computation proceeds as follows. One substitutes the sum in
the right--hand side of \tht{4.9} into the needed equality
\tht{4.13} and collects the coefficients of
$\wt\M_n(y;c,\xi(t))$ using the relation $\partial/{\partial
s}=\Dot\xi(s)\,\partial/\partial (\xi(s))$. Each such
coefficient has two parts: one of them does not involve $\Dot
\xi(s)$ while the other one is equal to $\Dot \xi(s)$ times an
expression not involving $\Dot\xi(s)$. It turns out that each of
these parts vanishes, for the first part the needed relation is
\tht{4.11}, and for the second part one uses \tht{4.12}. The
details are tedious but straightforward, and we omit them.

As was mentioned before, it remains to prove that $\widehat
P_{xy}(s,t)$ is always nonnegative.

Let us use the notation $P^\up(s,t)$, $P^\down(s,t)$ for $P(s,t)$
when we consider a pure birth or a pure death process (that is,
$\xi(\tau)=e^{\const+2\tau}$ or $\xi(\tau)=e^{\const-2\tau}$,
respectively).

\proclaim{Lemma 4.8} $P^\down(s,t)$ is the unique solutions of
the backward equation for the pure death process with the
initial condition $P(t,t)\equiv \operatorname{Id}$, and
$P^\up(s,t)$ is the unique solution of the forward equation for
the pure birth process with the same initial condition.
Furthermore, with the notation $\zeta=\xi(s)$, $\eta=\xi(t)$, we
have
$$
\gather P_{xy}^\down(s,t)=\cases
\left(\dfrac{(1-\zeta)\eta}{(1-\eta)\zeta}\right)^x
\left(\dfrac{\zeta-\eta}{(1-\zeta)\eta}\right)^{x-y}
\dfrac{x!}{(x-y)!y!}\,\,,&x\ge
y,\\0,&x<y,
\endcases
\tag4.14
\\
P_{xy}^\up(s,t)=\cases \left(\dfrac{1-\eta}{1-\zeta}\right)^{c+x}
\left(\dfrac{\eta-\zeta}{1-\zeta}\right)^{y-x}
\dfrac{(c+x)_{y-x}}{(y-x)!}\,\,,&x\le y,\\0,&x>y.
\endcases
\tag4.15
\endgather
$$
\endproclaim

\demo{Proof} Consider $P^\down(s,t)$ first. Since
$Q_{x,x+1}\equiv 0$, Kolmogorov's backward equation takes the
form
$$
\multline
 -\frac{\partial}{\partial s}
P^\down_{xy}(s,t)=-Q_{x,x-1}P^\down_{xy}(s,t)+
Q_{x,x-1}P^\down_{x-1,y}(s,t)\\=\frac{2x}{1-\xi(s)}
\left(-P^\down_{xy}(s,t)+P^\down_{x-1,y}(s,t)\right).
\endmultline
\tag4.16
$$
If we fix $y$ then these differential equations can be solved
recursively: we subsequently find $P^\down_{0,y}$,
$P^\down_{1,y}$, $P^\down_{2,y}$, \dots, using the initial
conditions $P^\down_{x,y}(t,t)=\delta_{xy}$. This shows that the
backward equation for the pure death process has a unique
solution. A straightforward calculation shows that the
expression in the right--hand side of \tht{4.14} satisfies this
equation (with $\xi(s)=e^{\const-2s}$).

The case of the pure birth process is completely analogous. \qed
\enddemo

Since for the pure death process the backward equation has a
unique solution, we have just shown that $\widehat
P(s,t)=P^\down(s,t)$ is the corresponding transition matrix.

In order to make a similar conclusion for the pure birth
process, we need to know that $\widehat P(s,t)$ satisfies the
forward equation. This fact can be proved directly using Lemma
4.7. It can also be reduced to the case of the backward equation
as follows.

Note that $\pi_{c,\xi(s)}(x)\widehat P_{xy}(s,t)$ remains
invariant under the changes
$$
s\mapsto -t,\quad t\mapsto -s,\quad \xi(\tau)\mapsto
\wt\xi(\tau):=\xi(-\tau),\quad x\leftrightarrow y. \tag4.17
$$
Thus, instead of computing $\frac\partial {\partial t} \widehat
P_{xy}(s,t)$ we may compute
$$
-\frac\partial{\partial u}\left(\pi_{c,\wt\xi(u)}(y)\widehat
P_{yx}(u,v)(\pi_{c,\wt\xi(v)}(x))^{-1}\right) \Bigl|_{u=-t,v=-s}
$$
with the new $\wt\xi(\tau)$ obtained from $\xi(\tau)$ by the
time inversion. Since we already know that $\widehat
P_{xy}(s,t)$ solves the backward equation, for
$\xi(\tau)=e^{\const+2\tau}$ we obtain,\footnote{The argument
goes through for any admissible curve $\xi(\cdot)$, it just
becomes more tedious.} cf. \tht{4.16},
$$
\multline
 -\frac\partial{\partial
u}\frac{\pi_{c,\wt\xi(u)}(y)\widehat
P_{yx}(u,v)}{\pi_{c,\wt\xi(v)}(x)}
=\left(-\frac{2c\,\wt\xi(u)}{1-\wt\xi(u)}+{2y}\right)
\frac{\pi_{c,\wt\xi(u)}(y)\widehat
P_{yx}(u,v)}{\pi_{c,\wt\xi(v)}(x)}\\
+\frac{2y}{1-\wt\xi(u)}\,\frac{\pi_{c,\wt\xi(u)}(y)}{\pi_{c,\wt\xi(v)}(x)}
\Bigl(-\widehat P_{yx}(u,v)+\widehat P_{y-1,x}(u,v)\Bigr).
\endmultline
$$
Using $\pi_{c,\wt\xi(u)}(y)=\pi_{c,\wt\xi(u)}(y-1)\cdot
(c+y-1)\wt\xi(u)/y$ and substituting $u=-t$, $v=-s$, we obtain the
needed forward equation
$$
\frac\partial {\partial t} \widehat
P_{xy}(s,t)=-\frac{2(c+y)\,\xi(t)}{1-\xi(t)}\,\widehat
P_{xy}(s,t)+ \frac{2(c+y-1)\,\xi(t)}{1-\xi(t)}\,\widehat
P_{x,y-1}(s,t).
$$

The conclusion is that in case of the pure birth process,
$\widehat P(s,t)$ satisfies the forward equation, and by Lemma
4.8 we have $\widehat P(s,t)=P^\up(s,t)$.

The nonnegativity of $\widehat P(s,t)$ for arbitrary admissible
curves $\xi(\cdot)$ follows from

\proclaim{Lemma 4.9} Let $\xi(\cdot)$ be an admissible curve and
$N_{c,\bxi}$ be the corresponding birth-death process. Then for
any $s<t$, $\widehat P(s,t)$ is a product of $P^\down(s,u)$ with
$\xi(\tau)=e^{-2(\tau-s)+\ln\xi(s)}$ and $P^\up(u,t)$ with
$\xi(\tau)=e^{2(\tau-t)+\ln\xi(t)}$ for a certain choice of $u$.
Specifically, $u$ is determined from the continuity condition:
$$
e^{-2(u-s)+\ln\xi(s)}=e^{2(u-t)+\ln\xi(t)}\quad\Longleftrightarrow
\quad u=\frac{s+t}2+\frac{\ln\xi(s)-\ln\xi(t)}4\,. \tag4.18
$$
\endproclaim

\demo{Proof} This statement follows from the Chapman--Kolmogorov
equation \tht{4.1}, which $\widehat P(s,t)$ obviously satisfies
due to the orthogonality of Meixner polynomials, and from the
fact that $\widehat P(s,t)$ does not depend on the specific form
of the curve $\xi(\cdot)$, see Comment 2 after the statement of
Proposition 4.6. Thus, we may just replace $\xi(\tau)$ by a
continuous combination of $e^{-2\tau+\const}$ and
$e^{2\tau+\const}$ and preserve $\xi(s)$, $\xi(t)$, and $t-s$.
Note that the fact that $u$ given by the formula above is
between $s$ and $t$ follows from the inequality
$|\Dot\xi/\xi|\le 2$. \qed
\enddemo

Lemma 4.9 implies that $\widehat P_{xy}(s,t)$ is always
nonnegative, and this completes the proof of Proposition
4.6.\qed
\enddemo

\proclaim{Corollary 4.10} The process obtained from $N_{c,\bxi}$
by the time reversion is also of the form $N_{c,\wt\xi(\cdot)}$
with $\wt\xi(\tau)=\xi(-\tau)$.
\endproclaim

\demo{Proof} $N_{c,\bxi}$ is {\it characterized} by the fact
that it is a Markov process with two-dimensional distributions
$$
\Prob\{N_{c,\xi(s)}=x,\, N_{x,\xi(t)}=y\}=\pi_{c,\xi(s)}(x)
P_{xy}(s,t).
$$
As was already mentioned above, the right--hand side of
\tht{4.9} multiplied by $\pi_{c,\xi(s)}(x)$ is invariant with
respect to \tht{4.17}. This implies the statement.\qed
\enddemo

Note that, in particular, time inversion turns our pure birth
process into the pure death process and {\it vice versa}
(essentially, we gave a proof of this fact before Lemma 4.9),
and the stationary process $N_{c,\xi}$ with $\xi\equiv\const$ is
reversible. This is well known; any stationary birth-death
process with an invariant measure is reversible with respect to
this measure).

\subhead 4.4. Markov processes on partitions
\endsubhead
Our next goal is to extend birth--death processes $N_{c,\bxi}$
to partitions in the following sense. We construct continuous
time Markov processes on the state space $\Y$ (the set of all
Young diagrams, see \S1) parametrized by admissible pairs
$(z,z')$, see \S1, and admissible curves $\xi(\cdot)$. The
projection of such a process on $\Z_+$ obtained by looking at
the number of boxes of the random Young diagrams, coincides with
$N_{zz',\bxi}$.

Let us fix a pair $(z,z')$ of admissible parameters and set
$c=zz'>0$. Given an admissible curve $\xi(\cdot)$, we define the
matrix $Q$ of jump rates of our future Markov process
$\Lambda_{z,z',\xi}$ on $\Y$ by (set $n=|\la|$)
$$
Q_{\la \mu}(s)=\cases
(c+n)\left(1+\dfrac{\Dot\xi(s)}{2\xi(s)}\right)\,
\dfrac{\xi(s)}{1-\xi(s)}\cdot
p_{zz'}^\up(n,\la;n+1,\mu),&\la\nearrow\mu,\\
n\left(1-\dfrac{\Dot\xi(s)}{2\xi(s)}\right)\dfrac 1{1-\xi(s)}\cdot
p^\down(n,\la;n-1,\mu),&\la\searrow\mu,\\
-(c+n)\left(1+\dfrac{\Dot\xi(s)}{2\xi(s)}\right)\,
\dfrac{\xi(s)}{1-\xi(s)}-n\left(1-\dfrac{\Dot\xi(s)}{2\xi(s)}\right)\dfrac
1{1-\xi(s)},&\mu=\la,
\endcases
\tag4.19
$$
and $Q_{\la\mu}\equiv 0$ in all other cases. Here $p^\up_{zz'}$
and $p^\down$ are transition and cotransition probabilities from
\S1, see \tht{1.9} and \tht{1.1}, and the expressions involving
$\xi$ come from \tht{4.8}. Note that under the projection
$\Y\to\Z_+$, $\la\mapsto |\la|$, this matrix $Q$ turns into the
matrix of jump rates for $N_{c,\bxi}$.

\proclaim{Proposition 4.11} The minimal solution $P(s,t)$ of
Kolmogorov's backward equation with the matrix $Q$ defined above
is stochastic.
\endproclaim

\demo{Proof} We apply Propositions 4.3, 4.4. In the proof of
Proposition 4.4 it was shown that for any $a\in\Z_+$ there
exists a set of the form $X=\{0,1,\dots,n-1\}$ such that the
probability of exiting $X$ during the time period from $s$ to
$t$ with the initial state $a$ is smaller than any given
positive number $\epsi$. This means that if we start at time $s$
from $\la\in\Y$ with $|\la|=a$ then the probability of exiting
$\Y_0\cup\Y_1\cup\dots\cup\Y_{n-1}$ before time $t$ is just the
same as for the birth-death process and, hence, is less than
$\epsi$. Proposition 4.3 concludes the proof.\qed
\enddemo

\proclaim{Proposition 4.12 (cf. Proposition 4.5)} For any $s<t$
$$
M_{z,z',\xi(s)}P(s,t)=M_{z,z',\xi(t)}, \tag4.20
$$
where $P(s,t)$ is the transition matrix of Proposition 4.11, and
$M_{z,z',\xi}$ is the mixed z-measure \tht{1.11} viewed as a
row-vector with coordinates marked by elements of $\Y$.
\endproclaim

\demo{Proof} Since the formula obviously holds for $s=t$, it
suffices to show that the derivative with respect to $s$ of the
left--hand side of \tht{4.20} vanishes. Thus, it suffices to
show that
$$
-\frac{\partial} {\partial
s}\,M_{z,z',\xi(s)}(\mu)+\sum_{\la\in\Y}M_{z,z',\xi(s)}(\la)
Q_{\la\mu}=0 \tag4.21
$$
for any $\mu\in\Y$.
 Recall that
$$
M_{z,z',\xi}(\la)=M_{z,z'}^{(n)}(\la)\pi_{c,\xi}(n)\quad\text{
with}\quad  c=zz',\quad n=|\la|.
$$
Substituting this relation into \tht{4.21} we notice that we can
perform the summation over $\la$ using \tht{1.5} and \tht{1.6}.
Factoring out $M_{z,z'}^{(|\mu|)}(\mu)$ leads to the formula
which states that the derivative of $\pi_{c,\xi(s)}P(s,t)$ with
$P(s,t)$ being the transition matrix for the birth-death process
$N_{c,\xi}$, with respect to $s$ vanishes. But this has already
been proved in Proposition 4.5. \qed
\enddemo

We conclude that given an admissible pair $(z,z')$ and an
admissible curve $\xi(\tau)$, there exists a unique continuous
time Markov process on $\Y$ with jump rates $Q$ defined above
and with one--dimensional distributions $M_{z,z',\xi(\tau)}$.
This Markov process will be denoted by $\La_{z,z',\bxi}$.

As for the birth--death processes, we single out three important
special cases: the stationary process $\xi\equiv\const$, the
{\it ascending process} $\xi(\tau)=e^{2\tau+\const}$ and the
{\it descending process} $\xi(\tau)=e^{-2\tau+\const}$. The
projections of these processes on $\Z_+$ are the stationary
birth--death process, the pure birth and the pure death
processes, respectively.

As in \S1, for $\la\in\Y$ we denote by $\dim\la$ the number of
ascending paths in the Young graph leading from $\varnothing$ to
$\la$. More generally, we denote by $\dim(\mu,\la)$ the number of
ascending paths in $\Y$ leading from $\mu$ to $\la$; if there are
no such paths we set $\dim(\mu,\la)=0$. Also, for $\mu,\,\la\in
\Y$ such that $\mu\subset\la$ we set
$$
(x)_{\la\setminus\mu}=\prod_{(i,j)\in\la\setminus \mu}(x+j-i),
\qquad x\in\C,
$$
where the product is taken over all boxes in $\la\setminus\mu$.

\proclaim{Proposition 4.13 (cf. Lemma 4.8)}  The transition
matrix of the descending process $\La_{z,z',\bxi}$ has the form
$$
P_{\la\mu}^\down(s,t)=\left(\dfrac{(1-\zeta)\eta}{(1-\eta)\zeta}\right)^x
\left(\dfrac{\zeta-\eta}{(1-\zeta)\eta}\right)^{x-y}
\dfrac{x!}{(x-y)!\,y!}\, \frac{\dim\mu\,\dim(\mu,\la)}{\dim\la}
\tag4.22
$$
and the transition matrix of the ascending process
$\La_{z,z',\bxi}$ has the form
$$
P_{\la\mu}^\up(s,t)=\left(\dfrac{1-\eta}{1-\zeta}\right)^{zz'+x}
\left(\dfrac{\eta-\zeta}{1-\zeta}\right)^{y-x}
\dfrac{x!}{(y-x)!\,y!}\,\frac{\dim\mu\,\dim(\la,\mu)
}{\dim\la}\cdot(z)_{\mu\setminus\la}(z')_{\mu\setminus\la}
\tag4.23
$$
where $\zeta=\xi(s)$, $\eta=\xi(t)$, $x=|\la|$, $y=|\mu|$.
\endproclaim

\demo{Proof} Let us consider the descending process first. It is
immediate to check that the matrix $P_{\la\mu}^\down(s,t)$
obtained from the transition matrix $P_{xy}^\down(s,t)$ of the
pure death process by
$$
\multline P_{\la\mu}^\down(s,t)=P_{xy}^\down(s,t)\\ \times\sum
p^\down(x,\la;x-1,\mu^{(x-y-1)})
p^\down(x-1,\mu^{(x-y-1)};x-2,\mu^{(x-y-2)})\cdots
p^\down(y+1,\mu^{(1)};y,\mu)
\endmultline
$$
where the sum is taken over all paths ${\mu=\mu^{(0)}
\nearrow\mu^{(1)}\nearrow\dots\nearrow \mu^{(x-y)}=\la}$ from
$\mu$ to $\la$, satisfies the backward equation. All terms in
the above sum are equal to $\dim\mu/\dim\la$, and the number of
terms is equal to $\dim(\mu,\la)$. Together with \tht{4.14} this
implies \tht{4.22}.

Similarly, for the ascending process one has
$$
\multline P_{\la\mu}^\up(s,t)=P_{xy}^\up(s,t)\\ \times\sum
p^\up_{zz'}(x,\la;x+1,\la^{(1)})
p^\up_{zz'}(x+1,\la^{(1)};x+2,\la^{(2)})\cdots
p^\up_{zz'}(y-1,\la^{(y-x-1)};y,\mu)
\endmultline
$$
where the sum is taken over all paths ${\la=\la^{(0)}
\nearrow\la^{(1)}\nearrow\dots\nearrow \la^{(x-y)}=\mu}$ from
$\la$ to $\mu$. Again, the product of transition probabilities
does not depend on the path and it is equal to
$$
\frac 1{(c+x)_{y-x}}\frac{x!}{y!}\frac{\dim\mu}{\dim\la}\cdot
(z)_{\mu\setminus\la}(z')_{\mu\setminus\la}
$$
while the number of paths is equal to $\dim(\la,\mu)$. Together
with \tht{4.15} this gives \tht{4.23}.\qed
\enddemo

\head 5. Transition matrix for integral values of $z$.
\endhead

Our main goal in this section is to obtain a formula for the
transition matrix of the process $\La_{z,z',\bxi}$ in the case
when $z$ is a nonnegative integer. For $z=1$, the process
$\La_{z,z',\bxi}$ coincides with the birth--death process
$N_{z',c}$ (because it lives on the Young diagrams with only one
row), and our formula is reduced to \tht{4.9}.

Fix $z=N\in\{1,2,\dots\}$. In order for $(z,z')$ to be an
admissible pair, we must have $z'\in\R$ and $z'>N-1$. We will
use the notation $z'=N+\al$, $\al>-1$. As before, we set
$c=zz'=N(N+\al)$.

As was mentioned in \S1, the support of $M_{N,N+\al,\xi}$ consists
of the Young diagrams with no more than $N$ rows. It is convenient
to parameterize such diagrams $\la$ by sequences of $N$ strictly
decreasing nonnegative integers $(x_1,\dots,x_N)$,
$$
x_i=\la_i+N-i,\qquad  i=1,\dots,N.
$$
Given an admissible curve $\xi(\cdot)$, set
$$
v_{s,t}(x,y)=\sum_{k=0}^\infty e^{k(s-t)}\wt\M_k(x;\al,\xi(s))\,
\wt\M_k(y;\al,\xi(t)),\qquad x,y\in\Z_+. \tag5.1
$$

\proclaim{Theorem 5.1} Let $\la,\,\mu$ be Young diagrams with no
more than $N$ rows, and let $(x_1,\dots,x_N)$, $(y_1,\dots,y_N)$
be the corresponding sets of decreasing nonnegative integers.
For any admissible curve $\xi(\cdot)$ the transition matrix of
the Markov process $\La_{N,N+\al,\bxi}$ has the form
$$
P_{\la\mu}(s,t)=e^{\frac{(t-s)N(N-1)}{2}}
\left(\frac{M_{N,N+\al,\xi(t)}(\mu)}
{M_{N,N+\al,\xi(s)}(\la)}\right)^{\frac12}
\det\bigl[v_{s,t}(x_i,y_j)\bigr]_{i,j=1}^N. \tag5.2
$$
\endproclaim

We will use the term {\it Karlin--McGregor representation} for
this formula.

\demo{Proof} The arguments follow the same pattern as in the
proof of Proposition 4.6 (which is a special case of this
theorem). The first step is to show that the right--hand side of
\tht{5.2} satisfies Kolmogorov's backward equation. After that
we prove that this solution is stochastic.

We will use the notation
$$
\gathered  x=(x_1,\dots,x_N),\quad
y=(y_1,\dots,y_N),\qquad
\epsi_r=(0,\dots,0,\overset{(r)}\to1,0,\dots,0),\quad 1\le r\le N,\\
\ze=\xi(s),\quad
\eta=\xi(t),\qquad n=|\la|=\sum_{i=1}^Nx_i-\frac{N(N-1)}2\,,\\
f_k(\,\cdot\,)=\wt\M_k(\,\cdot\,;\alpha,\xi(s)),\qquad
g_k(\,\cdot\,)=\wt\M_k(\,\cdot\,;\alpha,\xi(t)).
\endgathered
$$
Also, denote the right--hand side of \tht{5.2} by $\widehat
P_{xy}(s,t)$.

The formulas of \S1 imply
$$
\left(\frac{M_{N,N+\al,\xi(t)}(\mu)}
{M_{N,N+\al,\xi(s)}(\la)}\right)^{\frac12}=
\frac{\eta^{\frac{|\mu|}2}(1-\eta)^{\frac
c2}}{\ze^{\frac{|\la|}2}(1-\ze)^{\frac c2}} \left(\prod_{j=1}^N
\frac{\Gamma(y_j+\al+1)\Ga(x_j+1)}{\Ga(x_j+\al+1)\Ga(y_j+1)}
\right)^{\frac 12}\,\frac{V(y)}{V(x)} \tag5.3
$$
where
$V(u)=\prod_{1\le i<j\le N}(u_i-u_j)$ stands for the Vandermonde
determinant.

Note that adding to $\la$ one box or removing from $\la$ one box
is equivalent to adding $\epsi_r$ to $x$ or subtracting
$\epsi_r$ from $x$, where $r$ is the row number of the box. We
have, cf. \tht{1.9}, \tht{1.1},
$$
\gather p^\up_{N,N+\al}(n,x;n+1,x+\epsi_r)=\frac{x_r+\al+1}{c+n}\,
\frac{V(x+\epsi_r)}{V(x)}\,,\\
p^\down(n,x;n-1,x-\epsi_r)=\frac{x_r}{n}\,\frac{V(x-\epsi_r)}{V(x)}\,.
\endgather
$$
The right-hand sides of these relations conveniently vanish
exactly when $x\pm\epsi_r$ does not represent a Young diagram
(two of the coordinates of $x\pm\epsi_r$ are equal). Using
\tht{4.19} we can now write down the needed backward equation:
$$
\align -\frac{\partial}{\partial s}\,\widehat P_{xy}(s,t)=
&-\left(\left(1+\dfrac{\Dot\xi(s)}{2\xi(s)}\right)\,
\dfrac{\xi(s)(c+n)}{1-\xi(s)}+\left(1-\dfrac{\Dot\xi(s)}{2\xi(s)}\right)\dfrac
n{1-\xi(s)}\right)\widehat P_{xy}(s,t)\\
&+\left(1+\dfrac{\Dot\xi(s)}{2\xi(s)}\right)\frac{\xi(s)}{1-\xi(s)}\sum_{r=1}^N
\frac{(x_r+\al+1)V(x+\epsi_r)}{V(x)}\cdot \widehat
P_{x+\epsi_r,y}(s,t) \\&+
\left(1-\dfrac{\Dot\xi(s)}{2\xi(s)}\right)\dfrac 1{1-\xi(s)}
\sum_{r=1}^N\frac{x_rV(x-\epsi_r)}{V(x)}\cdot \widehat
P_{x-\epsi_r,y}(s,t).
\endalign
$$
It is time to use the definition of $\widehat P_{xy}(s,t)$. The
Cauchy--Binet identity (see, e.g., \cite{Ga, ch. I, \S2})
implies (in all determinants below the indices run from 1 to
$N$)
$$
\multline
\det\bigl[v_{s,t}(x_i,y_j)\bigr]=\det\left[\sum_{k=0}^\infty
e^{(s-t)k}f_k(x_i)g_k(y_j)\right]\\=\sum_{k_1>k_2>\dots>k_N\ge 0}
e^{(s-t)(k_1+\dots+k_N)}\det[f_{k_i}(x_j)]\det[g_{k_i}(y_j)].
\endmultline
\tag5.4
$$
Let us use this relation and \tht{5.2} for $\widehat
P_{xy}(s,t),\,\widehat P_{x\pm\epsi_r,y}(s,t)$ in the backward
equation and collect the coefficients of $\det[g_{k_i}(y_j)]$.
Factoring out
$$
\left(\frac{M_{N,N+\al,\xi(t)}(\mu)}
{M_{N,N+\al,\xi(s)}(\la)}\right)^{\frac12}
e^{(s-t)\left(k_1+\dots+k_N-\frac{N(N-1)}2\right)}
$$
we obtain (Vandermonde
determinants cancel out quite conveniently)
$$
\gathered
\left(\frac{N(N-1)}2-(k_1+\dots+k_N)+\frac{\Dot\xi(s)}{2}\left(\frac
n\ze-\frac c{1-\ze}
\right)\right)\det[f_{k_i}(x_j)]\\-\frac{\partial}{\partial
s}\det[f_{k_i}(x_j)]
=-\left(\left(1+\dfrac{\Dot\xi(s)}{2\ze}\right)\,
\dfrac{\ze(c+n)}{1-\ze}+\left(1-\dfrac{\Dot\xi(s)}{2\ze}\right)\dfrac
n{1-\ze}\right)\det[f_{k_i}(x_j)]\\+
\left(1+\dfrac{\Dot\xi(s)}{2\ze}\right)\frac{1}{1-\ze}\sum_{r=1}^N
\sqrt{(x_r+1)(x_r+\al+1)\ze}\,\det\bigl[f_{k_i}((x+\epsilon_r)_j)\bigr]\\
+\left(1-\dfrac{\Dot\xi(s)}{2\ze}\right)\frac{1}{1-\ze}\sum_{r=1}^N
\sqrt{x_r(x_r+\al)\ze}\,\det\bigl[f_{k_i}((x-\epsilon_r)_j)\bigr].
\endgathered
\tag5.5
$$
We claim that the part the left--hand side of this relation that
does not involve derivatives $\Dot\xi(s)$ and
$-\frac{\partial}{\partial s}$ equals the part of the
right--hand side without $\Dot\xi(s)$, and the part of the
left--hand side with derivatives equals that of the right--hand
side with $\Dot\xi(s)$.

The part without derivatives gives
$$
\multline
\left(\ze(c+n)+n+\frac{N(N-1)}2\,(1-\ze)-(k_1+\dots+k_N)(1-\xi)\right)
\det[f_{k_i}(x_j)]
\\=\sum_{r=1}^N
\Bigl(\sqrt{(x_r+1)(x_r+\al+1)\ze}\,\det\bigl[f_{k_i}((x+\epsilon_r)_j)\bigr]
\\+\sqrt{x_r(x_r+\al)\ze}
\,\det\bigl[f_{k_i}((x-\epsilon_r)_j)\bigr]\Bigr).
\endmultline
\tag5.6
$$
The right--hand side of this equality can be rewritten as
$$
\multline \sum_{\sigma\in S_N}\sum_{r=1}^N
\operatorname{sgn}\sigma\, f_{k_{\sigma(1)}}(x_1)\cdots
f_{k_{\sigma(N)}}(x_{N})\\
\times \frac{\sqrt{(x_r+1)(x_r+\al+1)\ze}\,
f_{\sigma(r)}(x_r+1)+\sqrt{x_r(x_r+\al)\ze}\,f_{\sigma(r)}(x_r-1)}
{f_{\sigma(r)}(x_r)}\,.
\endmultline
$$
By \tht{4.11} the last ratio equals
$x_r(1+\ze)+(\al+1)\ze-k_{\sigma(r)}(1-\ze)$, and the whole
expression equals
$$
\bigl((x_1+\dots+x_N)(1+\ze)+N(\al+1)\ze-(k_1+\dots+k_N)(1-\ze)\bigr)
\det[f_{k_i}(x_j)],
$$
which is exactly the left--hand side of \tht{5.6} with
$c=N(N+\al)$ and $n=\sum x_i-N(N-1)/2$.

The part of \tht{5.5} with derivatives gives
$$
\multline
-\frac{2\ze(1-\ze)}{\Dot\xi(s)}\,\frac{\partial}{\partial
s}\det[f_{k_i}(x_j)]= \sum_{r=1}^N\Bigl(
\sqrt{(x_r+1)(x_r+\al+1)\ze}\,\det\bigl[f_{k_i}((x+\epsilon_r)_j)\bigr]
\\-\sqrt{x_r(x_r+\al)\ze}
\,\det\bigl[f_{k_i}((x-\epsilon_r)_j)\bigr]\Bigr).
\endmultline
$$
The same operation with determinants and \tht{4.12} show that
this equality holds.

This concludes the proof of the fact that $\widehat
P_{\la\mu}(s,t)$ (the right--hand side of \tht{5.2}) satisfies
the backward equation for $\La_{z,z',\bxi}$. It remains to
verify that the $\Y\times\Y$ matrix $\widehat P_{\la\mu}(s,t)$
is stochastic.

Let us check that $\sum_\mu \widehat P_{\la\mu}(s,t)=1$ for any
$\la\in\Y$ and $s<t$.

Using elementary row operations on the Vandermonde matrix, we
obtain
$$
V(u)=\prod_{1\le i<j\le N}(u_i-u_j)=\det
\bigl[u_j^{N-i}\bigr]=(-1)^{\frac
{N(N-1)}2}\det\bigl[\M_{N-i}(u_j;\al,\xi)\bigr] \tag5.7
$$
where the sign $(-1)^{\frac {N(N-1)}2}$ appears because the
highest coefficient of $\M_{k}(u;\al,\xi)$ is $(-1)^k$. Then
\tht{5.2} and \tht{5.3} imply
$$
\widehat P_{\la\mu}(s,t)=e^{\frac{(t-s)N(N-1)}{2}}
\frac{\det[g_{N-i}(y_j)]}{\det
[f_{N-i}(x_j)]}\,\det[v_{s,t}(x_i,y_j)]\,.
$$
Apply \tht{5.4} and sum the result over all $y_1>\dots>y_N\ge
0$. The Cauchy--Binet identity  gives
$$
\sum_{y_1>\dots>y_N\ge 0}\det[g_{k_i}(y_j)]\det[g_{N-i}(y_j)]=
\det\left[\sum_{u=0}^\infty g_{k_i}(u)g_{N-j}(u)\right]. \tag5.8
$$
Orthogonality of Meixner polynomials means that $\sum_{u\ge
0}g_k(u)g_l(u)=\delta_{kl}$. Hence, the last determinant equals 1
if $k_i=N-i$ for all $i=1,\dots,N$ and vanishes otherwise. This
gives the desired result.

Finally, the nonnegativity of $\widehat P_{\la\mu}(s,t)$ follows
from Proposition 4.13 and

\proclaim{Lemma 5.2 (cf. Lemma 4.9)} For any admissible curve
$\xi(\cdot)$, the matrix $\widehat P(s,t)$ given by the
right--hand side of \tht{5.2} is the product of the transition
matrix $P^\down(s,u)$ for the descending process with
$\xi(\tau)=e^{-2(\tau-s)+\ln\xi(s)}$ and the transition matrix
$P^\up(u,t)$ for the ascending process with
$\xi(\tau)=e^{2(\tau-t)+\ln\xi(t)}$ with $u$ given by
\tht{4.18}:
$$
u=\frac{s+t}2+\frac{\ln\xi(s)-\ln\xi(t)}4\,.
$$
\endproclaim

The proof of this lemma is very similar to that of Lemma 4.9.
The Chapman--Kolmogorov equation for $\widehat P(s,t)$ is easily
verified by means of \tht{5.4}, the orthogonality of Meixner
polynomials, and the trick with the Cauchy--Binet formula used
above.

Since the transition matrices of the descending and ascending
processes have nonnegative matrix elements \tht{4.22},
\tht{4.23}, the matrix elements of $\widehat P(s,t)$ are also
nonnegative, and the proof of Theorem 5.1 is complete. \qed
\enddemo

\proclaim{Corollary 5.3} The kernel \tht{4.10} is totally
positive.
\endproclaim

Indeed, Theorem 5.1 shows that the minors of this kernel are, up
to positive factors, transition probabilities.

\head \S6. Analytic continuation
\endhead

The goal of this section is to employ the analytic continuation
in the parameters $z,z'$ in two directions. First, we show that
the factorization of the transition matrix for $\La_{z,z',\bxi}$
into the product of transition matrices for descending and
ascending processes is carried over from the case of integral
$z$ (Lemma 5.2 above) to the general case. Second, we prove an
analog of Lemma 3.11 for finite-dimensional distributions of
$\La_{z,z',\bxi}$. This result will be used later to compute the
dynamical correlation functions of $\La_{z,z',\bxi}$.

\proclaim{Proposition 6.1} The statement of Lemma 5.2 holds for
arbitrary admissible pair of parameters $(z,z')$. That is, the
transition matrix $P(s,t)$ for $\La_{z,z',\bxi}$ is equal to the
product $P^\down(s,u)P^\up(u,t)$ with the descending and
ascending processes and the time moment $u$ specified as in
Lemma 5.2.
\endproclaim
\demo{Comments} 1. This statement together with Proposition 4.13
allows to write $P(s,t)$ down rather explicitly. Namely, we
obtain
$$
\multline
P_{\la\nu}(s,t)=\frac{(\ze-\theta)^l(\eta-\theta)^n(1-\eta)^{zz'}}
{\ze^l(1-\theta)^{l+n+zz'}}\,\frac{l!\dim\nu}{n!\dim\la}\,
(z)_\nu(z')_\nu\\ \times \sum_{\mu\in\Y}
\frac{(1-\ze)^m(1-\theta)^m\eta^m}{(l-m)!(n-m)!}\,
\frac{\dim(\mu,\la)\dim(\mu,\nu)}{(z)_\mu(z')_\mu}\,
\endmultline
\tag6.1
$$
with the notation
$$
\ze=\xi(s),\quad \eta=\xi(t),\quad
\theta=e^{s-t}\sqrt{\ze\eta},\qquad l=|\la|,\quad m=|\mu|,\quad
n=|\nu|.
$$
The sum above ia actually finite; $\mu$ ranges over all Young
diagrams which are smaller than both $\la$ and $\nu$.

2. The formula for $P_{\la\mu}(s,t)$ above after the
multiplication by
$$
M_{z,z',\ze}(\la)=(1-\ze)^{zz'}\ze^l
(z)_\la(z')_\la\,\frac{\dim^2\la}{l!}
$$
becomes symmetric with respect to $(\la,\ze)\longleftrightarrow
(\nu,\eta)$. This shows that the time reversion of
$\La_{z,z',\bxi}$ is again a process of this form with new
$\wt\xi(\tau)=\xi(-\tau)$, cf. Corollary 4.10.
\enddemo
\demo{Proof} Set $\widehat P(s,t)=P^\down(s,u)P^\up(u,t)$.
Clearly, this is a stochastic matrix that satisfies the initial
condition $\widehat P(t,t)=\operatorname{Id}$. Thus, it suffices
to verify that it satisfies Kolmogorov's backward equation
$$
-\frac{\partial}{\partial s}\,\widehat
P_{\la\nu}(s,t)=Q_{\la\la}(s)\widehat
P_{\la\la}(s,t)+\sum_{\mu:\la\nearrow\mu}Q_{\la\mu}(s)\widehat
P_{\mu\nu}(s,t)+\sum_{\mu:\la\searrow\mu}Q_{\la\mu}(s)\widehat
P_{\mu\nu}(s,t)
$$
with $Q(s)$ given by \tht{4.19}. As we substitute the expression
in the right--hand side of \tht{6.1} for $\widehat P(s,t)$, we
see that both sides of the equality above, as function in $z$
and $z'$, have the form
$$
\left(\frac{1-\eta}{1-\theta}\right)^{zz'}\times\bigl\{\text{a
polynomial in $z,z'$}\bigr\}.
$$
Since the equality has already been established for
$(z,z')=(N,N+\al)$ with $N=1,2,\dots$ and $\al>-1$, it must hold
for arbitrary $(z,z')$. \qed
\enddemo

The statement that we prove next will be used in the derivation
of the dynamical correlation functions later in the next
section.

Take an admissible pair of parameters $(z,z')$ and an admissible
curve $\xi(\cdot)$, and consider the Markov process
$\La_{z,z',\bxi}$. Let $t_1<t_2<\dots<t_n$ be arbitrary time
moments. Set
$$
\xi_i=\xi(t_i), \qquad
\eta_{i,i+1}=e^{t_i-t_{i+1}}\sqrt{\xi_i\xi_{i+1}}.
$$

Proposition 6.1 (or formula \tht{6.1}) and the fact that the
process is Markovian imply that the finite-dimensional
distributions
$$
\Prob\{\La_{z,z',\bxi}(t_i)=\la(i);\quad i=1,\dots,n\}
$$
with given Young diagrams $\la_1,\dots,\la_n$, depend on
parameters $\xi_1,\dots,\xi_n$ and $\eta_{12}$,
$\eta_{23},\dots,\eta_{n-1,n}$ but they do not depend on the
behavior of $\xi(t)$ inside the intervals $(t_i,t_{i+1})$. Thus,
in order to compute these finite-dimensional distributions we may
replace our process by a sequence of alternating descending and
ascending processes: We start off at the time moment $t_1$ and go
down till time
$$
u_{12}=\frac{t_1+t_2}2+\frac{\ln\xi_1-\ln\xi_2}4\,,
$$
the value of $\xi$ at this moment is exactly $\eta_{12}$. Then we
go up till $\xi_2$ and then again down, etc. The time moments when
we change directions are
$$
t_1<u_{12}<t_2<u_{23}<\dots<u_{n,n+1}<t_n
$$
with $u_{i,i+1}=(t_i+t_{i+1})/2+(\ln\xi_i-\ln\xi_{i+1})/4$, and
the values of $\xi$ at these points are $\eta_{i,i+1}$. At $t_i$'s
we switch from going up to going down, and at $u_{i,i+1}$'s we
switch from going down to going up.

Fix arbitrary subsets $\Cal D_1,\Cal D_2,\dots,\Cal D_n$; $\Cal
D_{12},\Cal D_{23},\dots,\Cal D_{n-1,n}$ of $\Y$. Let us compute
the probability that our new descending-ascending process hits
all $\Cal D_i$, $\Cal D_{j,j+1}$ at the time moments $t_i$,
$u_{j,j+1}$, respectively. It is equal to the sum
$$
\multline
\sum_{\Sb \la(i)\in \Cal D_i,\, i=1,\dots,n\\
\mu(j,j+1)\in \Cal D_{j,j+1},\, j=1,\dots,n-1\endSb}
M_{z,z',\xi_1}(\la(1))\,P^\down_{\la(1),\,\mu(1,2)}(t_1,u_{12})
P^\up_{\mu(1,2),\,\la(2)}(u_{12},t_2)\cdots\\
\cdots P^\down_{\la(n-1),\,\mu(n-1,n)}(t_{n-1},u_{n-1,n})
P^\up_{\mu(n-1,n),\,\la(n)}(u_{n-1,n},t_n).
\endmultline
\tag6.2
$$
Proposition 4.13 shows that this is in fact a function in
$\xi_i,\eta_{i,i+1}$ which we will denote by $F(\xi,\eta)$ with
$\xi=(\xi_1,\dots,\xi_n)$,
$\eta=(\eta_{12},\dots,\eta_{n-1,n})$. The parameters $\xi_i$
and $\eta_{i,i+1}$ may take arbitrary values between 0 and 1
subject to the inequalities
$$
\xi_1\ge\eta_{12}\le\xi_2\ge\eta_{23}\le\xi_3\ge\dots
\le\xi_{n-1}\ge\eta_{n,n-1}\le\xi_n. \tag6.3
$$

\proclaim{Proposition 6.2 (cf. Lemma 3.11)} (i) The function
$F(\xi,\eta)$ is a real-analytic function in $\xi$ and $\eta$.

(ii) The function $\epsi\mapsto F(\epsi\xi;\epsi\eta)$ which has
been defined so far for $\epsi>0$, can be analytically continued
to a nonempty disc of the form $|\epsi|<\const$. The
coefficients of the Taylor decomposition of this function at
$\epsilon=0$ are polynomial functions in $z,z'$.
\endproclaim

\demo{Comment} This statement implies that the function
$F(\xi,\eta)$ viewed also as a function in $z,z'$, is uniquely
determined by its values on the arguments $(\xi,\eta,z,z')$ with
arbitrary $\xi,\eta$, $0<\xi_i,\eta_{j,j+1}<1$, satisfying
\tht{6.3} and $(z,z')=(N,N+\alpha)$ with $N=1,2,\dots$, and
$\al>-1$. This uniqueness will be used in the next section to
extend certain formulas derived in the case of integral $z$ to
the general case.
\enddemo

\demo{Proof} In order to prove (i), by Weierstrass' uniform
convergence theorem it suffices to check that the series
\tht{6.2} with $M_{z,z',\xi_1}(\la(1))$ and all $P^\down$,
$P^\up$ replaced by their expressions given by \tht{1.11},
\tht{4.22}, \tht{4.23}, converges absolutely and uniformly in
$\xi,\eta$ varying in small discs around their values. It is
more convenient to work with the case when all $\Cal D_i$ and
$\Cal D_{i,i+1}$ coincide with $\Y$; clearly, the needed
convergence of the restricted sum follows from that of the
unrestricted sum.

Set $l_i=|\la(i)|$, $m_{i,i+1}=|\mu(i,i+1)|$. As seen from the
proof of Proposition 4.13, the matrix elements
$P^\down_{\la\mu}$ and $P^\up_{\la\mu}$ split into products of
transition probabilities for pure death and pure birth processes
of Lemma 4.8 and (co)transition probabilities on the Young
graph. By \tht{1.11}, $M_{z,z',\xi}(\la)$ is also a product of
the negative binomial distribution $\pi_{c,\xi}(n)$ on $\Z_+$
and the probability distributions $M_{z,z'}^{(n)}(\la)$ on
$\Y_n$'s. The probabilities related to the Young graph do not
depend on $\xi$ and $\eta$. Thus, we can split the sum in
\tht{6.2} (remember that all $\Cal D_i$ and $\Cal D_{i,i+1}$'s
are equal to $\Y$) into two: the outer sum is taken over all
nonnegative integers $l_1,\dots,l_n$ and
$m_{12},\dots,m_{n-1,n}$ satisfying
$$
l_1\ge m_{12}\le l_2\ge m_{23}\le\dots\ge m_{n-1,n}\le l_n,
\tag6.4
$$
and the inner sum is taken over all Young diagrams
$\la(i)\in\Y_{l_i}$, $\mu(i,i+1)\in\Y_{m_{i,i+1}}$. The
relations \tht{1.5}, \tht{1.6} applied to the z-measures imply
that the inner sum is equal to
$$
\multline
\pi_{zz',\xi_1}(l_1)\,P^\down_{l_1,\,m_{12}}(t_1,u_{12})
P^\up_{m_{12},\,l_2}(u_{12},t_2) \cdots\\ \cdots
P^\down_{l_{n-1},\,m_{n-1,n}}(t_{n-1},u_{n-1,n})
P^\up_{m_{n-1,n},\,l_n}(u_{n-1,n},t_n) \endmultline \tag6.5
$$
with $P^\down$ and $P^\up$ given by Lemma 4.8. Thus, we need to
verify the uniform convergence of the sum of such products taken
over nonnegative integers satisfying \tht{6.4}. Denote
$$
\align
&u_i=\frac{\xi_i-\eta_{i-1,i}}{1-\eta_{i-1,i}},\qquad\qquad\quad
i=1,\dots,n,\quad\text{with}\quad \eta_{0,1}:=0,\\
&v_{i,i+1}=\frac{(1-\xi_i)\,\eta_{i,i+1}}
{(1-\eta_{i,i+1})\,\xi_i},\,\qquad i=1,\dots,n-1.
\endalign
$$

The inequalities \tht{6.3} imply that $u_i\in(0,1)$ and
$v_{i,i+1}\in (0,1]$ for all $i$. Clearly, small (complex)
variations of $\xi$ and $\eta$ lead to small (complex)
variations of $u_i$'s and $v_{i,i+1}$'s. What is important for
us here is that if the variations are small enough than $u_i$'s
are bounded away from 1 and $v_{i,i+1}$'s are bounded away from
0.

Take the absolute value of \tht{6.5} and sum it over $l_n$. We
have
$$
\gathered
 \sum_{l_n\ge
m_{n-1,n}}\left|P^\up_{m_{n-1,n},\,l_n}(u_{n-1,n},t_n)\right|=
|1-u_n|^{zz'+m_{n-1,n}} \\ \times\sum_{l_n\ge
m_{n-1,n}}|u_n|^{l_n-m_{n-1,n}}\,\frac{(c+m_{n-1,n})_{l_n-m_{n-1,n}}}
{(l_n-m_{n-1,n})!}=\left(\frac{|1-u_n|}{1-|u_n|}\right)^{c+m_{n-1,n}}.
 \endgathered
 \tag6.6
$$
We conclude that this expression can be estimated, as a function
of $m_{n-1,n}$, by a constant times a geometric progression of the
form $r^{m_{n-1,n}}$ with a suitable $r>0$. If the variation of
$u_n$ is small enough, it is close to the real axis. Hence, by
decreasing the variations we can take $r$ arbitrarily close to 1.

Let us substitute this estimate into \tht{6.5} and sum over
$m_{n-1,n}$. We obtain
$$
\gathered \const \sum_{m_{n-1,n}\le
l_{n-1}}\left|P^\down_{l_{n-1},\,m_{n-1,n}}(t_{n-1},u_{n-1,n})\right|=
\const|v_{n-1,n}|^{l_{n-1}}\\ \times\sum_{m_{n-1,n}\le
l_{n-1}}\left|\frac1{v_{n-1,n}}-1\right|^{l_{n-1}-m_{n-1,n}}r^{m_{n-1,n}}
\,\frac{l_{n-1}!}{(l_{n-1}-m_{n-1,n})!m_{n-1,n}!}\\
=\const
\left(\left|\frac1{v_{n-1,n}}-1\right|+r\right)^{l_{n-1}}
|v_{n-1,n}|^{l_{n-1}}.
 \endgathered
$$
Again, this is bounded by a constant times a geometric progression
$\wt r^{l_{n-1}}$ where $\wt r$ can be made arbitrarily close to 1
by considering small enough variations of $u_n$ and $v_{n-1,n}$.

The next step, summation over $l_{n-1}$, is performed similarly
to \tht{6.6}. The only difference is in the presence of the
additional geometric progression $\wt r^{l_{n-1}}$. The
summation yields
$$
\const \left(\frac{|1-u_{n-1}|}{1-\wt
r\,|u_{n-1}|}\right)^{a+m_{n-2,n-1}}.
$$
Once again, for small variations of $u_{n-1}, v_{n-1,n}, u_n$,
this is bounded by a constant times a geometric progression with
exponent $m_{n-2,n-1}$ and a ratio that is close to 1. Induction
on $n$ and the presence of $\xi_{l_1}^{l_1}$ in
$\pi_{zz',\xi}(l_1)$ complete the proof of the uniform
convergence of the series.

Let us prove (ii). The first step is the same: the sum \tht{6.2}
is split into the outer sum over nonnegative integers
$l_1,\dots,l_n$ and $m_{12},\dots,m_{n-1,n}$ satisfying
\tht{6.4}, and the inner sum  over Young diagrams $\la(i)$,
$\mu(i,i+1)$ with $|\la(i)|=l_i$, $|\mu(i,i+1)|=m_{i,i+1}$
restricted by $\la(i)\in \Cal D_i$ and $\mu(i,i+1)\in \Cal
D_{i,i+1}$. As was mentioned above, if all $\Cal D$'s are equal
to $\Y$, the inner sum yields \tht{6.5}. Since all the summands
are nonnegative, with arbitrary $\Cal D$'s the inner sum yields
\tht{6.5} multiplied by a constant which depends on $l$'s and
$m$'s, does not depend on $\xi$ and $\eta$, and is between 0 and
1. It is also worth noting that these constants are polynomials
in $z$ and $z'$ because the inner sums are always finite.

As we replace $\xi$ and $\eta$ by $\epsi\xi$ and $\epsi\eta$ in
the expression obtained by using the formulas of Lemma 4.8 in
\tht{6.5}, we use the estimates
$$
\gathered \left|\left(\dfrac{(1-\epsi\xi)\epsi\eta}
{(1-\epsi\eta)\epsi\xi}\right)^l
\left(\dfrac{\epsi\xi-\epsi\eta}{(1-\epsi\xi)\epsi\eta}\right)^{l-m}
\dfrac{l!}{(l-m)!l!}\right|\le \const^l,\\
\left|\left(\dfrac{1-\epsi\xi}{1-\epsi\eta}\right)^{zz'+m}
\left(\dfrac{\epsi\xi-\epsi\eta}{1-\epsi\eta}\right)^{l-m}
\dfrac{(c+x)_{l-m}}{(l-m)!}\right|\le |\epsi|^l\cdot\const^{l-m},
\endgathered
$$
where the bounds are uniform in $\epsi$ varying in the unit disc
$|\epsi|\le 1$. These estimates together with
$|\pi_{c,\epsi\xi}(l)|\le |\epsi|^l\const^l$ imply that
\tht{6.5} is bounded by
$$
|\epsi|^{l_1+\dots+l_n-m_{12}-\dots-{m_{n-1,n}}}\cdot
\const^{l_1+\dots+l_n}. \tag6.7
$$

Let us show that the radius of convergence of the power series
in $|\epsi|$ obtained by adding expressions \tht{6.7} with all
nonnegative $(l_1,\dots,l_n,m_{12},\dots,m_{n-1,n})$ satisfying
\tht{6.4}, is positive. Indeed, it is not hard to see that for
any numbers satisfying the system of inequalities \tht{6.4}, one
has
$$
l_1+\dots+l_n-m_{12}-\dots-{m_{n-1,n}}\ge \max\{l_1,\dots,l_n\}.
$$
We can split the power series into parts according to which of
$l_i$'s is the largest one. It suffices to check the convergence
of each part. Let us take one of such subseries, say, assume
that $l=l_j=\max\{l_1,\dots,l_n\}$ for some $j$. Then for
$|\xi|\le1$
$$
|\epsi|^{l_1+\dots+l_n-m_{12}-\dots-m_{m_{n-1,n}}}\le |\epsi|^l,
\qquad \const^{l_1+\dots+l_n}\le \max(1,\const^{n\,l}).
$$
Finally, the number of sets of $2(n-1)$ nonnegative numbers
$\{l_i\}$ (not including $l_j=l$) and $\{m_{i,i+1}\}$ bounded by
$l$ from above is $(l+1)^{2(n-1)}$. Thus, our series is
majorized by
$$
\sum_{l=0}^\infty (l+1)^{2(n-1)}\const^{n\,l}|\epsi|^l,
$$
and this series converges for small enough $|\epsi|$.

Thus, we have verified that the series of expressions \tht{6.5}
absolutely and uniformly converges when $|\epsi|$ is small
enough. Clearly, the multiplication of the terms of this series
by constants between 0 and 1 which we mentioned at the beginning
of the proof of (ii) (these are the inner sums over Young
diagrams) does not affect the convergence. This proves the first
statement of (ii).

The second statement of (ii) is easy: the terms of \tht{6.5}
with $\xi,\eta$ replaced by $\epsi\xi$, $\epsi\eta$, have a zero
of order at least $l_1+\dots+l_n-m_{12}-\dots-m_{n-1,n}$ at
$\epsi=0$; it comes from the factors
$(\epsi\xi_i-\epsi\eta_{i-1,1})^{l_i-m_{i-1,i}}$ in $P^\up$'s
and from $\xi^l_1$ in $\pi_{c,\xi}(l_1)$. Thus, only finitely
many terms contribute to a fixed Taylor coefficient. Each of
these terms involve polynomial expressions in $z, z'$ and
expressions of the form $(1-\epsi\xi_i)^{zz'}$,
$(1-\epsi\eta_{i,i+1})^{-zz'}$, and their Taylor coefficients at
$\epsi=0$ are also polynomials in $z,z'$. \qed

\enddemo

\head 7. Dynamical correlation functions \endhead

\subhead 7.1. Definitions
\endsubhead
Consider a continuous time stochastic process $\La(t)$ with the
state space $\Y$ (all Young diagrams). As in \S3 we view the
Young diagrams as point configurations (=subsets) of
$\Z'=\Z+\frac12$ via
$$
\la\in\Y\mapsto \unX(\la)=(x_1,x_2,\dots)\subset \Z',\qquad
x_i=\la_i-i+\tfrac 12\,,\quad i=1,2,\dots
$$
Then $\La(t)$ is equivalent to the corresponding process with
values in point configurations in $\Z'$; let us denote this
process by $\unX(t)$.

For any $n=1,2,\dots$ define the {\it $n$th dynamical
correlation function} of $n$ pairwise distinct arguments
$(t_1,x_1),\dots(t_n,x_n)\in (t_{min},t_{max})\times\Z'$ by
$$
\rho_n(t_1,x_1;t_2,x_2;\dots;t_n,x_n)=\Prob\left\{\unX(t_i)\text{
contains } x_i \text{  for all  } i=1,\dots,n\right\}.
$$

In other words, the dynamical correlation functions describe
probabilities of events of the following type: for given time
moments $s_1<\dots<s_m$ and given finite sets $Y_1,\dots,Y_m$,
the random point configurations $\unX(s_1),\dots,\unX(s_m)$
contain $Y_1,\dots,Y_m$, respectively. Thus, the notion of the
dynamical correlation functions is a hybrid of the
finite-dimensional distributions of a stochastic process and
standard correlation functions of probability measures on point
configurations.

Clearly, the dynamical correlation functions uniquely determine
the finite--di\-men\-sional distributions of the process and,
thus, the process itself. The reason why we are interested in
these quantities is the same as in the ``static'' (fixed time)
case: As we take scaling limits of our processes, for the
limiting object the notion of the weight of a point
configuration does not make sense anymore. Thus, the
probabilities of the form
$\Prob\left\{\unX(s_1)=\unX_1,\dots,\unX(s_m)=\unX_m\right\}$ do
not have any meaning in the limit while the scaling limits of
the correlation functions are well-defined and, moreover, carry
a lot of useful information about the limit process.

We say that the process $\unX(t)$ is {\it determinantal}  (cf.
\S3) if the exists a kernel
$$
K:((t_{min},t_{max})\times\Z')\times
((t_{min},t_{max})\times\Z')\to \C
$$
such that for any $n=1,2,\dots$
$$
\rho_n(t_1,x_1;\dots;x_n,t_n)=
\det\bigl[K(t_i,x_i;t_j,x_j)\bigr]_{i,j=1}^n.
$$
As in the ``static'' case, if such a kernel exists then it is not
unique. In particular, transformations of the form
$$
K(s,x;t,y)\longrightarrow \frac{f(s,x)}{f(t,y)}\,K(s,x;t,y)
\tag7.1
$$
do not change the correlation functions.

\subhead 7.2. Main results \endsubhead

\proclaim{Theorem 7.1} Let $(z,z')$ be a pair of admissible
parameters and $\xi(\cdot)$ be an admissible curve. Consider the
Markov process $\La_{z,z',\bxi}$ defined in \S4, and denote by
$\unX_{z,z',\bxi}$ the corresponding process with values in the
space of point configurations in $\Z'$. Then the process
$\unX_{z,z',\bxi}$ is determinantal.
\endproclaim

Recall that in \tht{2.1} we introduced the functions
$\psi_a(x;z,z',\xi)$ which form, for any $\xi\in(0,1)$, an
orthonormal basis in $\ell^2(\Z')$. These functions were defined
under the condition that $(z,z')$ belong to either principal or
complementary series.

\proclaim{Theorem 7.2} Assume that $(z,z')$ is either in
principal or complementary series. Then the kernel
$$
\unK_{z,z',\bxi}(s,x;t,y)=\pm\sum_{a\in\Z_+'}e^{\pm a(t-s)}
\psi_{\pm a}(x;z,z',\xi(s))\,\psi_{\pm a}(y;z,z',\xi(t)) \tag7.2
$$
with ``$+$'' taken for $s\ge t$ and ``$-$'' taken for $s<t$, is
a correlation kernel of the process $\unX_{z,z',\bxi}$.
\endproclaim

\proclaim{Theorem 7.3} The correlation kernel \tht{7.2} can also
be written in the form
$$
\unK_{z,z',\bxi}(s,x;t,y)
=e^{\frac12(s-t)}\varphi_{z,z'}(x,y)\,\hatK_{z,z',\bxi}(s,x;t,y)
\tag7.3
$$
where, as in \tht{2.6},
$$
\varphi_{z,z'}(x,y)=\frac{\sqrt{\Ga(x+z+\tfrac12)\Ga(x+z'+\tfrac12)
\Ga(y+z+\tfrac12)\Ga(y+z'+\tfrac12)}}
{\Ga(x+z'+\tfrac12)\Ga(y+z+\tfrac12)}
$$
and the kernel $\hatK_{z,z',\bxi}(s,x;t,y)$ can be written as a
double contour integral {\rm (}set $\zeta=\sqrt{\xi(s)}$,
$\eta=\sqrt{\xi(t)}${\rm )}
$$
\aligned &\hatK_{z,z',\bxi}(s,x;t,y)\\
&=\frac{\sqrt{(1-\zeta)(1-\eta)}}{(2\pi
i)^2}\oint\limits_{\{\om_1\}}\oint\limits_{\{\om_2\}}
\left(1-\zeta\om_1\right)^{-z'}\left(1-\zeta\,\om_1^{-1}\right)^{z}
\left(1-\eta\om_2\right)^{-z}\left(1-\eta\,\om_2^{-1}\right)^{z'}\\
&\qquad\qquad\qquad\qquad\qquad\times
\frac{\om_1^{-x-\frac12}\om_2^{-y-\frac12}}{e^{s-t}
\left(\om_1-\zeta\right)\left(\om_2-\eta\right)-
\left(1-\zeta\om_1\right)\left(1-\eta\om_2\right)}
\,d\om_1d\om_2
\endaligned
\tag7.4
$$
with the contours $\{\om_1\}$ and $\{\om_2\}$ of $\omega_1$ and
$\omega_2$ satisfying the following conditions:

$\bullet$ $\{\om_1\}$ goes around 0 in positive direction and
passes between  $\zeta$ and $\zeta^{-1}$;

$\bullet$ $\{\om_2\}$ goes around 0 in positive direction and
passes between  $\eta$ and $\eta^{-1}$;

$\bullet$ if $s\ge t$ then the image of $\{\om_1\}$ under the
fractional--linear map
$$
\omega\mapsto \frac{\om\left(e^{s-t}\eta-\zeta\right)+1-e^{s-t}
\zeta \eta} {\om\left(e^{s-t}-\zeta\eta\right)+\eta-
e^{s-t}\zeta} \tag7.5
$$
is contained inside $\{\om_2\}$;

$\bullet$ if $s<t$ then the domain bounded by $\{\om_2\}$ does
not intersect the image of $\{\om_1\}$ under the map above.

The kernels $\unK_{z,z',\bxi}(s,x;t,y)$ and
$\hatK_{z,z',\bxi}(s,x;t,y)$ are equivalent. Namely, they are
related by a ``gauge transformation'' \tht{7.1},
$$
\hatK_{z,z',\bxi}(s,x;t,y)
=\frac{f_{z,z'}(s,x)}{f_{z,z'}(t,y)}\,\unK_{z,z',\bxi}(x,y),
\qquad x,y\in\Z',
$$
where
$$
f_{z,z'}(s,x)=\frac{e^{-\frac12 s}\,\Ga(x+z'+\tfrac12)}
{\sqrt{\Ga(x+z+\tfrac12)\Ga(x+z'+\tfrac12)}}\tag7.6
$$

The kernel $\hatK_{z,z,\bxi}(s,x;t,y)$ can serve as a
correlation kernel for all admissible values of parameters
$(z,z')$, including the degenerate series.

\endproclaim

\demo{Comments} 1. The fractional-linear transformation
\tht{7.5} arises from the condition that the denominator in the
integral representation \tht{7.4} has to be nonzero. Solving the
equation denominator=0 with respect to $\om_1$ yields the
right--hand side of \tht{7.5} with $\om=\om_2$.

2. It is not {\it a priori\/} clear why the needed contours
$\{\om_1\}$ and $\{\om_2\}$ exist. Let us show that it is indeed
so. Set
 $q=e^{s-t}$. Note that
\tht{7.5} maps $\zeta\mapsto \eta^{-1}$ and $\zeta^{-1}\mapsto
\eta$.

Consider the case $q\ge 1$ (i.e. $s\ge t$) first. Let us take a
circle with center at the origin and radius $r$ slightly smaller
than $\zeta^{-1}$ as $\{\om_1\}$. Then its image is again a
circle which is symmetric with respect to the real axis and
which passes through the images of $r$ and $-r$. The image of
$r$ is close to the image of $\zeta^{-1}$ which is $\eta$, and
the image of $-r$ is close to the image of $-\zeta^{-1}$ which
is equal to
$$
\frac{-q\eta(\zeta+\zeta^{-1})+2}{-q(\zeta+\zeta^{-1})+2\eta}\,.
$$
Since $\zeta$ and $\eta$ are strictly between 0 and 1, we
immediately see that the denominator is negative, and the whole
expression is $<\eta$. Thus, the image of $\{\om_1\}$ is a
finite circle that lies to the left of $\eta$ plus a small
number. Clearly, there exists $\{\om_2\}$ that passes between
$\eta$ and $\eta^{-1}$ and encircles both 0 and the image of
$\{\om_1\}$.

Let us consider the case $q<1$ now. As $\{\om_1\}$ we again take
a circle with center at the origin but with radius slightly
greater than $\zeta$. Then its image is a circle which is
symmetric with respect to the real axis and which passes through
images of points that are close to the image of $\zeta$ which is
$\eta^{-1}$, and to the image of $-\zeta$ which is
$$
\frac{\zeta+\zeta^{-1}-2q\eta}{\eta(\zeta+\zeta^{-1})-2q}\,.
$$
If the denominator is negative then the whole expression is
negative, and there exists a contour $\{\om_2\}$ inside this
circle that passes between $\eta$ and $\eta^{-1}$ and goes
around the origin. If the denominator is positive then the whole
expression is $>\eta^{-1}$, and $\{\om_2\}$ can be a circle of
radius between $\eta$ and $\eta^{-1}$ with center at the origin.
\qed
\enddemo

Theorems 7.1, 7.2, and 7.3 are generalizations of Theorems 3.1,
3.2, and 3.3, respectively.

\subhead 7.3. Proof of Theorems 7.1, 7.2, and 7.3
\endsubhead
The ideas used in the proof are similar to those of \S3. The
first step is to consider the case $z=N\in\{1,2,\dots\}$,
$z'=N+\al$, $\al>-1$. Then the state space of our Markov process
$\La_{z,z',\xi}$ is smaller than the whole set $\Y$. Namely,
since $M_{N,N+\al,\xi}(\la)$ vanishes if $\ell(\la)$ (the number
of nonzero rows of $\la$) is greater than $N$, our process lives
on the set $\Y(N)$ of Young diagrams with no more than $N$ rows.
Consider an embedding of $\Y(N)$ into the set of $N$-point
subsets of $\Z_+$ given by
$$
\la\in\Y(N)\mapsto \wt X(\la)=(\wt x_1,\dots,\wt x_N),\qquad \wt
x_i=\la_i+N-i,\quad i=1,\dots,N,
$$
and denote by $\wt X_{N,N+\al,\xi}$ the corresponding stochastic
process on the space of $N$-point configurations in $\Z_+$.
Obviously, the processes $\unX_{N,N+\al,\xi}$ and $\wt
X_{N,N+\al,\xi}$ are equivalent: if $\la\in\Y(N)$ then
$\unX(\la)=(x_1,x_2,\dots)$ and $\wt X(\la)=(\wt x_1,\dots,\wt
x_N)$ are related by
$$
x_i=\cases \wt x_i-N+\tfrac12, &i=1,\dots,N,\\-i+\tfrac 12,&i\ge
N+1.
\endcases
$$
This implies that the dynamical correlation functions $\rho_l$
of $\unX_{N,N+\al,\xi}$ and $\wt\rho_l$ of $\wt X_{N,N+\al,\xi}$
are related by
$$
\rho_l(\tau_1,x_1;\dots;\tau_l,x_l)= \wt\rho_l(\tau_1,\wt
x_1;\dots;\tau_l,\wt x_l)
$$
with $\wt x_i=x_i+N-\frac12\in\Z_+$ for $i=1,\dots,l$.

Define the {\it extended Meixner kernel} $K^\m_{N,\al,\bxi}$ by,
cf. Lemma 3.4,
$$
K^\m_{N,\al,\bxi}(s,\wt x;t,\wt y)=\cases
K^{\m,+}_{N,\al,\bxi}(s,\wt
x;t,\wt y),&s\ge t,\\
K^{\m,-}_{N,\al,\bxi}(s,\wt x;t,\wt y),&s<t,\endcases
$$
with $\wt x,\wt y\in\Z_+$ and
$$
\gather K_{N,\al,\bxi}^{\m,+}(s,\wt x;t,\wt y)=\sum_{m=0}^{N-1}
e^{m(s-t)} \wt\M_m(\wt x;\al,\xi(s))\wt\M_m(\wt
y;\al,\xi(t)),\tag7.7
\\ K_{N,\al,\bxi}^{\m,-}(s,\wt x;t,\wt y)=-\sum_{m=N}^{\infty}
e^{m(s-t)} \wt\M_m(\wt x;\al,\xi(s))\wt\M_m(\wt y;\al,\xi(t)).
\tag7.8
\endgather
$$

\proclaim{Lemma 7.4} The process $\wt X_{N,N+\al,\bxi}$ is
determinantal. Its correlation functions have the form
$$
\wt\rho_l(\tau_1,\wt x_1;\dots;\tau_l,\wt x_l)=\det\Bigl[
K^\m_{N,\al,\bxi}(\tau_i,\wt x_i;\tau_j,\wt
x_j)\Bigr]_{i,j=1}^l,\qquad l=1,2,\dots
$$
\endproclaim
We postpone the proof of this lemma till \S7.4.

The next step is to connect the extended Meixner kernel $K^\m$
and the kernel $\unK$ of Theorem 7.2 (cf. Lemma 3.7).

\proclaim{Lemma 7.5} We have
$$
K^\m_{N,\al,\bxi}(s,\wt x;t,\wt y)=e^{(N-\frac 12)(s-t)}
\unK_{N,N+\al,\bxi}(s,x;t,y) \tag7.9
$$
with $\wt x=x+N-\frac12\in\Z_+$, $\wt y=y+N-\frac 12\in\Z_+$.
\endproclaim

\demo{Proof} We argue as in the proof of Lemma 3.7. For $s\ge t$
take \tht{7.7} and change the summation index $m=N-a-\frac 12$.
Then, see Proposition 2.8,
$$
e^{m(s-t)}=e^{(N-\frac12)(s-t)}\cdot e^{a(t-s)},\qquad
\wt\M_m(\wt x;\al,\xi)=\psi_a(x;N,N+\al,\xi)
$$
with $\wt x=x+N-\frac 12$. Furthermore,
$\psi_a(x;N,N+\al,\xi)\equiv 0$ if $a=N+\frac 12,N+\frac
32,\dots$ because of the factor $\Ga(z-a+\frac12)$ in \tht{2.1}.
This yields \tht{7.9}. For $s<t$ the argument is similar; it
uses \tht{7.8}, the summation index change $m=N-a-\frac 12$ and
$$
e^{m(s-t)}=e^{(N-\frac12)(s-t)}\cdot e^{-a(s-t)},\qquad \wt
\M_{m}(\wt x;\al,\xi)=\psi_{-a}(x;N,N+\al,\xi).
$$
\qed
\enddemo

Lemma 7.4 and  Lemma 7.5 imply that the correlation functions
$\rho_l$ of $X_{N,N+\al,\bxi}$ can be written as
$$
\rho_l(\tau_1, x_1;\dots;\tau_l, x_l)=\det\Bigl[
\hatK_{N,N+\al,\bxi}(\tau_i, x_i;\tau_j,
x_j)\Bigr]_{i,j=1}^l,\qquad l=1,2,\dots, \tag7.10
$$
if $x_i\ge -N+\frac12$ for all $i=1,\dots,l$ (indeed, the factor
$e^{(N-\frac 12)(s-t)}$ in \tht{7.9} does not affect the
correlation functions).

The next claim is a counterpart of Lemma 3.9.

\proclaim{Lemma 7.6} Assume that

$\bullet$ either $(z,z')$ is not in the degenerate series  and
$x,y\in\Z'$ are arbitrary

$\bullet$ or $z=N=1,2,\dots$, $z'>N-1$, and both $x,y$ are in
$\Z_+-N+\tfrac12$.

Then the kernel $\hatK_{z,z',\bxi}(s,x;t,y)$ of Theorem 7.3 is
related to the kernel $\unK_{z,z',\bxi}(s,x;t,y)$ of Theorem 7.2
by equality \tht{7.3}. Or, that is the same, by the ``gauge
transformation''
$$
\hatK_{z,z',\bxi}(s,x;t,y)=\frac{f_{z,z'}(s,x)}{f_{z,z'}(t,y)}\,
\unK_{z,z',\bxi}(s,x;t, y),
$$
where $f_{z,z'}$ is defined in \tht{7.6}.
\endproclaim

\demo{Proof} Applying formula \tht{2.4} of Proposition 2.3 to
$\psi_a$ and setting $a=k+\tfrac12$ we obtain
$$
\aligned \psi_{k+\frac12}(x;z,z',\xi(s))&=\left(\frac{
\Ga(x+z+\frac12)\Ga(x+z'+\frac12)} {\Ga(z-k)\Ga(z'-k)}
\right)^\frac
12\frac{\Ga(z'-k)(1-\xi(s))^{\frac{z'-z+1}2}}{\Ga(x+z'+\frac12)}\\
\times&\oint\left(1-\sqrt{\xi(s)}\om_1\right)^{-z'+k}
\left(1-\sqrt{\xi(s)}\,\om_1^{-1}\right)^{z-k-1}\om_1^{-x-k-\frac12}
\,\frac{d\om_1}{\om_1}\,,\\
\psi_{k+\frac12}(y;z,z',\xi(t))&=\left(\frac{
\Ga(x+z+\frac12)\Ga(x+z'+\frac12)} {\Ga(z-k)\Ga(z'-k)}
\right)^\frac
12\frac{\Ga(z-k)(1-\xi(t))^{\frac{z-z'+1}2}}{\Ga(x+z+\frac12)}\\
\times&\oint\left(1-\sqrt{\xi(t)}\om_2\right)^{-z+k}
\left(1-\sqrt{\xi(t)}\,\om_2^{-1}\right)^{z'-k-1}\om_2^{-y-k-\frac12}
\,\frac{d\om_2}{\om_2}\,.
\endaligned
$$
(As in the proof of Proposition 2.3, we used the symmetry
$\psi_a(x;z,z',\xi)=\psi_a(x;z',z,\xi)$ to get the second
relation.)

Note that both sides of \tht{7.3} are real--analytic functions
of $q=e^{s-t}$ on $q\ge 1$ and $0<q<1$ with all other parameters
$(z,z',\xi(s),\xi(t),x,y)$ being fixed. Thus, it suffices to
prove \tht{7.3} for $q$ large enough in the case of $q\ge 1$,
and for $q$ small enough in the case $q<1$.

Let us consider the case $q\ge 1$. Substituting the integral
representations above in \tht{7.2}, we observe that the
summation index $k=a-\frac 12\in\Z_+$ enters the resulting
expression only as the exponent in
$$
\left(\frac{\left(1-\sqrt{\xi(s)}\om_1\right)
\left(1-\sqrt{\xi(t)}\om_2\right)}
{q\left(\om_1-\sqrt{\xi(s)}\right)
\left(\om_2-\sqrt{\xi(t)}\right)} \right)^k.
$$
For large enough $q$ the geometric progression $\sum_{k\ge 0}$
with this ratio converges uniformly on any fixed contours
$\{\om_1\},\{\om_2\}$. Computing the sum yields \tht{7.3},
\tht{7.4}.

Similarly, in the case $q<1$ the computation reduces to summing
the geometric progression with the ratio
$$
\frac{q\left(\om_1-\sqrt{\xi(s)}\right)
\left(\om_2-\sqrt{\xi(t)}\right)}
{\left(1-\sqrt{\xi(s)}\om_1\right)
\left(1-\sqrt{\xi(t)}\om_2\right)}
$$
which always makes sense for small enough $q$.

For large $q$ the image of any finite contour under \tht{7.5} is
concentrated near $\eta=\sqrt{\xi(t)}$, and for small $q$ such
image is concentrated near $\eta^{-1}$. These points are always
inside/outside of any contour $\{\om_2\}$ provided by
Proposition 2.3. The conditions on contours in the statement of
Theorem 7.3 reflect the deformations of the contours of
Proposition 2.3. Note that the denominator of \tht{7.4} must
stay away from zero while deforming the contours, which means
that the image of $\{\om_1\}$ under \tht{7.5} is not allowed to
intersect $\om_2$. \qed
\enddemo

The relations \tht{7.10} and \tht{7.3} yield the following
formula for the correlation functions of $X_{N,N+\al,\xi}$:
$$
\rho_l(\tau_1, x_1;\dots;\tau_l, x_l)=\det\Bigl[
\hatK_{z,z',\bxi}(\tau_i, x_i;\tau_j,
x_j)\Bigr]_{i,j=1}^l,\qquad l=1,2,\dots, \tag7.11
$$
where $(z,z')=(N,N+\al)$ and $x_i+N-\frac12\in\Z_+$ for all
$i=1,\dots,l$. In order to prove Theorem 7.3, we need to extend
this formula to arbitrary admissible $(z,z')$ and arbitrary
$x_i\in\Z'$. We will do that by means of Proposition 6.2.

Up till now the time moments $\tau_1,\dots,\tau_l$ were
arbitrary, they were not ordered and some of them were allowed
to coincide. Let us denote by $t_1,\dots,t_n$, $n\le l$, the
same numbers but ordered and without repetitions. Thus, each
$\tau_i$ is equal to one and only one $\tau_j$. As in \S6, we
set
$$
\xi_i=\xi(t_i),\quad 1\le i\le n,\qquad
\eta_{i,i+1}=e^{t_i-t_{i+1}}\sqrt{\xi_i\xi_{i+1}}.
$$
Then in the notation of \S6, see \tht{6.2} and below,
$\rho_l(\tau_1,x_1;\dots;\tau_l,x_l)$ is equal to $F(\xi,\eta)$
with a suitable choice of the sets $\Cal D_1,\dots,\Cal D_n$ and
$\Cal D_{12},\dots,\Cal D_{n-1,n}$. Namely, we take all $\Cal
D_{i,i+1}$ equal to $\Y$, and the set $\Cal D_i$ is determined
according to the following recipe: take all numbers
$j\in\{1,\dots,l\}$ such that $\tau_j=t_i$, then the
corresponding points $x_j\in\Z'$ must be pairwise distinct. Then
$$
\Cal D_i=\bigl\{\la\in\Y\mid \unX(\la) \text{  contains all
}x_j\text{ such that  } \tau_j=t_i\bigr\}.
$$

Proposition 6.2 says that $\rho_l(\tau_1,x_1;\dots;\tau_l,x_l)$
is a real analytic function in $\xi,\eta$, and after the
substitution $(\xi,\eta)\mapsto (\epsi\xi,\epsi\eta)$ the
corresponding function in $\epsi$ can be analytically continued
in a neighborhood of $\epsi=0$. Moreover, its Taylor
coefficients at this point are polynomials in $z$ and $z'$.

Now let us look at the right--hand side of \tht{7.11} with
$N,N+\al$ replaced by $z,z'$. The values
$\hatK_{z,z',\bxi}(\tau_i, x_i;\tau_j, x_j)$ of the kernel are
given by \tht{7.4}. This formula involves $\sqrt{\xi(\tau_i)}$,
$\sqrt{\xi(\tau_j)}$, and $e^{\tau_i-\tau_j}$, which is
expressible through $\sqrt{\xi_i}$'s and $\eta_{i,i+1}$'s in a
polynomial fashion. Moreover, $e^{\tau_i-\tau_j}$ do not change
if we scale $\xi$ and $\eta$ by $\epsi$.

The integral representation \tht{7.4} implies that
$\hatK_{z,z',\bxi}(\tau_i, x_i;\tau_j, x_j)$ are real analytic
functions in $\sqrt{\xi_i}$'s and $\eta_{i,i+1}$'s. Further, if
we scale $\xi$ and $\eta$ by $\epsi$, then \tht{7.4} viewed as a
function in $\delta=\epsi^{\frac 12}$, extends to an analytic
function in a small enough disc
$\{\delta:|\delta|<\const\}$.\footnote{It is worth noting that
for small $\epsi$ we can choose the contours of integration in
\tht{7.4} which would be independent of $\epsi$; it suffices to
consider suitable circles centered at the origin.}
 Moreover, its Taylor coefficients at
$\delta=0$ are polynomials in $z$ and $z'$ because the Taylor
coefficients of $(1-u)^\kappa$ at $u=0$ are polynomials in
$\kappa$.

We conclude that both sides of \tht{7.11} are uniquely
determined by their values for $(z,z')=(N,N+\al)$ with any large
enough $N$ and any $\al>-1$. This completes the proof of Theorem
7.3.

Theorem 7.2 is a direct corollary of Theorem 7.3 and Lemma 7.6.
Theorem 7.1 follows from Theorem 7.3. \qed

\subhead 7.4. Eynard--Mehta theorem and the proof of Lemma 7.4
\endsubhead
Here we state the Eynard--Mehta theorem \cite{EM} in a form
which is convenient for our purposes and show that Lemma 7.4 is
a corollary of this theorem.

Let $m$ be a fixed natural number and let the index $k$ range over
$\{1,\dots,m\}$. Consider the Hilbert space $\ell^2(\Z_+)$ taken
with respect to the counting measure on the set
$\Z_+=\{0,1,2,\dots\}$. Assume that for each $k$ we are given an
orthonormal basis $\{\phi_{k,n}\}_{n=0,1,\dots}$ of real--valued
functions in  $\ell^2(\Z_+)$. Next, assume that for each
$k=1,\dots,m-1$ and each $n=0,1,\dots$ we are a given a number
$c_{k,k+1;n}>0$. As $n\to\infty$, these numbers have to decay fast
enough to make convergent certain infinite sums specified below.
Finally, we will impose on these data certain positivity
conditions, see below.

We aim to construct a probability measure on collections
$(X_1,\dots,X_m)$, where each $X_k$ is an arbitrary $N$--point
subset in $\Z_+$ and $N$ is a fixed natural number. This measure
can be regarded as a Markov process with ``time'' $k=1,\dots,m$,
the state space being the set of $N$--point subsets in $\Z_+$.

The construction goes as follows. For an arbitrary $N$--point
subset $X=(x_1<\dots<x_N)\subset\Z_+$ we introduce the $N\times N$
matrix $\phi_k(X)$ with the entries $\phi_{k,i}(x_j)$, where the
row index $i$ takes values in $\{0,\dots,N-1\}$, and the column
index $j$ takes values in $\{1,\dots,N\}$.

As $X$ ranges over all $N$--point subsets of $\Z_+$, one has
$$
\sum_{X}(\det\phi_k(X))^2=1.
$$
The proof follows from the Cauchy-Binet identity and
orthonormality of $\phi_{k,n}$'s, cf. \tht{5.8}.

Thus, for each $k=1,\dots,m$ we have a probability measure $\si_k$
on $N$--point subsets in $\Z_+$ which assigns to a subset $X$ its
weight $(\det\phi_k(X))^2$. The measures $\si_k$ are the
1--dimensional distributions for our future Markov process.

For each $k=1,\dots,m-1$ we set
$$
v_{k,k+1}(x,y)=\sum_{n=0}^\infty c_{k,k+1;n}
\phi_{k,n}(x)\phi_{k+1,n}(y), \qquad x,y\in\Z_+\,,
$$
where the sum is assumed to be convergent. Since $\{\phi_{k,n}\}$
is an orthonormal basis for each fixed $k$, we have
$$
\gathered \sum_{x\in\Z_+}\phi_{k,n}(x)v_{k,k+1}(x,y)
=c_{k,k+1;n}\phi_{k,n}(y),\\
\sum_{y\in\Z_+}v_{k,k+1}(x,y)\phi_{k+1,n}(y) =c_{k,k+1;n}
\phi_{k,n}(x).
\endgathered
\tag7.12
$$
For arbitrary subsets $X=(x_1<\dots<x_N)$ and $Y=(y_1<\dots<y_N)$
we form an $N\times N$ matrix $v_{k,k+1}(X,Y)$ with entries
$v_{k,k+1}(x_i,y_j)$, and we set
$$
\si_{k,k+1}(X,Y)=\frac{\det\phi_k(X)\det
v_{k,k+1}(X,Y)\det\phi_{k+1}(Y)}{\prod\limits_{n=0}^{N-1}c_{k,k+1;n}}
$$

Once again, using the Cauchy-Binet identity and \tht{7.12}, it
is not hard to show that for any $k=1,\dots,m-1$ one has
$$
\sum_{Y}\si_{k,k+1}(X,Y)=\si_k(X), \qquad
\sum_{X}\si_{k,k+1}(X,Y)=\si_{k+1}(Y).
$$

Assume that $\si_{k,k+1}(X,Y)\ge0$ for all $X$ and $Y$. Then we
may regard $\si_{k,k+1}$ as a probability measure on couples
$(X,Y)$ with marginal measures $\si_k$ and $\si_{k+1}$. This is
the ``one step'' 2--dimensional distribution of our Markov
process. We also assume that $\det\phi_k(X)$ does not vanish. Then
we define the ``one step'' transition probability function as
follows
$$
\PP_{X,Y}(k,k+1)=\frac{\si_{k,k+1}(X,Y)}{\si_k(X)} =\frac{\det
v_{k,k+1}(X,Y)\det\phi_{k+1}(Y)}
{\det\phi_k(X)\prod\limits_{n=0}^{N-1}c_{k,k+1;n}}\,. \tag7.13
$$
We regard this as a matrix $\PP(k,k+1)$ whose rows and columns are
labelled by $N$--point subsets.

Finally, we define a Markov process $X(k)$, where the ``time''
$k$ takes values from $1$ to $m$ and $X(k)$ is an $N$--point
subset of $\Z_+$, using the initial distribution $\si_1$ and the
``one step'' transition probabilities \tht{7.13}:
$$
\Prob(X(1),\dots,X(m))
=\si_1(X_1)\PP_{X(1),X(2)}(1,2)\dots\PP_{X(m-1),X(m)}(m-1,m).
$$

For arbitrary indices $k,l$ such that $1\le k<l\le m$ we set
$$
\gather c_{k,l;n}=c_{k,k+1;n}\,c_{k+1,k+2;n}\dots c_{l-1,l;n}\,,
\qquad n=0,1,\dots, \\
v_{k,l}(x,y)=\sum_{n=0}^\infty c_{k,l;n}
\phi_{k,n}(x)\phi_{l,n}(y), \qquad x,y\in\Z_+\,.
\endgather
$$

\proclaim{Theorem 7.7 (Eynard--Mehta \cite{EM})} Under the above
assumptions, let us regard the Markov process $X(k)$ as a
probability measure on $mN$--point configurations
$X=(X(1),\dots,X(m))$ in the space $\{1,\dots,m\}\times\Z_+$.
Then this measure is determinantal, and its correlation kernel
has the form
$$
K(k,x;l,y)=\sum_{i=0}^{N-1}
\frac1{c_{l,k;i}}\,\phi_{k,i}(x)\phi_{l,i}(y), \qquad k\ge l,
$$
{\rm(}where we agree that $c_{k,k;i}=1${\rm)} and
$$
\align K(k,x;l,y) &=\sum_{i=0}^{N-1}
c_{k,l;i}\,\phi_{k,i}(x)\phi_{l,i}(y)\;-v_{k,l}(x,y)\\
 &=-\sum_{i=N}^\infty c_{k,l;i}\,\phi_{k,i}(x)\phi_{l,i}(y),
\qquad k< l.
\endalign
$$
\endproclaim

In other words, for any $n=1,2, \dots$ we have
$$
\align \rho_n(k_1,x_1;\dots;k_n,x_n):&=\Prob\{X(k_i)\ni x_i\,
\quad \text{\rm for each $i=1,\dots,n$}\}\\
&=\det\left[K(k_i,x_i;k_j,x_j)\right]_{i,j=1}^n,
\endalign
$$
where $(k_i,x_i)\in\{1,\dots,m\}\times\Z$ and
$(k_i,x_j)\ne(k_j,x_j)$ for $i\ne j$.

\demo{Proof} See \cite{EM}, \cite{NF}, \cite{Jo3}, \cite{TW},
\cite{BR}. \qed
\enddemo

\demo{Proof of Lemma 7.4} The process $\La_{N,N+\al,\xi}$
restricted to any finite sequence of time moments
$t_1<\dots<t_m$ fits into this formalism perfectly. Indeed, we
take
$$
\phi_{k,n}(x)=\wt\M_n(x;\al,\xi(t_k)), \qquad
c_{k,k+1;n}=e^{n(t_{k}-t_{k+1})}.
$$
Then \tht{5.7} or Lemma 3.4 imply that $M_{N,N+\al,\xi(t_k)}$ is
exactly $\sigma_k$, and the transition matrix \tht{5.2}
coincides with \tht{7.13}. \footnote{Recall that the Young
diagrams with no more than $N$ rows are identified with
$N$-point subsets of $\Z_+$ via $\la\mapsto
(\la_1+N-1,\la_2+N-2,\dots,\la_N)$.} Lemma 7.4 is thus a direct
corollary of Theorem 7.7. \qed
\enddemo

\subhead 7.5. An interpretation via nonintersecting paths
\endsubhead
In this section we interpret the stationary process
$\La_{N,N+\al,\xi}$ in terms of $N$ nonintersecting trajectories
of independent birth--death processes. This is done using
formulas of Karlin--McGregor \cite{KMG3} and an idea of
Johansson \cite{Jo2}.

Instead of dealing with $\La_{N,N+\al,\xi}$ we will use the
associated process $\wt X_{N,N+\al,\xi}$ introduced in the
beginning of \S7.3. Recall that its state space consists of
$N$-point configurations in $\Z_+$.

In the special case $N=1$ the process $\wt X_{1,1+\al,\xi}$ is
just the birth--death process $N_{1+\al,\xi}$. We aim to
construct $\wt X_{N,N+\al,\xi}$ directly in terms of
$N_{1+\al,\xi}$.

Let us take a large $T>0$ and consider a new process
$Y_{N,\al,\xi,T}$ introduced as follows. This process is defined
on the time interval $[-T,T]$. Let us take $N$ independent
birth--death processes which start at the moment $-T$ at the
points $a_1<\dots<a_N$ and end up at the moment $T$ at the
points $b_1<\dots<b_N$ conditioned on the event that the
trajectories $x_i(t)$ do not intersect on $[-T,T]$:
$$
x_1(t)<x_2(t)<\dots<x_N(t), \qquad -T\le t\le T.
$$
The boundary conditions $\{a_i\}$ and $\{b_i\}$ are arbitrary
but fixed while the parameter $T$ will vary.

\proclaim{Theorem 7.8} In the above notation, as $T\to\infty$
the processes $Y_{N,\al,\xi,T}$ converge to $\wt
X_{N,N+\al,\xi}$ in the sense of convergence of the finite
dimensional distributions.
\endproclaim

\demo{Proof} Let us fix arbitrary time moments $t_1<\dots<t_k$
inside $(-T,T)$. Then by \cite{KMG3} and Theorem 5.1, the
corresponding finite-dimensional distribution of
$Y_{N,\al,\xi,T}$ has the form
$$
\multline
\Prob\{Y_{N,\al,\xi,T}(t_i)=(y_1^{(i)}<\dots<y_N^{(i)}) \text{
for all } i=1,\dots,k\}=\\
\frac{\det[v_{-T,t_1}(a_i,y_j^{(1)})]\det[v_{t_1,t_2}({y_i^{(1)},y_j^{(2)}})]
\cdots \det[v_{t_{k-1},t_k}({y_i^{(k-1)},y_j^{(k)}})]
\det[v_{t_k,T}({y_i^{(k)},b_j})]}{\det[v_{-T,T}({a_i,b_j})]}
\endmultline
$$
where $v_{s,t}(x,y)$ is given by \tht{5.1} with
$\xi(\,\cdot\,)\equiv \xi$.

On the other hand, the finite-dimensional distribution of $\wt
X_{N,N+\al,\xi}$ is given by
$$
\multline \Prob\{\wt
X_{N,N+\al,\xi}(t_i)=(y_1^{(i)}<\dots<y_N^{(i)}) \text{ for all
}
i=1,\dots,k\}\\=e^{\frac{(t_k-t_1)N(N-1)}2}\frac{\det[\phi_i(y_j^{(1)})]}
{\det[\phi_i(y_j^{(k)})]}\,
{\det[v_{t_1,t_2}({y_i^{(1)},y_j^{(2)}})] \cdots
\det[v_{t_{k-1},t_k}({y_i^{(k-1)},y_j^{(k)}})]}
\endmultline
$$
with $\phi_{n}(x)=\wt\M_n(x;\al,\xi)$, see Theorem 5.1.

Note that we have the following asymptotic relation: for
arbitrary $x_1',x_1''$, $\dots,$ $x_N',x_N''\in\Z_+$
$$
\det\left[\sum_{n=0}^\infty
\epsilon^n\phi_n(x_i')\phi_n(x_j'')\right]=
\epsilon^{\frac{N(N-1)}2}\det[\phi_{i-1}(x_j')]
\det[\phi_{i-1}(x_j'')]+O\left(\epsilon^{\frac{N(N-1)}2+1}\right)
$$
as $\epsilon\to 0$, cf. \tht{5.4}. Applying this asymptotic
relation to $v_{-T,t_1}$, $v_{t_k,T}$, $v_{-T,T}$, we obtain
$$
\align \det[v_{-T,t_1}(a_i,y_j^{(1)})]&\sim
e^{-\frac{(t_1+T)N(N-1)}2}\det[\phi_{i-1}(a_j)]
\det[\phi_{i-1}(y_j^{(1)})],\\
\det[v_{t_1,T}(y_i^{(k)},b_j)]&\sim
e^{-\frac{(T-t_k)N(N-1)}2}\det[\phi_{i-1}(y_j^{(k)})]
\det[\phi_{i-1}(b_j)],\\
\det[v_{-T,T}(a_i,b_j)]&\sim
e^{-TN(N-1)}\det[\phi_{i-1}(a_j)]\det[\phi_{i-1}(b_j)],
\endalign
$$
as $T\to+\infty$. This completes the proof. \qed
\enddemo

\head 8. Particle--hole involution
\endhead

For any set $\x$ and its subset $\y$ one can define an
involution on point configurations $X\subset \x$ by $X\mapsto
X\,\triangle\,\y$. This map leaves intact the ``particles'' of
$X$ outside of $\y$, and inside $\y$ it picks the ''holes''
(points of $\y$ free of particles). This involution is called
the {\it particle-hole involution} on $\y$.

The goal of this section is to give a different description of
the z-measures using a new identification of Young diagrams and
point configurations on $\Z'$. Instead of using the
configurations
$$
\unX(\la)=\{\la_i-i+\tfrac 12\mid i=1,2,\dots\}
$$
we will use the configurations
$$
X(\la)=(\unX(\la)\cap \Z'_+)\cup (\Z'_-\setminus\unX(\la))=
\unX(\la)\,\triangle\, \Z'_-
$$
which are obtained from $\unX(\la)$ by applying the
particle-hole involution on $\Z'_-$.

The parametrization of Young diagrams $\la$ by configurations
$X(\la)$ corresponds to considering the {\it Frobenius
coordinates} of $\la$, see \cite{BOO, \S1.2} for details. The
reason for passing to $X(\la)$ is very simple: in the continuous
limit
 $\xi\nearrow 1$ which will be considered below in \S9,
the point process generated by $\unX(\la)$ does not survive, while
the process corresponding to $X(\la)$ has a well defined limit.

Observe that  $X(\la')=-X(\la)$ for any $\la\in\Y$.

Given an arbitrary kernel $K(x,y)$ on $\x\times\x$, and a subset
$\y$ of $\x$, we assign to it another kernel,
$$
K^\circ(x,y)=\cases K(x,y), & x\notin \y,\\
\de_{xy}- K(x,y), & x\in \y,\endcases
$$
where $\de_{xy}$ is the Kronecker symbol. Slightly more
generally, given an arbitrary map $\epsi: \x\to\R^*$, we set
$$
K^{\circ,\epsi}(x,y)=\epsi(x) K^\circ(x,y)\epsi(y)^{-1}.
$$

\proclaim{Proposition 8.1} Let $P$ be a probability measure in
point configurations on a discrete space $\x$ and let $P^\circ$
be the image of $P$ under the particle-hole involution on
$\y\subset \x$. Assume that the correlation functions of $P$
have determinantal form with a certain kernel $K(x,y)$,
$$
\rho_m(x_1,\dots,x_m\mid P)=\det \Sb 1\le i,j\le m\endSb
[K(x_i,x_j)], \qquad m=1,2,\dots\,.
$$
Then the correlation functions of the measure $P^\circ$ also
have a similar determinantal form, with the kernel
$K^\circ(x,y)$ as defined above or, equally well, with the
kernel $K^{\circ,\epsi}(x,y)$, where the map $\epsi:\x\to\R^*$
may be chosen arbitrarily,
$$
\gathered \rho_m(x_1,\dots,x_m\mid P^\circ)=\det \Sb 1\le i,j\le
m\endSb [K^\circ(x_i,x_j)]=\det \Sb 1\le i,j\le m\endSb
[K^{\circ,\epsi}(x_i,x_j)],\\
 m=1,2,\dots\,.
\endgathered
$$
\endproclaim
\demo{Proof} The factor $\epsi(\,\cdot\,)$ does not affect the
values of determinants in right--hand side of the above formula,
so that we may take $\epsi(\,\cdot\,)\equiv1$. Then the result
is obtained by applying the inclusion/exclusion principle, see
Proposition A.8 in  \cite{BOO}. \qed
\enddemo

Later on we choose the function $\varepsilon(\,\cdot\,)$ in a
specific way (see \tht{8.3} below) which is appropriate for the
limit transition of \S9.

The main result of this section is a determinantal formula for
the dynamical correlation functions of $\La_{z,z',\xi}$ computed
in terms of $X(\la)$. For any $n=1,2,\dots$ define the {\it
$n$th dynamical correlation function} of $n$ pairwise distinct
arguments $(t_1,x_1),\dots(t_n,x_n)\in
(t_{min},t_{max})\times\Y$ by
$$
\multline
\rho_n(t_1,x_1;t_2,x_2;\dots;t_n,x_n)\\=\Prob\left\{X(\la)\text{
at the moment $t_i$ contains } x_i \text{  for all  }
i=1,\dots,n\right\}.
\endmultline
$$

Here and in what follows we denote by $\La_{z,z',\xi}$ the {\it
stationary\/} Markov process corresponding to the constant curve
$\xi(t)\equiv\xi$, where $\xi\in(0,1)$ is fixed.

\proclaim{Theorem 8.2} Let $(z,z')$ be a pair of admissible
parameters not from the degenerate series. Consider the Markov
process $\La_{z,z',\xi}$, and denote by $X_{z,z',\xi}$ the
process with values in the space of point configurations in
$\Z'$, obtained from $\La_{z,z',\xi}$ via $\la\mapsto X(\la)$.

Then the process $X_{z,z',\xi}$ is determinantal and its
correlation kernel has the form (in all the formulas below $x$
and $y$ are positive, in $\pm$ and $\mp$ the upper sign
corresponds to the case $s\ge t$, and the lower sign corresponds
to $s<t$)
$$
\aligned &K_{z,z',\xi}(s,x;t,y)=\pm\sum_{a\in\Z'_+}
e^{-a|s-t|}\psi_{\pm a}(x;z,z',\xi)\psi_{\pm a}(y;z,z',\xi),\\
&K_{z,z',\xi}(s,x;t,-y)=\pm\sum_{a\in\Z'_+} (-1)^{\pm a-\frac
12}e^{-a|s-t|}\psi_{\pm a}(x;z,z',\xi)\psi_{\mp
a}(y;-z,-z',\xi),\\
&K_{z,z',\xi}(s,-x;t,y)=\mp\sum_{a\in\Z'_+}(-1)^{\pm a-\frac 12}
e^{-a|s-t|}\psi_{\mp a}(x;-z,-z',\xi)\psi_{\pm
a}(y;z,z',\xi),\\
&K_{z,z',\xi}(s,-x;t,-y)=\mp\sum_{a\in\Z'_+}
e^{-a|s-t|}\psi_{\mp a}(x;-z,-z',\xi)\psi_{\mp a}(y;-z,-z',\xi),
\endaligned \tag8.1
$$
where the fourth formula is valid for $s\ne t$, and for $s=t$ we
have
$$
K_{z,z',\xi}(s,-x;s,-y)=\sum_{a\in\Z'_+} \psi_{
a}(x;-z,-z',\xi)\psi_{ a}(y;-z,-z',\xi). \tag8.2
$$
\endproclaim

\demo{Comments} 1. For $s=t$ this kernel coincides with the
hypergeometric kernel derived in \cite{BO2}, see also \cite{BO4},
\cite{BO5}. In those papers the kernel was written in another,
so-called ``integrable form'', see Remark 8.3 below.

2. The kernel $K_{z,z',\xi}$ has the following symmetries
($x,y\in\Z'$):
$$
\gathered K_{z,z',\xi}(s,x;t,y)=(-1)^{\sgn x\cdot \sgn
y}K_{z,z',\xi}(s,y;t,x),
\\
K_{z,z',\xi}(s,x;t,y)=\cases K_{-z,-z',\xi}(t,-x;s,-y),&s\ne t,
\\
(-1)^{\sgn x\cdot \sgn y} K_{-z,-z',\xi}(t,-x;s,-y),&s=t.\endcases
\endgathered
$$
\enddemo

\demo{Proof} We use Proposition 8.1. As the initial kernel we
take the expression for $\unK_{z,z',\bxi}$ given in Theorem 7.2,
the set $\x$ is the union of finitely many copies of $\Z'$ which
correspond to times at which we evaluate the dynamical
correlation function, and $\y$ is the union of the same number
of copies of $\Z'_-$. On each copy of $\Z'$ the function
$\varepsilon(\,\cdot\,)$ is chosen in the following way:
$$
\varepsilon(x)=\cases 1,&x>0,\\
(-1)^{-x-\frac12},&x<0.\endcases \tag8.3
$$
The statement then follows from Proposition 2.7. The last
formula (for $s=t$) arises from the relation
$$
\multline \delta_{x,y}-\sum_{a\in\Z'_+}\psi_{-a}(x;-z,-z',\xi)
\psi_{-a}(y;-z,-z',\xi)\\=\sum_{a\in\Z'_+}\psi_{a}(x;-z,-z',\xi)
\psi_{a}(y;-z,-z',\xi),
\endmultline
$$
which follows from the fact that $\psi_a$ form an orthonormal
basis, see Proposition 2.4.\qed
\enddemo

\example{Remark 8.3} Denote by $K_{z,z',\xi}(x,y)$ the kernel
that is obtained from the kernel $\unK_{z,z',\xi}(x,y)$ of \S3
by the procedure described above. That is,
$K_{z,z',\xi}=(\unK_{z,z',\xi})^{\circ,\varepsilon}$ with
$\varepsilon$ given by \tht{8.3}. This is a correlation kernel
for the z--measure $M_{z,z',\xi}$, corresponding to the map
$\la\mapsto X(\la)$. Clearly, $K_{z,z',\xi}(x,y)$ coincides with
the specialization of the kernel of Theorem 8.2 at $s=t$. Let us
abbreviate
$$
\psi_{\pm\frac12}(x)=\psi_{\pm\frac12}(x;z,z',\xi), \qquad
\wt\psi_{\pm\frac12}(x)=\psi_{\pm\frac12}(x;-z,-z',\xi).
$$
We have for $x,y\in\Z'_+$ (cf. \tht{3.13})
$$
\gathered K_{z,z',\xi}(x,y)=\frac{\sqrt{zz'\xi}}{1-\xi}\;
\frac{\psi_{-\frac12}(x)\psi_{\frac12}(y)
-\psi_{\frac12}(x)\psi_{-\frac12}(y)}{x-y} \\
K_{z,z',\xi}(x,-y)=\frac{\sqrt{zz'\xi}}{1-\xi}\;
\frac{\psi_{-\frac12}(x)\wt\psi_{-\frac12}(y)
+\psi_{\frac12}(x)\wt\psi_{\frac12}(y)}{x+y}\\
K_{z,z',\xi}(-x,y)=-\,\frac{\sqrt{zz'\xi}}{1-\xi}\;
\frac{\wt\psi_{\frac12}(x)\psi_{\frac12}(y)
+\wt\psi_{-\frac12}(x)\psi_{-\frac12}(y)}{x+y}\\
K_{z,z',\xi}(-x,-y)=\frac{\sqrt{zz'\xi}}{1-\xi}\;
\frac{\wt\psi_{-\frac12}(x)\wt\psi_{\frac12}(y)
-\wt\psi_{\frac12}(x)\wt\psi_{-\frac12}(y)}{x-y}
\endgathered \tag8.4
$$
Indeed, the first three formulas are easily obtained from
\tht{3.13}, and for the last formula we use the symmetry
relation
$$
K_{z,z',\xi}(-x,-y)=K_{-z,-z',\xi}(x,y), \qquad x,y\in\Z'_+\,.
$$
These four formulas coincide with the expressions obtained in
\cite{BO2, Theorem 3.3}.
\endexample

\head 9. Limit transition to the Whittaker kernel
\endhead
In this section we compute the scaling limit of the kernel
$K_{z,z',\xi}$ of \S8 as $\xi\nearrow 1$ and the arguments $x$
and $y$ are scaled by $(1-\xi)$. In this limit the lattice $\Z'$
turns into the punctured real line $\R^*=\R\setminus \{0\}$.

Let us introduce the continuous analogs of the functions
$\psi_a$. These new functions $w_a(u;z,z')$ are indexed by
$a\in\Z'$ and the argument $u$ varies in $\R_{>0}$. They are
expressed through the classical Whittaker functions
$W_{\kappa,\mu}(u)$, see \cite{Er, ch. 6} for the definition, as
follows:
$$
w_a(u;z,z')=\left(\Gamma(z-a+\tfrac 12)\Gamma(z'-a+\tfrac
12)\right)^{-\frac 12}\,u^{-\frac 12}\,
W_{\frac{z+z'}2-a,\frac{z-z'}2}(u). \tag9.1
$$
Since $W_{\kappa,\mu}(u)=W_{\kappa,-\mu}(u)$, this expression is
symmetric with respect to $z\leftrightarrow z'$.

It will be convenient for us to use the following integral
representation of $w_a$:
$$
\multline w_a(u;z,z') =\frac{\Gamma(z'-a+\frac 12)e^{\pi
i(z'-a)}u^{\frac{z-z'}2}}{2\pi\left(\Gamma(z-a+\frac
12)\Gamma(z'-a+\frac 12)\right)^{\frac12}}\\ \times
\int^{0-}_{+\infty} \zeta^{-z'+a-\frac 12}(1+\zeta)^{z-a-\frac
12}e^{-u(\zeta+\frac 12)}d\zeta.
\endmultline
\tag9.2
$$
The (standard) notation for the contour of integration means
that we start at $+\infty$, go along the real axis, then around
the origin in the clockwise direction, and back to $+\infty$
along the real axis. On the last part of the contour we choose
the principal branch of $\zeta^{-z+a-\frac 12}$, which uniquely
determines the values of this function on the whole contour.

This formula is easily seen to be equivalent to one of the
classical integral representations for the confluent
hypergeometric function $\Psi$, see \cite{Er, 6.11.2(9)}.

\proclaim{Proposition 9.1} If $\xi\nearrow 1$ and $x\in\Z'_+$
goes to $+\infty$ so that $(1-\xi)x\to u>0$, then
$$
\psi_a(x;z,z',\xi)\sim (1-\xi)^{\frac 12} w_a(u;z,z').
$$
\endproclaim

\demo{Proof} This statement can be proved in a number of ways,
see e.g. \cite{Er, 6.8(1)}. We will give an argument which uses
the integral representations of $\psi_a$ and $w_a$. A similar
argument will also be employed in the proof of Theorem 9.2
below.

We start with the integral representation \tht{2.4} for
$\psi_a$. Let us choose as $\{\omega\}$ the following contour
$C(R,r,\xi)$, where $r>0$ is small enough (smaller than the
distance between $\sqrt\xi$ and $1/\sqrt\xi$) and $R>0$ is big
enough (greater than $1/\sqrt\xi+r$): The contour starts at the
point $\om=R$, goes along the full circle $|\omega|=R$ in the
positive direction, then along the real line until the point
$\om=1/\sqrt{\xi}+r$, further along the full circle
$|\om-1/\sqrt{\xi}|=r$ in the negative direction, and back along
the real line to $\om=R$. Thus, $C(R,r,\xi)$ consists of a ``big
circle'' of radius $R$, a ``small circle'' of radius $r$, and a
``bridge'' between them.

We now fix $R$, pick $r$ of order $(1-\xi)$, and take the limit
$\xi\nearrow 1$ of the integral. The integration over the ``big
circle'' $|\om|=R$ converges to zero exponentially in
$(1-\xi)^{-1}$ thanks to the factor $\om^{-x-a}$. To take care
of the rest of the integral, we make the change of the
integration variable
$$
\om=1/\sqrt{\xi}+(1-\xi)\,\zeta.
$$
Then we have
$$
1-\sqrt{\xi}\om= -(1-\xi)\zeta\cdot\sqrt{\xi},\quad
1-\sqrt{\xi}/\om= (1-\xi)(1+\zeta)\cdot
\frac{(1/\sqrt{\xi}-\sqrt{\xi})(1-\xi)^{-1}+\zeta}
{(1+\zeta)(1/\sqrt{\xi}+(1-\xi)\zeta)}.
$$
Note that the second factors in the formulas above are
asymptotically equal to 1 for $\xi$ close to 1 and $\zeta$
bounded, and are uniformly bounded away from 0 and $\infty$ for
$\xi$ close to 1 and $\zeta$ corresponding to arbitrary $\om$ on
the contour. Hence, the rest of the integral is asymptotically
equal to the absolutely convergent integral
$$
\frac{(1-\xi)^{z-z'}}{2\pi
i}\,\int_{+\infty}^{0-}(-\zeta)^{-z'+a-\frac
12}(1+\zeta)^{z-a-\frac 12}e^{-u(\zeta+\frac 12)}d\zeta.
$$
Taking into account the convention stated in Comment
1 to Lemma 2.2 one can check that $\arg(-\zeta)=-\pi i$ on the
last part of the contour. Therefore, changing
$(-\zeta)^{-z'+a-\frac 12}$ to $\zeta^{-z'+a-\frac 12}$ produces
the factor
$$
e^{\pi i(z'-a+\frac 12)}=i\, e^{\pi i(z'-a)}\,.
$$
Finally, the prefactor in \tht{2.4} asymptotically equals
$$
(1-\xi)^{z'-z+\frac12}\,\frac{\Gamma(z'-a+\frac
12)u^{\frac{z-z'}2}}{\left(\Gamma(z-a+\frac 12)\Gamma(z'-a+\frac
12)\right)^{\frac12}}\,.
$$
Thus, \tht{2.4} asymptotically equals
$$
\multline (1-\xi)^{\frac12}\,\frac{\Gamma(z'-a+\frac 12)e^{\pi
i(z'-a)}u^{\frac{z-z'}2}}{2\pi\left(\Gamma(z-a+\frac
12)\Gamma(z'-a+\frac 12)\right)^{\frac12}}\\ \times
\int^{0-}_{+\infty} \zeta^{-z'+a-\frac 12}(1+\zeta)^{z-a-\frac
12}e^{-u(\zeta+\frac 12)}d\zeta.
\endmultline
$$
\qed
\enddemo

\proclaim{Theorem 9.2} Consider the extended hypergeometric
kernel $K_{z,z',\xi}(s,x;t,y)$ as described in Theorem 8.2. Let
$\xi\nearrow 1$ and assume that $x,y\to \infty$ inside $\Z'$ so
that $(1-\xi)x\to u$, $(1-\xi)y\to v$, where $u,v\in\R^*$.

Then there exists a limit kernel $K^W_{z,z'}(s,u;t,v)$ on
$\R^*\times \R^*$:
$$
\lim_{\xi\nearrow1}(1-\xi)^{-1}K_{z,z',\xi}(s,x;t,y)=K^W_{z,z'}(s,u;t,v).
\tag9.3
$$

For $s\ne t$ the formulas for the limit kernel are obtained from
formulas \tht{8.1} for the kernel $K_{z,z',\xi}$ by replacing
$\psi_a$'s with $w_a$'s and setting $\xi=1$: for $u,v>0$
$$
\align &K_{z,z'}^W(s,u;t,v)=\pm\sum_{a\in\Z'_+}
e^{-a|s-t|}w_{\pm a}(u;z,z')w_{\pm a}(v;z,z'),\\
&K_{z,z'}^W(s,u;t,-v)=\pm\sum_{a\in\Z'_+} (-1)^{\pm a-\frac
12}e^{-a|s-t|}w_{\pm a}(u;z,z')w_{\mp
a}(v;-z,-z'),\\
&K_{z,z'}^W(s,-u;t,v)=\mp\sum_{a\in\Z'_+}(-1)^{\pm a-\frac 12}
e^{-a|s-t|}w_{\mp a}(u;-z,-z')w_{\pm
a}(v;z,z'),\\
&K_{z,z'}^W(s,-u;t,-v)=\mp\sum_{a\in\Z'_+} e^{-a|s-t|}w_{\mp
a}(u;-z,-z')w_{\mp a}(v;-z,-z').
\endalign
$$
\endproclaim

\demo{Comments} 1. The prefactor $(1-\xi)^{-1}$ in \tht{9.3} is
due to rescaling of the state space $\Z'$ by $(1-\xi)$.

2. The reason of the restriction $s\ne t$ in above formulas is
the divergence of the series for $K_{z,z'}^W(s,u;s,-v)$ and
$K_{z,z'}^W(s,-u;s,v)$. The series for $K_{z,z'}^W(s,u;s,v)$ and
$K_{z,z'}^W(s,-u;s,-v)$ do converge and give the correct answer.
For $s=t$ there exist analogs of formulas \tht{8.4}:
$$
\gathered K^W_{z,z'}(u,v)=\sqrt{zz'}\;
\frac{w_{-\frac12}(u)w_{\frac12}(v)
-w_{\frac12}(u)w_{-\frac12}(v)}{u-v} \\
K^W_{z,z'}(u,-v) =\sqrt{zz'}\; \frac{w_{-\frac12}(u)\wt
w_{-\frac12}(v)
+w_{\frac12}(u)\wt w_{\frac12}(v)}{u+v}\\
K^W_{z,z'}(-u,v)=-\,\sqrt{zz'}\; \frac{\wt
w_{\frac12}(u)w_{\frac12}(v)
+\wt w_{-\frac12}(u)w_{-\frac12}(v)}{u+v}\\
K^W_{z,z'}(-u,-v)=\sqrt{zz'}\; \frac{\wt w_{-\frac12}(u)\wt
w_{\frac12}(v) -\wt w_{\frac12}(u)\wt w_{-\frac12}(v)}{u-v}
\endgathered \tag9.4
$$
Here we abbreviate
$$
w_{\pm\frac12}(u)=w_{\pm\frac12}(u;z,z'), \qquad \wt
w_{\pm\frac12}(u)=w_{\pm\frac12}(u;-z,-z').
$$
Formulas \tht{9.4} can be derived from \tht{8.4} using
Proposition 9.1. They were previously obtained in \cite{B1},
\cite{BO2, \S5}.

3. In accordance with the terminology of these papers (where the
kernel \tht{9.4} was called the Whittaker kernel) we call the
limit kernel $K^W_{z,z'}(s,u;t,v)$ the {\it extended Whittaker
kernel}.
\enddemo

\demo{Proof} We use Proposition 9.1. In order to prove the
theorem, we need to justify the interchange of the summation and
the limit transition in \tht{8.1}. To do this it suffices to
show that the series converge uniformly in $\xi$.

We will prove that each of the four expressions
$$
\gather |\psi_{\pm a}(x;z,z',\xi)\psi_{\pm a}(y;z,z',\xi)|,\quad
|\psi_{\mp a}(x;-z,-z',\xi)\psi_{\mp a}(y;-z,-z',\xi)|,\tag9.5\\
|\psi_{\pm a}(x;z,z',\xi)\psi_{\mp a}(y;-z,-z',\xi)|,\quad
|\psi_{\mp a}(x;-z,-z',\xi)\psi_{\pm a}(y;z,z',\xi)|,\tag9.6
\endgather
$$
is estimated from above by $\const(u,v)(1-\xi)\cdot q^{|a|}$,
where $q>1$ can be chosen arbitrarily close to 1, and
$\const(u,v)$ does not depend on $a$ and $\xi$. Together with
the factors $e^{-a|s-t|}$ in \tht{8.1} this ensures the needed
uniform convergence.

Both expressions \tht{9.5} are estimated in the same way, let us
handle the first one. We apply formula \tht{2.5} of Proposition
2.3 and we get a double contour integral, in which we single out
the terms involving $a$; we observe that all together they can
be written in the form $(F(\om_1,\om_2;\xi))^{k}$, where
$k:=a-\frac12$ ranges over $\Z_+$, and $\om_1$ and $\om_2$ are
the variables of integration. Let us write down precisely the
whole expression separately for the upper and the lower choice
of sign in the subscript $\pm a$:
$$
\gathered \psi_{a}(x;z,z',\xi)\psi_{a}(y;z,z',\xi)\\
= \frac{(\Gamma(x+z+\frac 12)\Gamma(x+z'+\frac
12)\Gamma(y+z+\frac 12)\Gamma(y+z'+\frac 12))^{\frac
12}}{\Gamma(x+z'+\frac 12)\Gamma(y+z+\frac 12)}\, (1-\xi)\\
\times\frac1{(2\pi i)^2}\,
\oint_{\{\om_1\}}\oint_{\{\om_2\}}(F_{++}(\om_1,\om_2;\xi))^k
\left(1-\sqrt{\xi}\om_1\right)^{-z'}
\left(1-\frac{\sqrt{\xi}}{\om_1}\right)^{z-1}\\
\times\left(1-\sqrt{\xi}\om_2\right)^{-z}
\left(1-\frac{\sqrt{\xi}}{\om_2}\right)^{z'-1}\,
\om_1^{-x-\frac12}\om_2^{-y-\frac12}\,\frac{d\om_1}{\om_1}\,
\frac{d\om_2}{\om_2}
\endgathered \tag9.7
$$
with
$$
F_{++}(\om_1,\om_2;\xi)=F_+(\om_1;\xi)F_+(\om_2,\xi),
$$
where
$$
F_+(\om,\xi)=\frac{1-\sqrt\xi\om}{\om-\sqrt\xi}\,.\tag9.8
$$

For $\psi_{-a}(x;z,z',\xi)\psi_{-a}(y;z,z',\xi)$ we obtain a
similar expression:
$$
\gathered \psi_{-a}(x;z,z',\xi)\psi_{-a}(y;z,z',\xi)\\
= \frac{(\Gamma(x+z+\frac 12)\Gamma(x+z'+\frac
12)\Gamma(y+z+\frac 12)\Gamma(y+z'+\frac 12))^{\frac
12}}{\Gamma(x+z'+\frac 12)\Gamma(y+z+\frac 12)}\, (1-\xi)\\
\times\frac1{(2\pi i)^2}\,
\oint_{\{\om_1\}}\oint_{\{\om_2\}}(F_{--}(\om_1,\om_2;\xi))^k
\left(1-\sqrt{\xi}\om_1\right)^{-z'-1}
\left(1-\frac{\sqrt{\xi}}{\om_1}\right)^{z}\\
\times\left(1-\sqrt{\xi}\om_2\right)^{-z-1}
\left(1-\frac{\sqrt{\xi}}{\om_2}\right)^{z'}\,
\om_1^{-x+\frac12}\om_2^{-y+\frac12}\,\frac{d\om_1}{\om_1}\,
\frac{d\om_2}{\om_2}
\endgathered \tag9.9
$$
with
$$
F_{--}(\om_1,\om_2;\xi)=F_-(\om_1;\xi)F_-(\om_2,\xi),
$$
where
$$
F_-(\om,\xi)=(F_+(\om,\xi))^{-1}=\frac{\om-\sqrt\xi}{1-\sqrt\xi\om}\,.
\tag9.10
$$

Now we need a lemma.

\proclaim{Lemma 9.3} Let $F_\pm(\om,\xi)$ be defined by
\tht{9.8} and \tht{9.10}. For any $q>1$ there exists a contour
$C_\pm(\xi, q)$ which is of the same kind as in the proof of
Proposition 9.1, and such that
$$
|F_\pm(\om,\xi)|\le q \qquad \forall \om\in C_\pm(\xi, q).
\tag9.11
$$
\endproclaim

\demo{Proof of Lemma 9.3} Recall that in the proof of
Proposition 9.1 we used a specific family $\{C(R,r,\xi)\}$ of
contours. We will show that it is possible to take $C_\pm(\xi,
q)=C(R,r,\xi)$ with an appropriate choice of parameters $R$ and
$r$ (the radii of the ``big circle'' and the ``small circle'' in
$C(R,r,\xi)$).

Consider first the case of $F_+$. Fix $q>1$ and let $\wt \om$ be
related to $\om$ by the equivalent relations
$$
\wt\om=F_+(\om,\xi)/q=\frac{1-\sqrt\xi\om}{q\om-q\sqrt\xi}\,,
\qquad \om=\frac{1+q\sqrt\xi\wt\om}{q\wt\om+\sqrt\xi}.
$$
To fulfill inequality \tht{9.11} the contour $C_+(\xi,q)$ must
be contained in the image of the unit disk $|\wt\om|\le1$ under
the conformal map $\wt\om\mapsto\om$. Let $S_+(\xi,q)$ denote
the image of the circle $|\wt\om|=1$; $S_+(\xi,q)$ is the circle
that is symmetric with respect to the real axis and passes
through the real points
$$
\frac{q\sqrt\xi-1}{q-\sqrt\xi}\,, \qquad
\frac{q\sqrt\xi+1}{q+\sqrt\xi}
$$
(these are the images of $-1$ and $1$, respectively). Since we
are interested in the limit transition as $\xi\nearrow1$ we may
assume that $\xi$ is so close to 1 that $q>1/\sqrt\xi$. Then we
have
$$
\frac{q\sqrt\xi-1}{q-\sqrt\xi}\,<\sqrt\xi\,<\,
\frac{q\sqrt\xi+1}{q+\sqrt\xi}\,<\,\frac1{\sqrt\xi}\,. \tag9.12
$$
Observe that the image of the disk $|\wt\om|\le1$ is the
exterior of $S_+(\xi,q)$ (for instance, this follows from the
fact that the image of 0 is the point $1/\sqrt\xi$ which is
outside $S_+(\xi,q)$). Now we take $C_+(\xi,q)=C(R,r,\xi)$,
where $R$ and $r$ are chosen so that  a both the ``big circle''
and the ``small circle'' in $C(R,r,\xi)$ are in the exterior of
the circle $S_+(\xi,q)$: the ``big circle'' surrounds
$S_+(\xi,q)$ while the ``small circle'' lies to the right of
$S_+(\xi,q)$. This is possible due to inequalities \tht{9.12}
and the fact that the distance between the points
$\frac{q\sqrt\xi+1}{q+\sqrt\xi}$ and $\frac1{\sqrt\xi}$ is of
order $1-\xi$.

The case of $F_-$ is handled analogously. We define $\wt\om$ by
the equivalent relations
$$
\wt\om=F_-(\om,\xi)/q=\frac{\om-\sqrt\xi}{q-q\sqrt\xi\om}\,,
\qquad \om=\frac{q\wt\om+\sqrt\xi}{q\sqrt\xi\om+1}\,.
$$
Instead of the circle $S_+(\xi,q)$ we have another circle,
denoted by $S_-(\xi,q)$, which is symmetric with respect to the
real axis and passes through the points
$$
\frac{q+\sqrt\xi}{q\sqrt\xi+1}\,, \qquad
\frac{q-\sqrt\xi}{q\sqrt\xi-1}\,.
$$
We note that
$$
\sqrt\xi\,<\,\frac{q+\sqrt\xi}{q\sqrt\xi+1}\,<\,\frac1{\sqrt\xi}\,<\,
\frac{q-\sqrt\xi}{q\sqrt\xi-1}.
$$
The contour $C_-(\xi,q)$ must lie in the exterior of
$S_-(\xi,q)$. We again can take $C_-(\xi,q)=C(R,r,\xi)$ with
appropriate $R$ and $r$. But, in contrast to the case of $F_+$,
now the ``small circle'' in $C(R,r,\xi)$ must surround
$S_-(\xi,q)$ (because $1/\sqrt\xi$ is inside $S_-(\xi,q)$, see
the inequalities above). This requirement can be satisfied
because the diameter of $S_-(\xi,q)$ (the distance between the
points $\frac{q+\sqrt\xi}{q\sqrt\xi+1}$ and
$\frac{q-\sqrt\xi}{q\sqrt\xi-1}$) is of order $(1-\xi)$.

This completes the proof of Lemma 9.3.\qed
\enddemo

We return to the proof of Theorem 9.2. Let us estimate
\tht{9.7}. The product of the prefactors is asymptotically
$$
(1-\xi)u^{\frac12(z-z')}v^{\frac12(z'-z)}\,.
$$
To estimate the integral we take as contours $\{\om_1\}$ and
$\{\om_2\}$ the contour $C_+(\xi,q^{\frac12})$ as described in
Lemma 9.3. According to Lemma 9.3, on the product of these
contours, $|F_{++}(\om_1,\om_2;\xi)|\le q$, whence we get
$$
\gather \oint\limits_{\{\om_1\in C_+(\xi,
q^{\frac12}\}}\oint\limits_{\{\om_2\in
C_+(\xi,q^{\frac12)}\}}\bigg|F_{++}(\om_1,\om_2;\xi)^k
\left(1-\sqrt{\xi}\om_1\right)^{-z'}
\left(1-\frac{\sqrt{\xi}}{\om_1}\right)^{z-1}\\
\times\left(1-\sqrt{\xi}\om_2\right)^{-z}
\left(1-\frac{\sqrt{\xi}}{\om_2}\right)^{z-1}\,
\om_1^{-x-\frac12}\om_2^{-y-\frac12}\,\frac{d\om_1}{\om_1}\,
\frac{d\om_2}{\om_2}\bigg|\\
\le q^k \oint\limits_{\{\om_1\in C_+(\xi,
q^{\frac12})\}}\oint\limits_{\{\om_2\in
C_+(\xi,q^{\frac12})\}}\bigg|
\left(1-\sqrt{\xi}\om_1\right)^{-z'}
\left(1-\frac{\sqrt{\xi}}{\om_1}\right)^{z-1}\\
\times\left(1-\sqrt{\xi}\om_2\right)^{-z}
\left(1-\frac{\sqrt{\xi}}{\om_2}\right)^{z-1}\,
\om_1^{-x-\frac12}\om_2^{-y-\frac12}\,\frac{d\om_1}{\om_1}\,
\frac{d\om_2}{\om_2}\bigg|
\endgather
$$
Arguing as in the proof of Proposition 9.1 we check that the
last integral is uniformly bounded as $\xi\nearrow1$. This
yields for \tht{9.7} the required estimate of the form $\const
(1-\xi)q^k$ with arbitrary $q>1$. The estimate for \tht{9.9} is
obtained in exactly the same way, by using the contours
$\{\om_1\}=\{\om_2\}=C_-(\xi,q^{\frac12})$.

The quantities in \tht{9.6} are estimated similarly. We leave
the details to the reader, and only point out two minor
differences. First, while writing the double integral
representation, we do not need to switch parameters $z$ and $z'$
(as in the proof of Proposition 2.3) to get rid of the gamma
factors containing $a$. Second, we have to use as $\{\om_1\}$
and $\{\om_2\}$ two distinct contours: either
($C_+(\xi,q^{\frac12})$ and $C_-(\xi,q^{\frac12})$) or
($C_-(\xi,q^{\frac12})$ and $C_+(\xi,q^{\frac12})$). \qed
\enddemo

Our next goal is to give an integral representation for the
extended Whittaker kernel computed in Theorem 9.2. Let us
introduce contours $C_\pm(q)$ for $q>1$. They can be viewed as
limits of the images of the contours $C_\pm(\xi,q)$ as
$\xi\nearrow 1$ in the new variable $\zeta$ where
$$
\omega= 1/\sqrt{\xi}+(1-\xi)\zeta.
$$

The contour $C_+(q)$ starts at $+\infty$, goes along the real
axis, circles around 0 in the negative direction, and returns to
$+\infty$ along the real axis. It has to leave on its left the
point $-1$ together with the circle which is symmetric with
respect to the real axis and passes through the points
$-q/(q-1)$ and $-q/(q+1)$ (this circle contains $-1$).

The contour $C_-(q)$ also starts at $+\infty$, goes along the
real axis, circles around 0 in the negative direction, and
returns to $+\infty$ along the real axis. It has to leave on its
left the point $-1$, and it has to leave on its right the circle
which is symmetric with respect to the real axis and passes
through the points $-1/(q+1)$ and $1/(q-1)$ (this circle
contains $0$).

Note that if $\zeta\in C_+(q)$ then $|\zeta/(1+\zeta)|<q$, and
if $\zeta\in C_-(q)$ then $|(1+\zeta)/\zeta|<q$.

\proclaim{Theorem 9.4} The extended Whittaker kernel
$K_{z,z'}^W(s,u;t,v)$ of Theorem 9.2 for $s\ne t$ has the
following integral representation $(u>0,v>0)$:
$$
\multline
 K^W_{z,z'}(s,u;t,v)=e^{\pi i(z+z')}(u/v)^{\frac
{z-z'}2}e^{\frac 12(s-t)}\\
\times\frac 1{(2\pi i)^2}\oint\limits_{+\infty}^{0-}
\oint\limits_{+\infty}^{0-}\zeta_1^{-z'}(1+\zeta_1)^z\zeta_2^{-z}(1+\zeta_2)^{z'}
\frac{e^{-u(\zeta_1+\frac12)-v(\zeta_2+\frac12)}\,d\zeta_1
d\zeta_2} {e^{s-t}(1+\zeta_1)(1+\zeta_2)-\zeta_1\zeta_2}
\endmultline
$$
where for $s>t$ both contours $\{\zeta_1\}$ and $\{\zeta_2\}$
are of the form $C_+(e^{\frac 12(s-t)})$, and for $s<t$ both
contours are of the form $C_-(e^{\frac 12(t-s)})$;
$$
\multline
 K^W_{z,z'}(s,u;t,-v)=\frac{(\sin(\pi z) \sin(\pi z'))^{\frac 12}}{\sin(\pi z')}
 \,(u/v)^{\frac
{z-z'}2}e^{\frac 12(s-t)}\\
\times\frac 1{(2\pi)^2}\oint\limits_{+\infty}^{0-}
\oint\limits_{+\infty}^{0-}\zeta_1^{-z'}(1+\zeta_1)^z\zeta_2^{-z'}(1+\zeta_2)^{z}
\frac{e^{-u(\zeta_1+\frac12)-v(\zeta_2+\frac12)}\,d\zeta_1
d\zeta_2} {e^{s-t}(1+\zeta_1)\zeta_2-\zeta_1(1+\zeta_2)}
\endmultline
$$
where for $s> t$, the contour $\{\zeta_1\}$ is of the form
$C_+(e^{\frac 12(s-t)})$ and the contour $\{\zeta_2\}$ is of the
form $C_-(e^{\frac 12(s-t)})$, while for $s<t$ the contour
$\{\zeta_1\}$ is of the form $C_-(e^{\frac 12(t-s)})$ and the
contour $\{\zeta_2\}$ is of the form $C_+(e^{\frac 12(t-s)})$;
$$
\gather K^W_{z,z'}(s,-u;t,v)=-K^W_{z,z'}(s,v;t,-u);\\
K^W_{z,z'}(s,-u;t,-v)=K^W_{-z,-z'}(t,u;s,v).
\endgather
$$
The contours may be chosen differently by deforming the contours
above so that the denominators of the integrands do not vanish.
\endproclaim

\demo{Proof} We take the integral representation \tht{9.2} for
the functions $w_a$ and plug it in into the series of Theorem
9.2. Computing the sum of geometric progression under the
integral yields the formulas above. The contours are chosen in
such a way that the absolute values  of the ratios of geometric
progressions involved are less than one.

As in the proof of Proposition 2.3, in the derivation of the
first formula we switch $z$ and $z'$ in the integral
representation of the second factor, which cancels the gamma
factors involving the summation index. In the derivation of the
second formula we do not need to do that, the gamma factors
disappear thanks to the relation
$\Gamma(x)\Gamma(1-x)=\pi/\sin(\pi x)\,.$\qed
\enddemo

\head 10. Limit transition to the gamma kernel
\endhead

In this section we compute the limit of the extended
hypergeometric kernel $K_{z,z',\xi}(s,x;t,y)$ as $\xi\nearrow 1$
and scaling of time $s=(1-\xi)\sigma$, $t=(1-\xi)\tau$ with
finite $\sigma, \tau\in\R$.

\proclaim{Theorem 10.1} There exists a limit of the extended
hypergeometric kernel
$$
\unK^{\g}_{z,z'}(\sigma,x;\tau,y)=\lim_{\xi\to 1}
\unK_{z,z',\xi}((1-\xi)\sigma,x;(1-\xi)\tau,y)
$$
where $x,y\in\Z'$, $\sigma,\tau\in\R$.

For $\sigma\ge \tau$, the correlation kernel can be written in
two different ways: as a double contour integral
$$
\multline \unK^{\g}_{z,z'}(\sigma,x;\tau,y)\\=
 \frac{\Ga(-z'-x+\frac12)\Ga(-z-y+\frac12)e^{-\pi i(z+z')}(-1)^{x+y+1}}
 {\bigl(\Ga(-z-x+\frac12)\Ga(-z'-x+\frac12)
 \Ga(-z-y+\frac12)\Ga(-z'-y+\frac12)\bigr)^{\frac12}}\\
\times\frac 1{(2\pi i)^2}\oint\limits_{+\infty}^{0-}
\oint\limits_{+\infty}^{0-}
\frac{\zeta_1^{z'+x-\frac12}(1+\zeta_1)^{-z-x-\frac12}
\zeta_2^{z+y-\frac12}(1+\zeta_2)^{-z'-y-\frac12}\,d\zeta_1
d\zeta_2} {1+(\sigma-\tau)+\zeta_1+\zeta_2}
\endmultline
\tag10.1
$$
and as a single integral
$$
\unK^{\g}_{z,z'}(\sigma,x;\tau,y)=\int_0^{+\infty}
e^{-u(\sigma-\tau)}w_x(u;-z,-z') w_y(u;-z,-z')du. \tag10.2
$$
The values of the kernel for $\sigma< \tau$ are obtained from
the above formulas using the symmetry property
$$
\unK^{\g}_{z,z'}(\sigma,x;\tau,y)
=(-1)^{x+y+1}\unK^{\g}_{-z,-z'}(\tau,-x;\sigma,-y),\qquad
\sigma\ne \tau.
$$
\endproclaim

\demo{Comment} For $\sigma=\tau$ the kernel $\unK_{z,z'}^{\g}$
has a simpler ``integrable'' expression
$$
\multline \unK_{z,z'}^{\g}(x;y)=\frac{\sin(\pi z)\sin(\pi
z')}{\pi\sin(\pi(z-z'))} \\
\times \left\{\Ga(z+x+\tfrac12)\Ga(z'+x+\tfrac12)
\Ga(z+y+\tfrac12)\Ga(z'+y+\tfrac12)\right\}^{-1/2}\\
\times \frac{\Ga(z+x+\tfrac12)\Ga(z'+y+\tfrac12)-
\Ga(z'+x+\tfrac12)\Ga(z+y+\tfrac12)}{x-y}\,,
\endmultline \tag10.3
$$
see \cite{BO5, Theorem 2.3}. We called this kernel the {\it
gamma kernel}. The more general kernel
$\unK_{z,z'}^{\g}(\sigma,x;\tau,y)$ of Theorem 10.1 will be
called the {\it extended gamma kernel}.
\enddemo

\demo{Proof} We start with the series representation of the
extended hypergeometric kernel, Theorem 7.2. Using the symmetry
relations of Propositions 2.5 and 2.7, we rewrite the kernel in
the following form
$$
\multline \unK_{z,z',\xi}(s,x;t,y)\\=\cases
\sum\limits_{a\in\Z_+'}e^{-a|s-t|}
\psi_{x}(a;-z,-z',\xi)\,\psi_{y}(a;-z,-z',\xi),&s\ge t,\\
(-1)^{x+y+1}\sum\limits_{a\in\Z_+'}e^{-a|s-t|}
\psi_{-x}(a;z,z',\xi)\,\psi_{-y}(a;z,z',\xi),&s< t.
\endcases
\endmultline
$$
This formula implies that if we prove the statement for
$\sigma\ge \tau$, then the case $\sigma<\tau$ will follow by
symmetry. Thus, we continue with the assumption $s\ge t$ and
therefore $\sigma\ge \tau$.

Formula \tht{10.2} is the limit variant of the formula for
$\unK_{z,z',\xi}(s,x;t,y)$ above. Indeed, by virtue of
Proposition 9.1,
$$
\psi_{x}(a;-z,-z',\xi)\,\psi_{y}(a;-z,-z',\xi)\sim
(1-\xi)w_{x}(u;-z,-z')\,w_{y}(u;-z,-z')
$$
provided that $a\sim (1-\xi)^{-1}u$. The factor $(1-\xi)$ is
responsible for turning the sum into an integral over $u$; this
sum is just an approximation to the integral.

This empirical argument needs a rigorous justification. It is
simpler to turn the series representation for
$\unK_{z,z',\xi}(s,x;t,y)$ into a double contour integral and
then pass to the limit in the integral. The limit integral will
be identified with the right-hand side of \tht{10.2}.

Using formula \tht{2.5} with appropriately changed parameters,
we obtain
$$
\multline \unK_{z,z',\xi}(s,x;t,y)\\
=\frac{\Gamma(-z'-x+\frac12)\Gamma(-z-y+\frac12)}
{\bigl(\Gamma(-z-x+\frac12)\Gamma(-z'-x+\frac12)
\Gamma(-z-y+\frac12)\Gamma(-z'-y+\frac12)\bigr)^{\frac12}} \\
\times\frac{1-\xi}{(2\pi i)^2}\sum_{a\in\Z'_+}\oint\oint
e^{-a(s-t)}\left(1-\sqrt{\xi}\om_1\right)^{z'+x-\tfrac12}
\left(1-\frac{\sqrt{\xi}}{\om_1}\right)^{-z-x-\tfrac12}\\
\times\left(1-\sqrt{\xi}\om_2\right)^{z+y-\tfrac12}
\left(1-\frac{\sqrt{\xi}}{\om_2}\right)^{-z'-y-\tfrac12}
\om_1^{-x-a}\om_2^{-y-a}
\,\frac{d\om_1}{\om_1}\frac{d\om_2}{\om_2}
\endmultline
$$
with contours $\{\om_1\}$ and $\{\om_2\}$ chosen as in
Proposition 2.3. Now we want to sum the geometric progression
inside the integrals. The ratio of the geometric progression is
$e^{t-s}\om_1^{-1}\om_2^{-1}$. In order to justify the
interchange of summation and integration we need to ensure that
the absolute value of this ratio, as a function in $\om_1$,
$\om_2$, is bounded from above by a constant strictly less than
one. This is easy to arrange by requiring, for example, that
both contours contain the unit circle.

Performing the summation, we obtain
$$
\multline \unK_{z,z',\xi}(s,x;t,y)\\
=\frac{e^{\frac
12(s-t)}\Gamma(-z'-x+\frac12)\Gamma(-z-y+\frac12)}
{\bigl(\Gamma(-z-x+\frac12)\Gamma(-z'-x+\frac12)
\Gamma(-z-y+\frac12)\Gamma(-z'-y+\frac12)\bigr)^{\frac12}} \\
\times\frac{1-\xi}{(2\pi i)^2}\oint\oint
\left(1-\sqrt{\xi}\om_1\right)^{z'+x-\tfrac12}
\left(1-\frac{\sqrt{\xi}}{\om_1}\right)^{-z-x-\tfrac12}\\
\times\left(1-\sqrt{\xi}\om_2\right)^{z+y-\tfrac12}
\left(1-\frac{\sqrt{\xi}}{\om_2}\right)^{-z'-y-\tfrac12}
\frac{\om_1^{-x-\frac 12}\om_2^{-y-\frac
12}}{e^{s-t}\omega_1\omega_2-1} \,{d\om_1}{d\om_2}.
\endmultline
$$
Recall the notation $C(R,r,\xi)$ for certain type of contours
introduced in the proof of Proposition 9.1. We assume that both
integration variables range over such a contour with $R$ being a
fixed number greater than 1, and $r$ being of order $1-\xi$, and
such that $1/\sqrt{\xi}-r>1$.

Let us split each of the contours into two parts: the first one
is the ``big'' circle $|\om|=R$, and the second part is the
rest. If both $\om_1$ and $\om_2$ range over their big circles
then the integrand is uniformly bounded, and the prefactor
$(1-\xi)$ sends the whole expression to zero as $\xi\to 1$.

If one of the variables, say, $\om_2$ ranges over its big circle
and $\om_1$ ranges over the second part of its contour, we
observe that all the factors of the integrand involving $\om_2$
are uniformly bounded. The absolute value of the remaining part
of the integrand
$$
\left(1-\sqrt{\xi}\om_1\right)^{z'+x-\tfrac12}
\left(1-\frac{\sqrt{\xi}}{\om_1}\right)^{-z-x-\tfrac12}
$$
is uniformly bounded by
$$
\const\cdot\left|
(1-\xi)^{z'-z-1}\zeta_1^{z'+x-\tfrac12}(1+\zeta_1)^{-z-x-\tfrac12}\right|
$$
with $\omega_1=1/\sqrt{\xi}+(1-\xi)\zeta_1$, where we used the
same argument as in the proof of Proposition 9.1. Thus, our
double integral is bounded in absolute value by the following
one-dimensional integral in $\zeta_1$:
$$
\const\cdot(1-\xi)^{\Re(z'-z)} \oint_{\wt
R}^{0-}\left|\zeta_1^{z'+x-\tfrac12}(1+\zeta_1)^{-z-x-\tfrac12}d\zeta_1\right|
$$
with
$$
\wt R=1/\sqrt{\xi}+R(1-\xi)^{-1}\sim R(1-\xi)^{-1}.
$$
Hence, our expression is bounded by
$$
\const\cdot (1-\xi)^{\Re(z'-z)}\int_1^{\wt
R}\zeta_1^{\Re(z'-z)-1}d\zeta_1
$$
which is either bounded by a constant (if $\Re(z-z')\ne 0$) or
by $|\ln(1-\xi)|$ (if $\Re(z-z')=0$). In both cases, the
prefactor $(1-\xi)$ in the integral representation for
$\unK_{z,z',\xi}(s,x;t,y)$ sends the whole expression to zero.

The only asymptotically significant part of the integral comes
from the case when both $\om_1$ and $\om_2$ vary over the second
parts of their contours. Making the change of variables
$$
\omega_1=1/\sqrt{\xi}+(1-\xi)\zeta_1,\qquad
\omega_2=1/\sqrt{\xi}+(1-\xi)\zeta_2,
$$
and arguing as in the proof of Proposition 9.1, we conclude,
using the asymptotic relation
$$
\frac{1-\xi}{e^{s-t}\om_1\om_2-1}\sim \frac
1{1+(\sigma-\tau)+\ze_1+\ze_2},
$$
that the limit value of the kernel is given by the right-hand
side of \tht{10.1}. Note that the integral in \tht{10.1} is
absolutely convergent. To see this we use the estimate
$$
|1+(\sigma-\tau)+\zeta_1+\zeta_2|\ge \const\cdot
|\zeta_1|^\nu|\zeta_2|^{1-\nu}
$$
which holds for any $\zeta_1,\zeta_2$ on our contours and any
$\nu\in (0,1)$. We apply this inequality with
$\nu=\frac12+\Re(z'-z)/2$. The fact that $\nu\in(0,1)$ follows
from our basic assumptions on $z,z'$, see \S1.

Thus, we have proved the integral representation \tht{10.1}. To
see the equivalence of \tht{10.1} and \tht{10.2} we substitute
the integral representation \tht{9.2} into \tht{10.2} and
integrate explicitly over $u$. \qed
\enddemo

\bigskip

{\smc A.~Borodin}: Mathematics 253-37, Caltech, Pasadena, CA
91125, U.S.A.,

\medskip

E-mail address: {\tt borodin\@caltech.edu}

\bigskip

{\smc G.~Olshanski}: Dobrushin Mathematics Laboratory, Institute
for Information Transmission Problems, Bolshoy Karetny 19,
127994 Moscow GSP-4, RUSSIA.

\medskip

E-mail address: {\tt olsh\@online.ru}

\enddocument
\end